\documentclass[aps,pra,twocolumn,superscriptaddress]{revtex4}
\usepackage{graphicx}
\usepackage{dcolumn}
\usepackage{bm}
\usepackage{physics}
\usepackage{amsmath}
\usepackage{amssymb}
\usepackage{array}
\usepackage{color}
\usepackage{xcolor}
\newcolumntype{P}[1]{>{\centering\arraybackslash}p{#1}}
\usepackage[colorlinks=true, breaklinks=true, linkcolor=blue, citecolor=blue, urlcolor=blue]{hyperref}
\allowdisplaybreaks

\newcommand{\geneva}{Department of Quantum Matter Physics, University of Geneva, Quai Ernest-Ansermet 24, 1211 Geneva, Switzerland}

\usepackage{ifthen}
\usepackage{booktabs}
\usepackage{multirow}

\begin{document}

\title{Interaction dependence of the Hall response for the Bose-Hubbard triangular ladder}

\date{\today}

\begin{abstract}
We explore the behavior of the Hall response of a Bose-Hubbard triangular ladder in a magnetic field as a function of the repulsive on-site atomic interactions.
We consider a wide range of interaction strengths, from the weakly interacting limit to the hardcore regime.
This is realized by computing the Hall polarization following the quench of a weak linear potential which induces the flow of a current through the system, using time-dependent matrix product state numerical simulations.
We complement our understanding in the regime of small magnetic fields by analytical calculations of the equilibrium value of the Hall polarization for non-interacting bosonic atoms, or under a mean-field assumption.
The Bose-Hubbard triangular flux ladder exhibits a rich phase diagram, containing Meissner, vortex and biased-chiral superfluid phases.
We show that the Hall response can be employed to fingerprint the various chiral state, the frustration effects occurring in the limit of strong interactions, and the phase boundaries of the equilibrium phase diagram. 
\end{abstract}
\author{Catalin-Mihai Halati}
\affiliation{\geneva}
\author{Thierry Giamarchi}
\affiliation{\geneva}
\maketitle

\section{Introduction\label{sec:intro}}

Strongly correlated topological matter exhibits exotic properties like particles with fractional quantum numbers and anyonic exchange statistics, which offer a promising avenue for quantum computing applications \cite{Kitaev2003, NayakDasSarma2008}.
One of the paradigmatic example of topological quantum states is the fractional quantum Hall state \cite{TsuiGossard1982, Laughlin1983, StormerGossard1999}, stemming from the interplay of strong interactions and magnetic fields.
The realization of such topological non-trivial states has been an important goal for ultracold atoms platforms.
As in these systems the atoms are neutral the magnetic fields are artificially realized, e.g.~by coupling to laser light via Raman processes \cite{DalibardOehberg2011, GoldmanSpielman2014, HaukeCarusotto2022}.
This technique lead to the experimental realization of the artificial magnetic fields for atoms confined to quasi-one-dimensional ladders, or two-dimensional geometries systems, for both bosonic and fermionic atomic species \cite{AidelsburgerBloch2011, StruckWindpassinger2012, AidelsburgerBloch2013, MiyakeKetterle2013, AtalaBloch2014, AidelsburgerGoldman2015, ManciniFallani2015, TaiGreiner2017, GenkinaSpielman2019, ChalopinNascimbene2020, ZhouFallani2023}.
Furthermore, recently a Laughlin-type fractional quantum Hall state of two-atoms has been prepared \cite{LeonardGreiner2023}.

One of the central questions in the field is related to the design of experimentally relevant probes that can unravel the non-trivial topological properties of the prepared quantum states. 
In solid state materials, the Hall effect, i.e.~monitoring the induced transverse current upon the application of a force, has been a widely employed transport measurement.
More recently, the Hall response has also accessible for ultracold atoms in optical lattice and for weakly interacting gases has been measured from the center-of-mass drifts or local currents \cite{AidelsburgerGoldman2015, GenkinaSpielman2019, ChalopinNascimbene2020}.
Furthermore, theoretical proposals relate the quantized Hall response to topological invariants for small interacting ensembles for identifying the fractional states \cite{RepellinGoldman2020, PeraltaGavenskyGoldman2023}.
However, a complete understanding of the behavior of the Hall response when strong interactions are present is still lacking. 
Theoretical progress is being made in the case of ladder systems \cite{PrelovsekZotos1999, ZotosPrelovsek2000, GreschnerGiamarchi2019,BuserGiamarchi2021, CitroOrignac2024, HalatiGiamarchi2024}, the minimal setups for the study of the interplay of interactions and orbital effects, or making use of special geometries \cite{LopatinGiamarchi2001,LeonGiamarchi2007,Auerbach2018, FilipponeGiamarchi2019}.
In particular, the theoretical prediction of an universal Hall response occurring for certain parameters for interacting fermionic ladders \cite{GreschnerGiamarchi2019} has been experimentally confirmed \cite{ZhouFallani2023}. For the case of ladders a universal relation between the Hall resistance and the charge stiffness has also been proposed \cite{CitroOrignac2024}. 

In this work, we explore the Hall response for a Bose-Hubbard triangular ladder under the action of a magnetic field, focusing on the behavior of the Hall polarization for a wide range of on-site atomic interactions.
This is motivated by recent studies which showed that the Hall response can be employed as a sensitive probe for the features of the underlying phase diagram, either in the case of hardcore bosons in the triangular geometry \cite{HalatiGiamarchi2024}, or in the limit of small magnetic fields for square ladders \cite{CitroOrignac2024}.
Furthermore, the triangular flux ladders have proven to exhibit rich phase diagrams
\cite{MishraParamekanti2013, AnisimovasJuzeliunas2016, AnGadway2018, RomenLaeuchli2018, GreschnerMishra2019, CabedoCeli2020, LiLi2020, RoyHauke2022, HalatiGiamarchi2023, BarbieroCeli2023, BeradzeNersesyan2023, BeradzeNersesyan2023b, BaldelliBarbiero2024}, with frustration-induced effects and phases without an equivalent in the unfrustrated square geometry \cite{HalatiGiamarchi2023}.

The plan of the paper is as in the following, in Sec.~\ref{sec:model} we describe the model we investigate and the protocol employed for the numerical calculation of the Hall polarization, afterwards we briefly present the main message of our results in Sec.~\ref{sec:summary}. In Sec.~\ref{sec:methods_numerics} we briefly present the numerical method based on matrix product states employed in this work, while in Sec.~\ref{sec:PH_analytics} we perform analytical calculations in the non-interacting and mean-field limits for computing the equilibrium Hall polarization.
The results are presented in Sec.~\ref{sec:results}, focusing first on the behavior in the Meissner superfluid phase, Sec.~\ref{sec:Meissner}, followed by an analysis around the phase transitions boundaries, Sec.~\ref{sec:response_PT}, in the biased-chiral superfluid phase, Sec.~\ref{sec:response_BCSF}, and finalizing by discussing the commensurability effects occurring in the vortex superfluid phase, Sec.~\ref{sec:commensurability}.
We discuss our results and conclude in Sec.~\ref{sec:conclusions}.

\begin{figure}[!hbtp]
\centering
\includegraphics[width=0.45\textwidth]{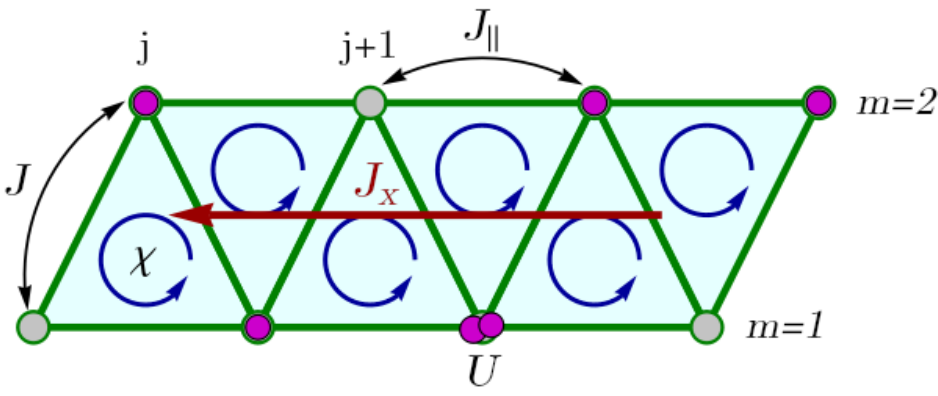}
\caption{\label{fig:sketch_PH}
Sketch of the triangular flux ladder model. The legs are denoted by $m=1,2$ and the sites on each leg numbered by $j$. The bosonic atoms can tunnel along the rungs with amplitude $J$ and along the legs with amplitude $J_\|$. We take into account repulsive on-site interaction between the atoms of strength $U$, and a flux $\chi$ pierces each triangular plaquette. Due to the presence of a linear potential $V_x$ a current $\boldsymbol{J}_x$ passes through the ladder.
}
\end{figure}

\section{Model and protocol\label{sec:model}}

We consider interacting bosonic atoms confined to a triangular ladder under the action of a magnetic field, as sketched in Fig.~\ref{fig:sketch_PH}. The Hamiltonian of the Bose-Hubbard model in a magnetic field is given by \cite{HalatiGiamarchi2023}
\begin{align} 
\label{eq:Hamiltonian}
 H=& H_\parallel+H_\perp+H_{\text{int}},\\
H_\parallel = & -J_\| \sum_{j=1}^{L-1} \left( e^{-i\chi}b^\dagger_{j,1}b_{j+1,1} + \text{H.c}. \right) \nonumber \\
& -J_\| \sum_{j=1}^{L-1} \left(e^{i\chi} b^\dagger_{j,2}b_{j+1,2} + \text{H.c}. \right), \nonumber \\
H_\perp = & -J \sum_{j=1}^L \left( b^\dagger_{j,1}b_{j,2} + \text{H.c}. \right) \nonumber \\
&-J \sum_{j=1}^{L-1} \left(b^\dagger_{j+1,1}b_{j,2} + \text{H.c}. \right),\nonumber \\
H_{\text{int}}=&\frac{U}{2} \sum_{j=1}^L \sum_{m=1}^2 n_{j,m}(n_{j,m}-1). \nonumber
\end{align}
We denote by $b_{j,m}$ and $b^\dagger_{j,m}$ the bosonic annihilation and creation operators at position $j$ and leg $m=1,2$. 
$\rho=N/(2L)$ represents the atomic filling, with the total number of particles $N=\sum_{j=1}^L \sum_{m=1}^2 n_{j,m}$ and $L$ the number of sites on each leg of the ladder.  
$H_\|$ gives the tunneling along the two legs of the ladder, with amplitude $J_\|$. The complex value of the tunneling amplitude stems from the presence of a magnetic field, with strength characterized by the flux $\chi$ \cite{DalibardOehberg2011, GoldmanSpielman2014}.
The tunneling along the rungs of the ladder is described by $H_\perp$ with amplitude $J$. 
The atoms interact repulsively if on the same lattice site, with the interaction strength $U>0$. We assume $\hbar=1$ in the following.
This model has a rich phase diagram of chiral quantum phases, as discussed in Ref.~\cite{HalatiGiamarchi2023, HalatiGiamarchi2024}, we give an overview of the phase diagrams for the considered parameter regimes in Sec.~\ref{sec:results}.

We are interested in the Hall response of the system and its dependence on the on-site interaction strength for the different phases present.
To realize this we monitor the dynamics of the system following the quench of a linear potential in the $x$-direction
\begin{align} 
\label{eq:potential}
V_x=\mu \sum_{j=1}^L\sum_{m=1}^2 \left[j+\frac{1}{2}(m-1)\right]n_{j,m}.
\end{align}
This protocol has been investigated for square ladders in Refs.~\cite{GreschnerGiamarchi2019,BuserGiamarchi2021} and we analyzed it for the triangular ladder in the limit of hardcore interactions in Ref.~\cite{HalatiGiamarchi2024}.
Furthermore, it has been experimentally implemented for interacting fermionic atoms on a square ladder in Ref.~\cite{ZhouFallani2023}.
In order to compute the Hall response, we begin with the system in its ground state at time $tJ=0$.
Following the quench with the potential $V_x$ at $t>0$ a total current, $\boldsymbol{J}_x$, is present in the $x$-direction and, due to the presence of the magnetic flux, between the two legs of the ladder a density imbalance, $P_y$, develops. 
These observables are defined for the triangular ladder as 
\begin{align} 
\label{eq:observables}
P_y=&\sum_{j} \left(n_{j,1}-n_{j,2}\right), \\
\boldsymbol{J}_x=&-i\sum_{j} \Big[\frac{J}{2} \left( b^\dagger_{j,1}b_{j,2}+b^\dagger_{j,2}b_{j+1,1} -\text{H.c}.\right)\nonumber  \\
&+ J_\| \left(e^{-i\chi}b^\dagger_{j,1}b_{j+1,1}+e^{i\chi}b^\dagger_{j,2}b_{j+1,2}-\text{H.c.} \right)\Big], \nonumber
\end{align}
where in the current $J_x$ we have contributions from the two legs of the ladder and also from the rungs due to the triangular geometry, for its derivation see Refs.~\cite{HalatiGiamarchi2024, LeonMillis2008}.
A current flowing towards smaller values of the index $j$ corresponds to negative values of $\boldsymbol{J}_x$. 

The Hall response of the system is quantified by the Hall polarization, defined as the ratio of the two observables defined in Eq.~(\ref{eq:observables}) \cite{GreschnerGiamarchi2019,BuserGiamarchi2021}
\begin{align} 
\label{eq:PH}
P_H(t)=\frac{\left\langle P_y \right\rangle(t)}{\left\langle \boldsymbol{J}_x \right\rangle(t)/J}.
\end{align}
In the numerator we usually consider the imbalance difference with respect to the ground state value, $\left\langle P_y \right\rangle(t)-\left\langle P_y \right\rangle(0)$, as phases like the biased-chiral superfluid exhibit a finite value of the imbalance in equilibrium.

The usefulness of employing the Hall polarization as a measure of the response stems from the fact that even though the magnitudes of the density imbalance and total current grow under the action of the linear potential, $P_H(t)$ stabilizes to a transient steady value at intermediate times \cite{GreschnerGiamarchi2019,BuserGiamarchi2021, HalatiGiamarchi2024}.
We compute $P_H(t)$ numerically using time-dependent matrix product states methods as described in Sec.~\ref{sec:methods_numerics} and its equilibrium value analytically in the non-interacting or mean-field limits in Sec.~\ref{sec:PH_analytics}.
The steady value of the Hall polarization we denote by $\langle\langle P_H\rangle\rangle$, where we performed the average over $P_H(t)$ for a time interval of at least $10/J$.
We work in a regime of small values of the linear potential $\mu/J$, such that the results shown are independent on its value. However, if we decrease the value of the potential we have access to longer times in the dynamics before the finite size effects become relevant. We note that we discussed the influence of $\mu/J$ in more details for the hardcore case in Ref.~\cite{HalatiGiamarchi2024}

\section{Key results\label{sec:summary}}

Our work aims at understanding the behavior of the Hall response of the quantum chiral phases occurring for interacting bosonic particle on a triangular ladder geometry. We show that the Hall polarization, computed numerically following the quench of a linear potential, is a reliable observable for quantifying the Hall response for the triangular ladder for all strengths of interactions considered.
We focus our discussion on the most striking features observed in the Hall polarization, how they correlate with the underlying ground state phase diagram and the behavior obtained for varying the local interactions.

We show that the Hall response is extremely sensitive to the phase boundaries, in particular, as we approach the transition thresholds within the Meissner superfluid phase, where a divergent-like behavior is observed.
By comparing the numerical results with analytical expressions obtained in the non-interacting limit, we can attribute the divergence to the expected vanishing of the current at the phase transition due to changes in the structure of the lower energy band.
While one can expect that we would observe a large negative Hall response even if we add weak interactions, interestingly we obtain large values close to the transition threshold also in the strongly correlated regimes.

We find several instances in which the Hall polarization changes sign. In the Meissner phase the sign can be changed by varying the interaction strength for certain values of the atomic filling, stemming from the presence of a particle-hole symmetry in the hardcore limit for half filling.
While in the vortex phase the sign of the Hall response can depend on both the hopping amplitudes and the magnetic flux.

The presence of frustration induced vortex commensurability effects in the vortex phase determines a large positive Hall polarization. Previously we discussed that in the hardcore case having a second commensurate vortex density in the current pattern of the vortex state is correlated with a strong Hall response \cite{HalatiGiamarchi2024}. Here we extend these results by showing that the competition between incommensurate and commensurate vortices is present also for finite, but large, values of the on-site interactions and that the Hall polarization is sensitive to these non-trivial effects.

\section{Matrix Product States Numerical Methods \label{sec:methods_numerics}}

In the following, we briefly describe the numerical approaches used in this work.
To numerically obtain the ground state of the Hamiltonian $H$, Eq.~(\ref{eq:Hamiltonian}), we employed a finite-size density matrix renormalization group (DMRG) algorithm in the matrix product state (MPS) representation \cite{White1992, Schollwoeck2005, Schollwoeck2011, Hallberg2006, Jeckelmann2002}, implemented using the ITensor Library \cite{FishmanStoudenmire2020}. 
We consider ladders with a number of sites on each leg of $L=60$ and $L=90$, and with a maximal bond dimension up to 500, ensuring that the truncation error is at most $10^{-10}$. 
Since we are considering a bosonic model with finite interactions the local Hilbert space is large, thus, a cutoff for its dimension is needed. 
We use a maximal local dimension of at least four or five states per site, depending on the value of the interactions. 

The time evolution with the additional potential $H+V_x$, Eq.~(\ref{eq:potential}), is performed using the time-dependent matrix product state method (tMPS) based on Trotter-Suzuki decomposition \cite{DaleyVidal2004, WhiteFeiguin2004, Schollwoeck2011}. 
The convergence was ensured with a time step of $\mathrm{d}t J/\hbar=0.01$ 
and the measurements were performed every tenth time step.
We maintain the same bond dimension as for the ground state search, this ensures that up to the times considered in this work, the truncation error is at most $10^{-9}$.

\section{Analytical calculation of the equilibrium Hall Polarization \label{sec:PH_analytics}}

In Refs.~\cite{GreschnerGiamarchi2019,BuserGiamarchi2021} it was shown that the transient steady value $\langle\langle P_H\rangle\rangle$ agrees with the equilibrium Hall polarization obtained for periodic boundary conditions upon the threading of a flux through the system.
In the following, we derive the equilibrium value analytically for the non-interacting case for small values of the flux $\chi$ and using a mean-field approach in the Meissner phase.

The Hamiltonian of the system for periodic boundary conditions in the $x$-direction, upon the threading of a flux $\Tilde{\Phi}$ through the cylinder and under the action of a potential difference between the two legs of the ladder is given by
\begin{align} 
\label{eq:Hamiltonian_PBC_phi}
 \Tilde{H}= & -J_\| \sum_{j=1}^{L} \left( e^{-i\chi-i\Tilde{\Phi}/L}b^\dagger_{j,1}b_{j+1,1} + \text{H.c}. \right)  \\
 & -J_\| \sum_{j=1}^{L} \left(e^{i\chi-i\Tilde{\Phi}/L} b^\dagger_{j,2}b_{j+1,2} + \text{H.c}. \right) \nonumber \\
 & -J \sum_{j=1}^L \left(e^{-i\Tilde{\Phi}/2L} b^\dagger_{j,1}b_{j,2} +e^{i\Tilde{\Phi}/2L}b^\dagger_{j+1,1}b_{j,2}+ \text{H.c}. \right) \nonumber \\
&+\frac{U}{2} \sum_{j=1}^L \sum_{m=1}^2 n_{j,m}(n_{j,m}-1) \nonumber\\
&+E_y \sum_{j=1}^L (n_{j,1}-n_{j,2}), \nonumber
 \end{align}
where $E_y$ is the energy difference between the two legs of the ladder.
The Hall polarization can be derived in terms of the ground state energy derivatives \cite{GreschnerGiamarchi2019, PrelovsekZotos1999, ZotosPrelovsek2000}.
In this sense, following the notations of Ref.~\cite{GreschnerGiamarchi2019}, we expand the ground state energy $\mathcal{E}_0(\Phi,\chi,E_y)$ to the third order in $\Phi,\chi,E_y$ around zero
\begin{align} 
\label{eq:E0}
\mathcal{E}_0(\Tilde{\Phi},\chi,E_y)&=\mathcal{E}_0(0,0,0)+\frac{\Tilde{\Phi}^2}{2}\frac{\partial^2\mathcal{E}_0}{\partial \Tilde{\Phi}^2}+\frac{\chi^2}{2}\frac{\partial^2\mathcal{E}_0}{\partial \chi^2}\\
& +\frac{E_y^2}{2}\frac{\partial^2\mathcal{E}_0}{\partial E_y^2}+\Tilde{\Phi}\chi E_y\frac{\partial^3\mathcal{E}_0}{\partial \Tilde{\Phi}\partial\chi\partial E_y}, \nonumber
\end{align}
where we considered only the terms that do not vanish due to symmetries.
The current and density imbalance can be computed as derivatives of the energy as 
\begin{align} 
\label{eq:E0_Jx_Py}
\left\langle \boldsymbol{J}_x\right\rangle^\text{eq}=L\frac{\partial\mathcal{E}_0}{\partial \Tilde{\Phi}},\quad\left\langle P_y \right\rangle^\text{eq}=\frac{\partial\mathcal{E}_0}{\partial E_y}.
\end{align}
Around the expansion point we have
\begin{align} 
\label{eq:E0_Jx_Py_2}
\left\langle \boldsymbol{J}_x\right\rangle^\text{eq}&=L\Tilde{\Phi}\left.\frac{\partial^2\mathcal{E}_0}{\partial \Tilde{\Phi}^2}\right\rvert_{\chi,\Tilde{\phi},E_y=0},\\
\quad\left\langle P_y \right\rangle^\text{eq}&=E_y\frac{\partial^2\mathcal{E}_0}{\partial E_y^2}+\Tilde{\Phi}\chi\left.\frac{\partial^3\mathcal{E}_0}{\partial \Tilde{\Phi}\partial\chi\partial E_y}\right\rvert_{\chi,\Tilde{\phi},E_y=0}.\nonumber
\end{align}
As we do not require a finite value of $E_y$ to compute the Hall polarization we obtain the following expression for the equilibrium value \cite{GreschnerGiamarchi2019}
\begin{align} 
\label{eq:P_Hall_analytic}
P_H^\text{eq}&=\frac{\chi}{L}\left.\frac{\frac{\partial^3\mathcal{E}_0}{\partial\Tilde{\phi}\partial\chi\partial E_y}}{\frac{\partial^2\mathcal{E}_0}{\partial\Tilde{\phi}^2}}\right\rvert_{\chi,\Tilde{\phi},E_y=0}. 
\end{align}

\subsection{Non-interacting limit}

In the non-interacting limit, $U=0$, for the Hamiltonian in Eq.~(\ref{eq:Hamiltonian_PBC_phi}) we can compute the dispersion relation of the bosonic atoms 
\begin{align} 
\label{eq:dispersion}
E_\pm(k)=&-2J_\|\cos(k+\Tilde{\Phi}/L)\cos(\chi)\pm \\
&\Big\{2J^2\left[1+\cos(k+\Tilde{\Phi}/L)\right] \nonumber\\
&\quad+\left[E_y+2J_\|\sin(k+\Tilde{\Phi}/L)\sin(\chi)\right]^2\Big\}^{1/2} .\nonumber
\end{align}
At small values of the flux $\chi$, where Eq.~(\ref{eq:P_Hall_analytic}) is valid, we are in the Meissner phase and the minimum of the lower band of the dispersion is at momentum $k=0$ \cite{HalatiGiamarchi2023}. Thus, using $\mathcal{E}_0=E_-(k=0)$ in Eq.~(\ref{eq:P_Hall_analytic}) we obtain
\begin{align} 
\label{eq:P_Hall_analytic_U0}
P_H^{\text{eq},U=0}&=\frac{-2 (J_\|/J) \chi\cos(\chi)}{1+4(J_\|/J)\cos(\chi)-4(J_\|/J)^2\sin(\chi)^2}. 
\end{align}
We note that even though this relation is derived only for small $\chi$, we kept in the ground state energy the full dependence on the flux $\chi$.
We compare this expression with numerical results at small interaction strengths in Sec.~\ref{sec:Meissner} and we find a very good agreement.

\subsection{Mean-field approach in the Meissner phase \label{sec:mf_limit}}

In the limit of large atomic fillings $\rho$ in the Meissner phase we can approximate the bosonic operator $b_{j,m}=\sqrt{\rho+\frac{1}{2}(-1)^m \delta\rho}~e^{i\phi}$. The values of $\delta\rho$ and $\phi$ can be computed from the minimization of the energy. Furthermore, from the approximate value of the ground state energy we can obtain the Hall polarization \cite{GreschnerGiamarchi2019}.
In this mean-field approximation the Hamiltonian $\Tilde{H}$, Eq.~(\ref{eq:Hamiltonian_PBC_phi}), for $E_y=0$ reads
\begin{align} 
\label{eq:Hamiltonian_PBC_phi_MF}
 \Tilde{H}=  -2J_\| L&\Bigg[ \cos(\chi+\Tilde{\Phi}/L)\left(\rho-\frac{1}{2}\delta\rho\right)  \\
&+ \cos(\chi+\Tilde{\Phi}/L)\left(\rho+\frac{1}{2}\delta\rho\right) \Bigg] \nonumber \\
  -4JL& \cos(\Tilde{\Phi}/2L) \left(\rho^2-\frac{\delta\rho^2}{4}\right)^{1/2} \nonumber \\
+UL &\left(\rho^2-\rho+\frac{\delta\rho^2}{4}\right). \nonumber
\end{align}
The energy is minimized for a local atomic imbalance of 
\begin{align} 
\label{eq:imbalance_MF}
\delta\rho=\frac{4J_\|\rho\sin(\chi)\sin(\Tilde{\Phi}/L)}{2J\cos(\Tilde{\Phi}/L)+\rho U}. 
\end{align}
By computing the total current as the derivative of the ground state energy, Eq.~(\ref{eq:E0_Jx_Py}), and using that $\left\langle P_y \right\rangle^\text{eq}=L\delta\rho$, we obtain the following mean-field value of the equilibrium Hall polarization
\begin{align} 
\label{eq:P_Hall_analytic_mf}
P_H^\text{eq,MF}&=\frac{-2 (J_\|/J) \sin(\chi)}{1+  \frac{\rho U}{2J}+4\frac{J_\|}{J}\cos(\chi)\left(1+\frac{\rho U}{2J}\right)-4\frac{J_\|^2}{J^2}\sin(\chi)^2}. 
\end{align}
We can notice that in the limit of $\chi\to 0$ and $\rho\to 0$ it agrees with the non-interacting result $P_H^{\text{eq},U=0}$, Eq.~(\ref{eq:P_Hall_analytic_U0}).

\section{Results \label{sec:results}}

In the following, we present the results for the Hall polarization throughout the phase diagram of the Hamiltonian, Eq.~(\ref{eq:Hamiltonian}). We initially focus on the filling $\rho=0.25$, considering the dependence of $\langle\langle P_H\rangle\rangle$ as a function of both the flux $\chi$ and the tunneling amplitude $J_\|/J$ for different values of the on-site interaction $U/J$.
For this value of the filling we obtain the same quantum phases in the ground state of the model for all values of the interaction considered, as discussed in Sec.~\ref{sec:ground_state} for $U/J=1$ and $U/J=10$.
This allows us to study the dependence of the Hall response within the same phase as a function of $U/J$.

In the rest of the Results section we present the behavior of the Hall response throughout the phase diagram, considering the interaction dependence from the weakly to the strongly interacting limits.

\begin{figure}[!hbtp]
\centering
\includegraphics[width=.44\textwidth]{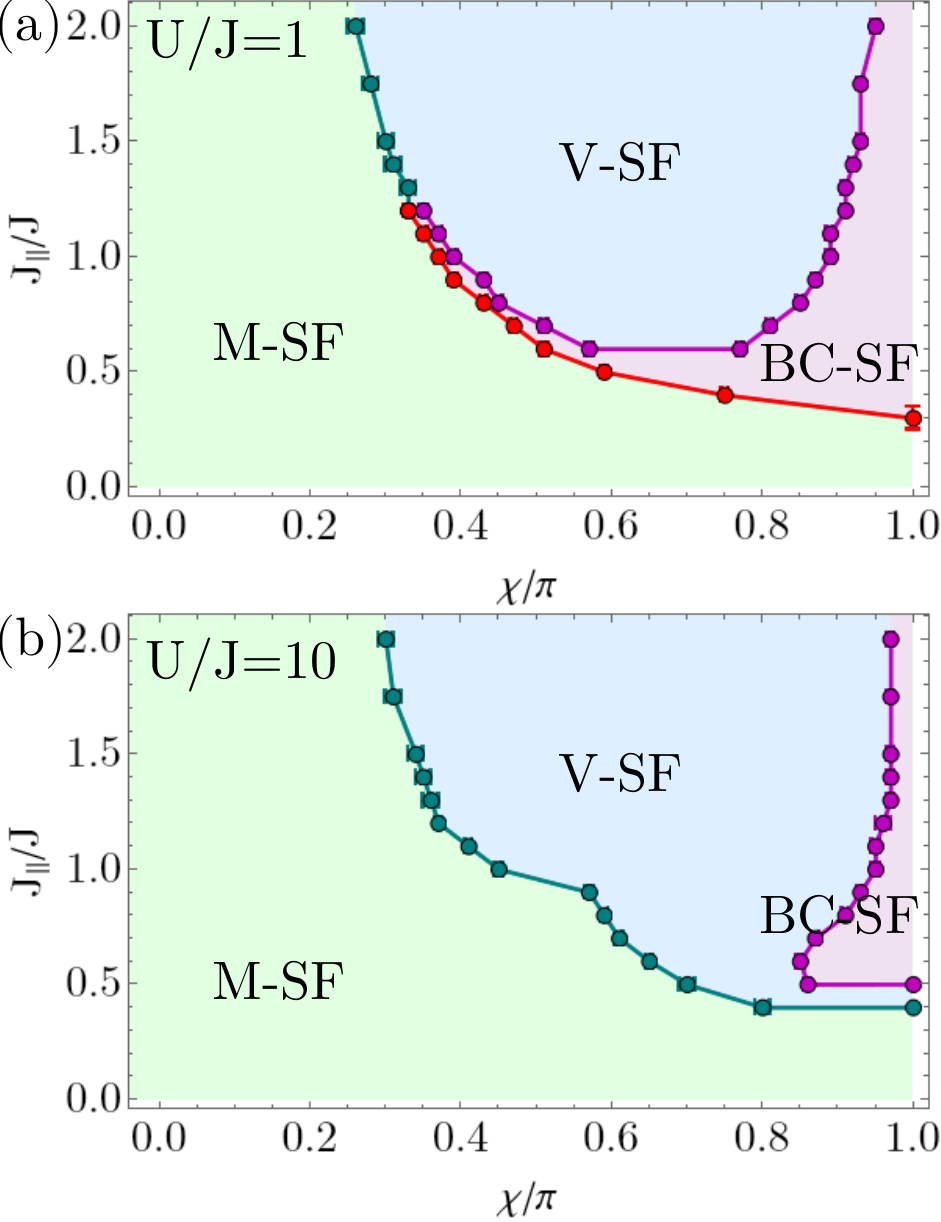}
\caption{Sketch of the phase diagram for $\rho=0.25$ with (a) $U/J=1$, (b) $U/J=10$.
The phases present for these parameters are the Meissner superfluid (M-SF), the vortex superfluid (V-SF) and biased chiral superfluid (BC-SF).
The nature of the quantum phases was identified based on DMRG numerical simulations analyzed similarly as in Ref.~\cite{HalatiGiamarchi2023} for a system size of $L=120$ sites on each leg. The symbols depicted correspond to the values of the flux $\chi/\pi$ at which the phase transitions occur determined numerically for a fixed value of $J_\|/J$.
The main characteristics of the phases are described in Sec.~\ref{sec:ground_state}.
}
\label{fig:phasediag}
\end{figure}

\subsection{Overview of the ground state phase diagram \label{sec:ground_state}}

\begin{figure}[!hbtp]
\centering
\includegraphics[width=.48\textwidth]{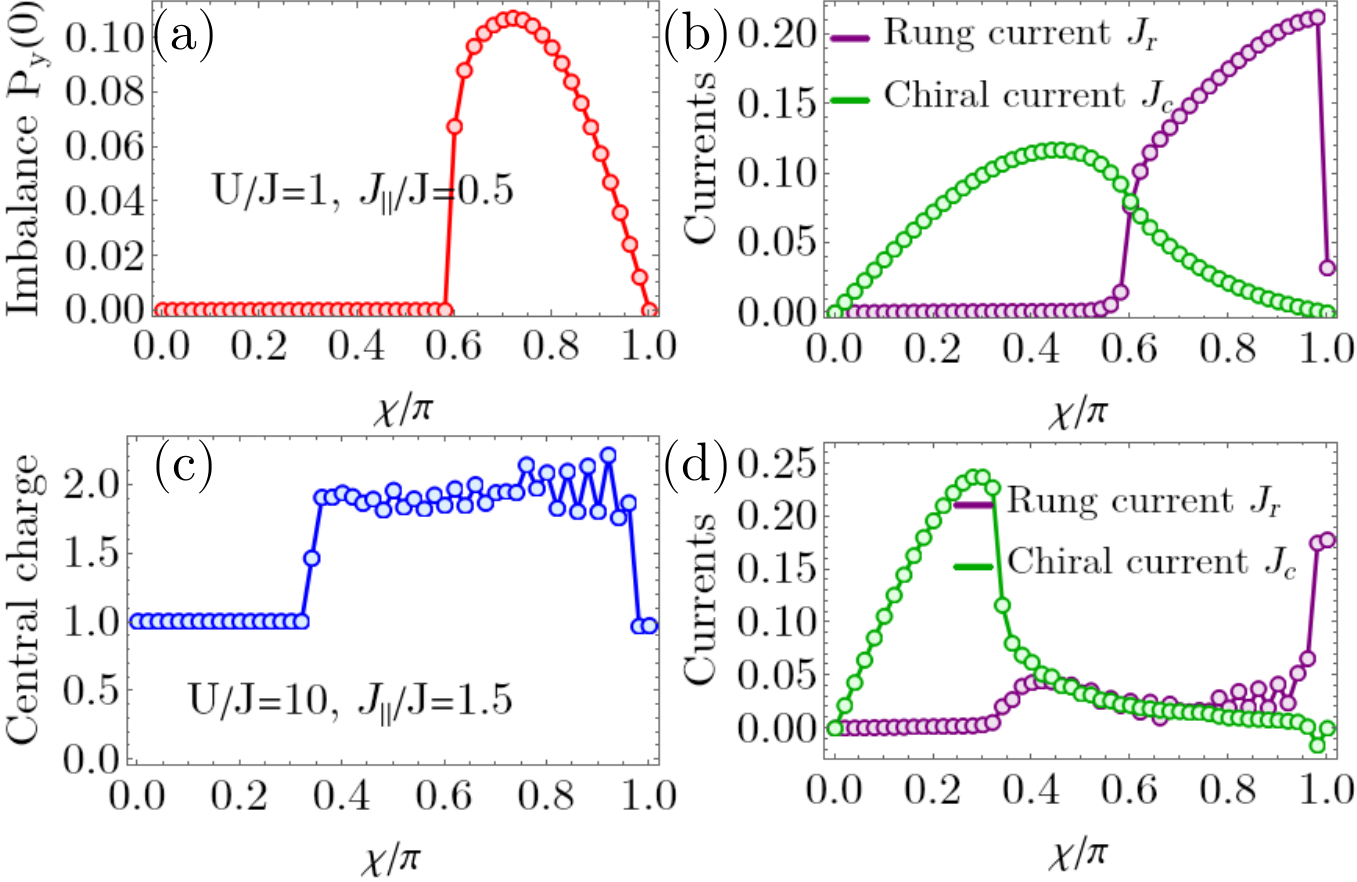}
\caption{
Ground state observables employed for the determination of the phase diagram shown in Fig.~\ref{fig:phasediag}, corresponding to the Hamiltonian given in Eq.~(\ref{eq:Hamiltonian}), for (a)-(b) $U/J=1$ and $J_\|/J=0.5$, and (c)-(d) $U/J=10$ and $J_\|/J=1.5$. We plot in (a) the ground state value of the density imbalance $P_y$, Eq.~(\ref{eq:observables}), in (b) and (d) the average rung current $J_r$ and the chiral current $J_c$, defined in Eqs.~(\ref{eq:cur})-(\ref{eq:cur2}), and in (c) the central charge, Eq.~(\ref{eq:cc}).
For the ground state simulations we considered a system of $L=120$ sites on each leg and a bond dimension $m=750$.
Based on the behavior of the observables we identified for $U/J=1$ and $J_\|/J=0.5$ a phase transition between M-SF and BC-SF at $\chi\approx0.59\pi$, and $U/J=10$ and $J_\|/J=1.5$ a phase transition between M-SF and V-SF at $\chi\approx0.34\pi$ and between V-SF and BC-SF at $\chi\approx0.97\pi$.
}
\label{fig:phasediag_obs}
\end{figure}

For a filling of $\rho=0.25$ in the phase diagram we observe three distinct quantum phases \cite{HalatiGiamarchi2023}, as shown in Fig.~\ref{fig:phasediag},
At small values of the flux $\chi$, or small values of the tunneling $J_\|/J$, we have the \emph{Meissner superfluid} (M-SF). The Meissner superfluid is characterized by vanishing values of currents on the rungs and by chiral currents on the legs of the ladder, and the presence of a single gapless mode.
For larger values of the flux $\chi$, a phase breaks the $\mathbb{Z}_2$ symmetry of the ladder is present, namely the \emph{biased chiral superfluid} (BC-SF), which is characterized by a finite density imbalance and a single gapless mode. The ground state manifold is spanned by two states exhibiting finite values of density imbalance between the two legs of opposite signs.
Increasing $J_\|/J$ we enter the \emph{vortex superfluid} (V-SF), characterized by finite values of the currents both on the legs and the rungs of the ladder, and two gapless modes. The current pattern determines a vortex density incommensurate with the ladder geometry, scaling linearly with the flux $\rho_v=\chi/\pi$. Furthermore, at large values of the interaction strength, for certain values of the flux additional vortex periodicities arise, determined by the following relation between the atomic filling and vortex density $\rho_v=1-\rho$ \cite{HalatiGiamarchi2023, HalatiGiamarchi2024}.
In Sec.~\ref{sec:commensurability} we further discuss the presence of the commensurability effects present at large values of the interaction strength. 

By increasing the interaction strength, from $U/J=1$ in Fig.~\ref{fig:phasediag}(a) to $U/J=10$ in Fig.~\ref{fig:phasediag}(b) we can see that the main changes occurring to the phase boundaries are due to the sensitivity on interactions of the biased chiral superfluid phase.
For $U/J=1$ the BC-SF extends to smaller values of the flux, $\chi<0.5\pi$ and we can even trace it as a narrow intermediate phase between the M-SF and V-SF up to at least $J_\|/J\approx1.3$. However, by increasing the interactions to $U/J=10$ the BC-SF only occurs for $\chi\gtrsim0.85\pi$ and we do not find a direct transition from the M-SF to the BC-SF, only via the V-SF.

In Fig.~\ref{fig:phasediag_obs} we show the behavior of the ground state observables employed to determine the phase boundaries shown in Fig.~\ref{fig:phasediag}.
We computed the ground state value of the density imbalance $P_y$, Eq.~(\ref{eq:observables}) [see Fig.~\ref{fig:phasediag_obs}(a)], which can identify the BC-SF phase, the values of the average rung $J_r$ and chiral currents $J_c$, defined as
\begin{align}
\label{eq:cur}
&J_c = \frac{1}{2 (L-1)} \sum_j \left\langle j^\|_{j,1} - j^\|_{j,2} \right\rangle,~\text{with}\\ 
&j^\|_{j,m} = -i J_\|\left[ e^{i\chi (-1)^m} b_{j,m}^\dagger b_{j+1,m} -\text{H.c.} \right].\nonumber
\end{align}
\begin{align}
\label{eq:cur2}
&J_r=\frac{1}{2 L-1} \sum_j \left|\left\langle j^\perp_{j}\right\rangle\right|,~\text{with}\\ 
&j^\perp_{2j-1} = -i J (b_{j,1}^\dagger b_{j,2} -\text{H.c.}), \nonumber \\
&j^\perp_{2 j} = -i J (b_{j+1,1}^\dagger b_{j,2} -\text{H.c.}). \nonumber
\end{align}
The behavior of the currents [see Fig.~\ref{fig:phasediag_obs}(b),(d)] can be used to distinguish the Meissner and vortex phases. Furthermore, we determine the number of gapless modes by calculating the central charge $c$ [see Fig.~\ref{fig:phasediag_obs}(c)], extracted from the scaling of the von Neumann entanglement entropy $S_{vN}(l)$ of subsystem of length $l$ embedded in the chain of length $L$.
The entanglement entropy for the ground state of gapless phases in the case of open boundary conditions is \cite{VidalKitaev2003,CalabreseCardy2004, HolzeyWilczek1994}
\begin{equation}
\label{eq:cc}
S_{vN}=\frac{c}{6}\log\left(\frac{L}{\pi}\sin\frac{\pi l}{L}\right)+s_1,
\end{equation}
where $s_1$ is a non-universal constant.

\begin{figure}[!hbtp]
\centering
\includegraphics[width=.43\textwidth]{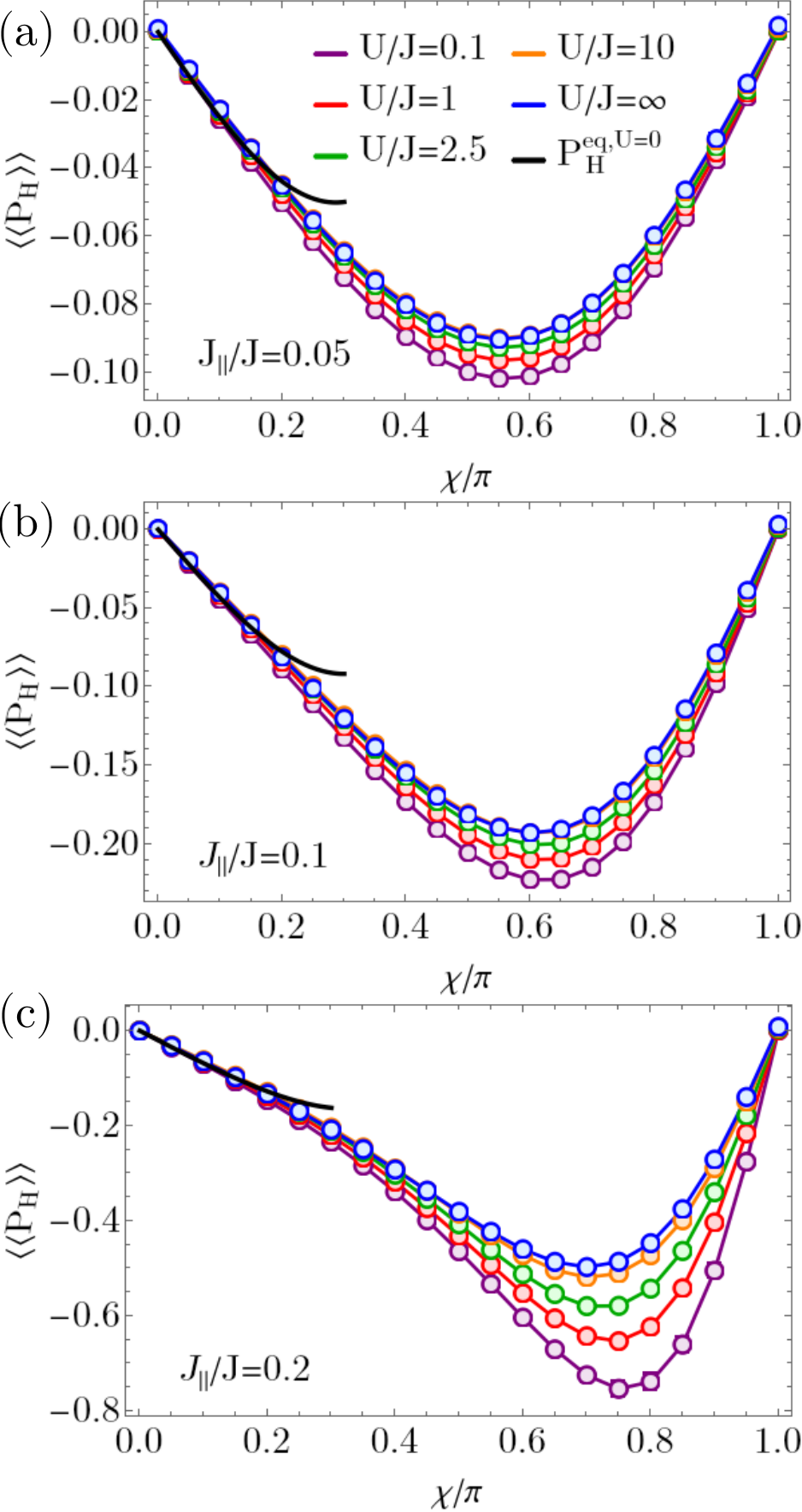}
\caption{
Time-averaged Hall polarization $\langle\langle P_H\rangle\rangle$ in the Meissner superfluid phase as a function of $\chi$ for (a) $J_\|/J=0.05$, (b) $J_\|/J=0.1$, (c) $J_\|/J=0.2$, for different values of the interaction strength $U$. The system size used is $L=60$, filling $\rho=0.25$ and the strength of the linear potential $\mu/J=0.01$.
The black curve at small values of the flux corresponds to the analytical value $P_H^{\text{eq},U=0}$, Eq.~(\ref{eq:P_Hall_analytic_U0}).
 }
\label{fig:PH_vs_flux}
\end{figure}

\begin{figure}[!hbtp]
\centering
\includegraphics[width=.48\textwidth]{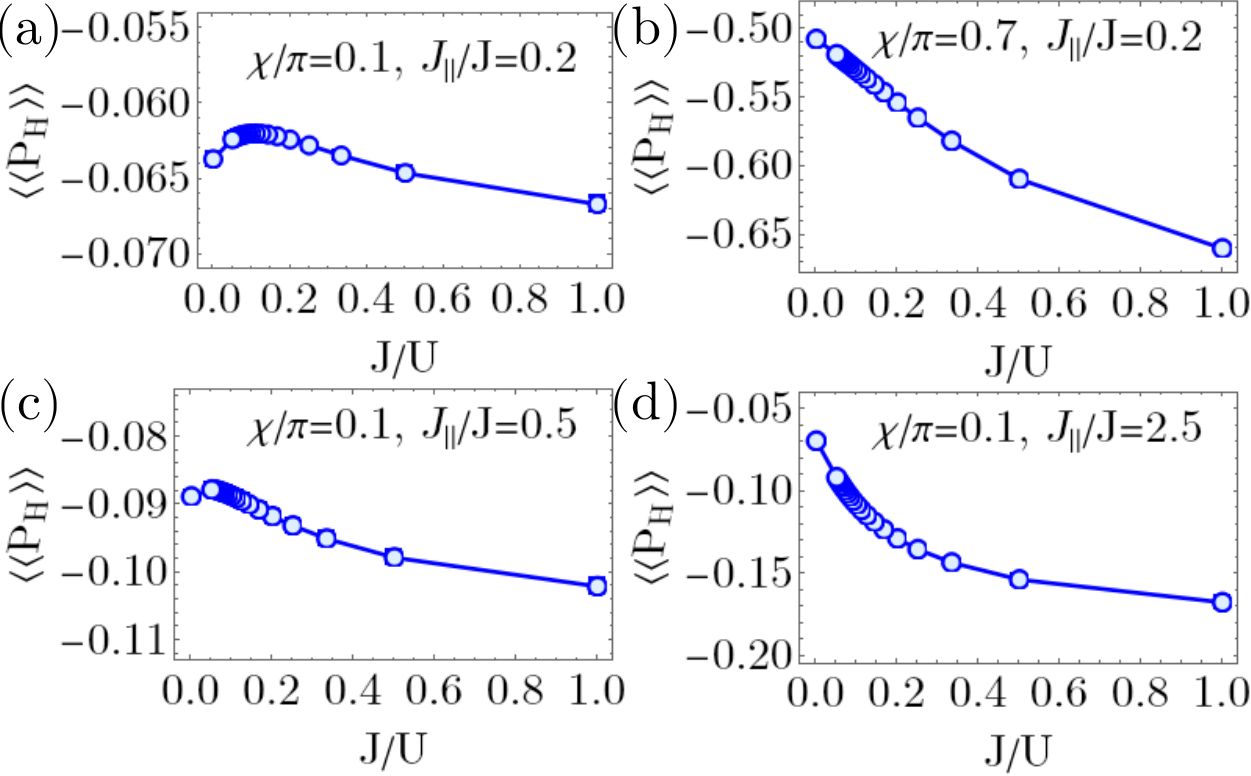}
\caption{
Time-averaged Hall polarization $\langle\langle P_H\rangle\rangle$ in the Meissner superfluid phase as a function of $J/U$ for (a) $\chi=0.1\pi$, $J_\|/J=0.2$, (b) $\chi=0.7\pi$, $J_\|/J=0.2$, (c) $\chi=0.1\pi$, $J_\|/J=0.5$, (d) $\chi=0.1\pi$, $J_\|/J=2.5$. The system size used is $L=90$, filling $\rho=0.25$ and the strength of the linear potential $\mu/J=0.001$.
 }
\label{fig:PH_vs_U}
\end{figure}

\begin{figure}[!hbtp]
\centering
\includegraphics[width=.43\textwidth]{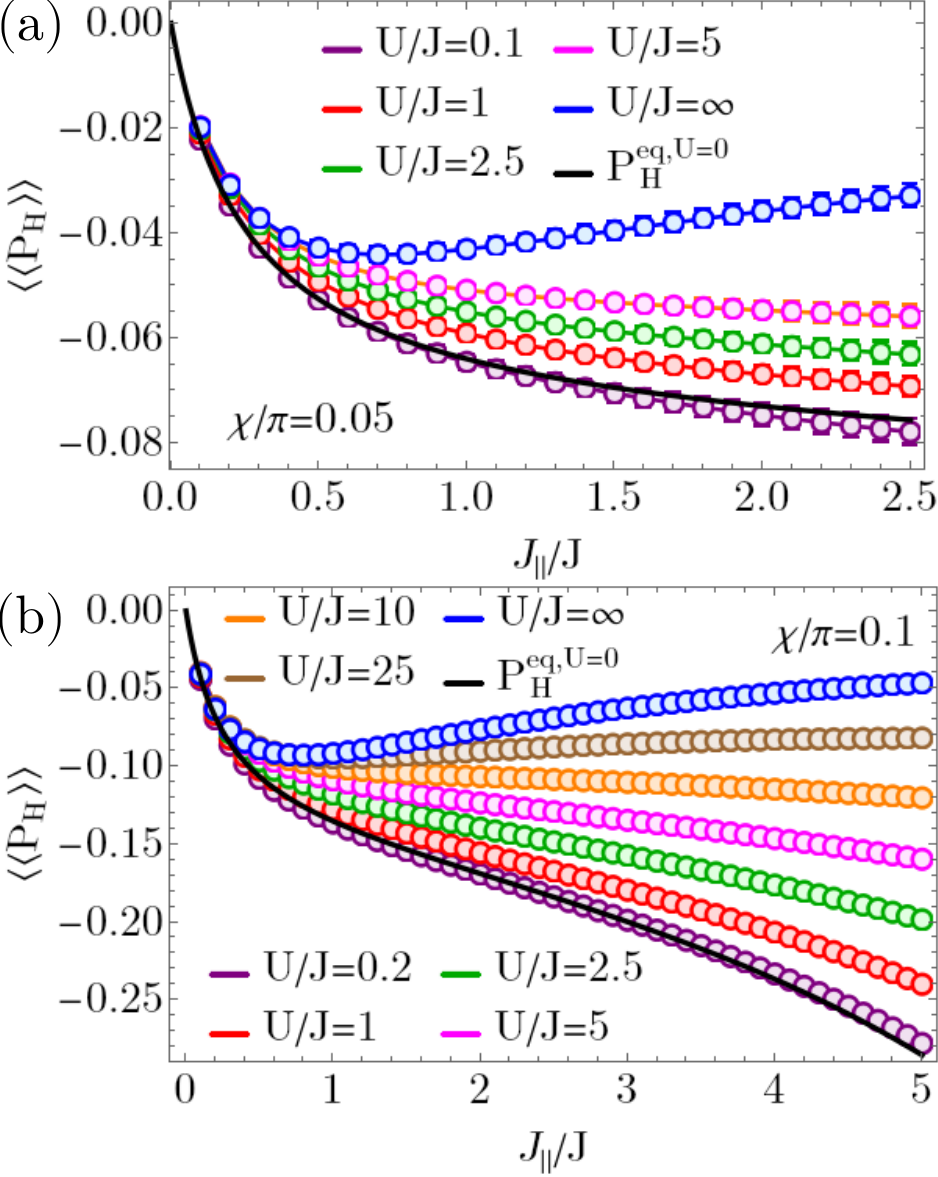}
\caption{
Time-averaged Hall polarization $\langle\langle P_H\rangle\rangle$ in the Meissner superfluid phase as a function of $J_\|/J$ for (a) $\chi=0.05\pi$, (b) $\chi=0.1\pi$, for different values of the interaction strength $U$. The system size used is (a) $L=60$, (b) $L=90$, filling $\rho=0.25$ and the strength of the linear potential (a) $\mu/J=0.01$, (b) $\mu/J=0.001$.
The black curve at small values of the flux corresponds to the analytical value $P_H^{\text{eq},U=0}$, Eq.~(\ref{eq:P_Hall_analytic_U0}).
 }
\label{fig:PH_vs_Jp}
\end{figure}

\begin{figure}[!hbtp]
\centering
\includegraphics[width=.43\textwidth]{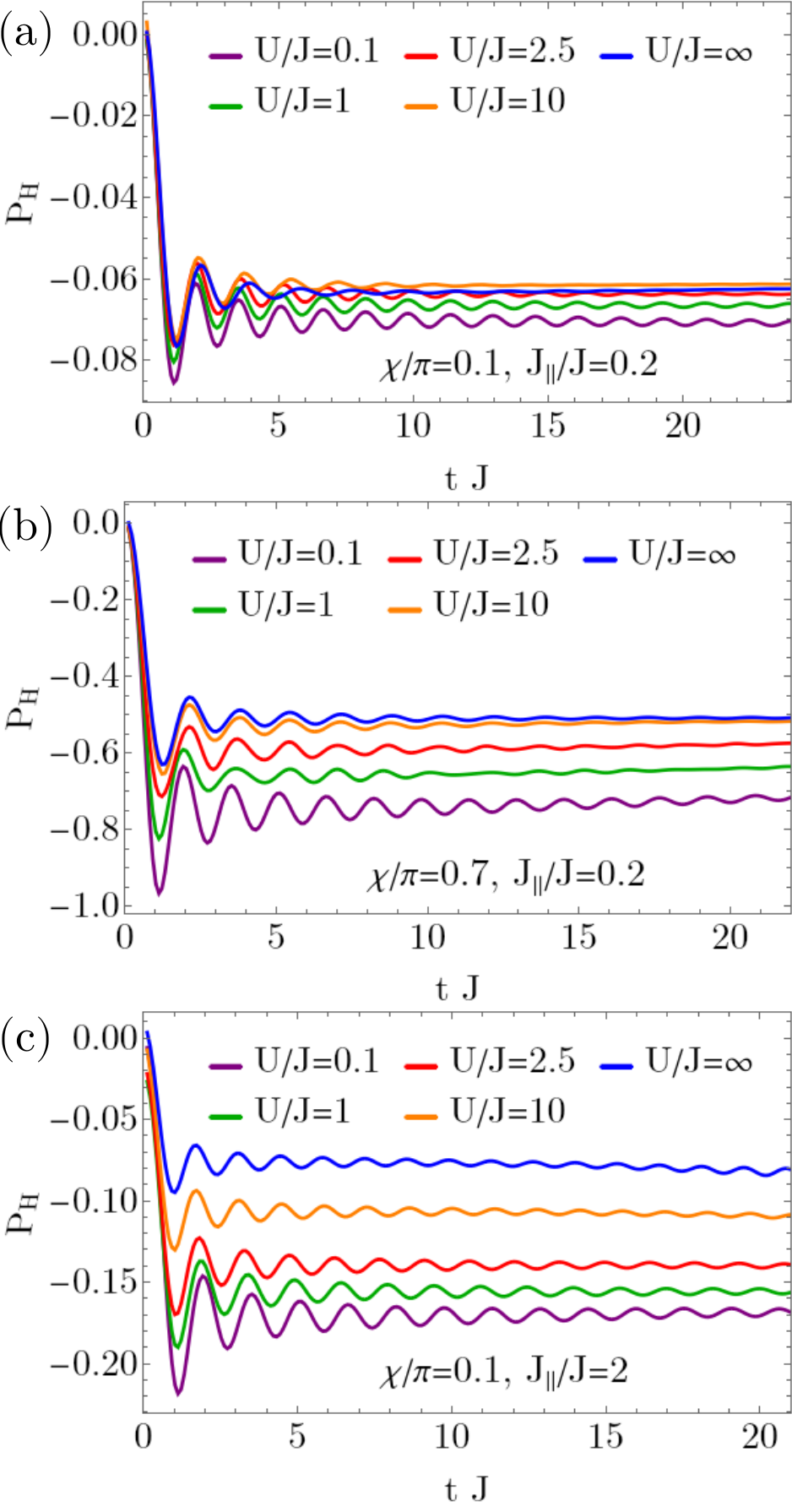}
\caption{
Time evolution of the Hall polarization $ P_H(t)$ in the Meissner superfluid phase for (a) $\chi=0.1\pi$, $J_\|/J=0.2$, (b) $\chi=0.7\pi$, $J_\|/J=0.2$, (c) $\chi=0.1\pi$, $J_\|/J=2$, for different values of the interaction strength $U$. 
The system size used is (a), (c) $L=90$ and (b) $L=60$, and the strength of the linear potential (a), (c) $\mu/J=0.001$ and (b) $\mu/J=0.01$.
 }
\label{fig:PH_dynamics_U}
\end{figure}

\subsection{The Hall response of the Meissner phase \label{sec:Meissner}}

We begin by analyzing the behavior of the Hall polarization in the Meissner superfluid phase.
In Fig.~\ref{fig:PH_vs_flux} we show the dependence of the steady value of the Hall polarization $\langle\langle P_H\rangle\rangle$ as a function of the flux for small values of $J_\|/J<0.2$ and different values of the interaction, from almost non-interacting atoms $U/J=0.1$, to the hardcore limit $U/J=\infty$, within the Meissner superfluid phase.
For $J_\|/J=0.05$, Fig.~\ref{fig:PH_vs_flux}(a), the dependence on the flux is almost symmetric with a maximum around the value $\chi/\pi=0.5$.
However, we can observe that by increasing $J_\|/J$ the maximum of $\langle\langle P_H\rangle\rangle$ increases in magnitude and moves to higher values of $\chi$.
As will be discussed in Sec.~\ref{sec:response_PT}, this is due to the proximity of the phase transitions to either to the vortex phase or to the biased chiral superfluid, effect which is more prominent at lower values of the interaction strength.
At low values of the flux, $\chi/\pi\lesssim0.15$, we have a very good agreement with the equilibrium value of the Hall polarization computed for non-interacting bosons, $P_H^{\text{eq},U=0}$. This is also due to the fact that the on-site interaction does not seem to play an important role in this regime, as seen in Fig.~\ref{fig:PH_vs_U}(a) for $J_\|/J=0.2$ and $\chi/\pi=0.1$, where in between $U/J=2$ and the hardcore limit $\langle\langle P_H\rangle\rangle$ is mostly independent on the value of the interaction.
However, for larger values of the magnetic flux, e.g.~in Fig.~\ref{fig:PH_vs_U}(b) for $J_\|/J=0.2$ and $\chi/\pi=0.7$, the value of $U/J$ is much more important, with the magnitude of $\langle\langle P_H\rangle\rangle$ decreasing with increasing the interaction strength.

\begin{figure}[!hbtp]
\centering
\includegraphics[width=.48\textwidth]{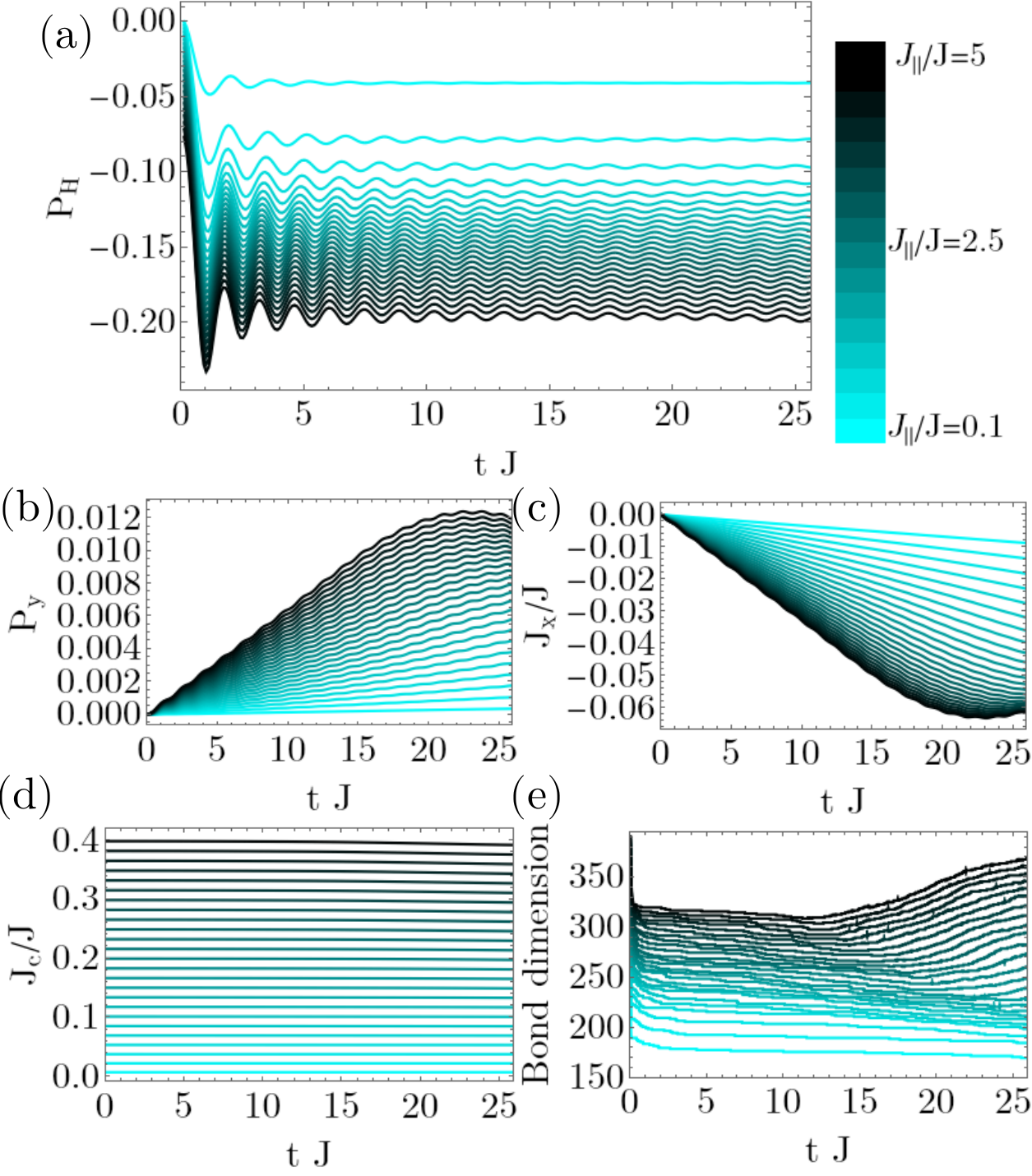}
\caption{
Time evolution in the Meissner superfluid phase of the (a)  Hall polarization $P_H$, (b) density imbalance $P_y$, (c) current $\boldsymbol{J}_x/J$, (d) chiral current $J_c/J$, and (e) bond dimension, for $\chi=0.1\pi$, $U/J=2.5$, for different values of the tunneling amplitude $J_\|/J$, in between $J_\|/J=0.1$ and $J_\|/J=5$. 
The system size used is $L=90$, and the strength of the linear potential $\mu/J=0.001$.
 }
\label{fig:PH_dynamics_Jp}
\end{figure}

\begin{figure}[!hbtp]
\centering
\includegraphics[width=.48\textwidth]{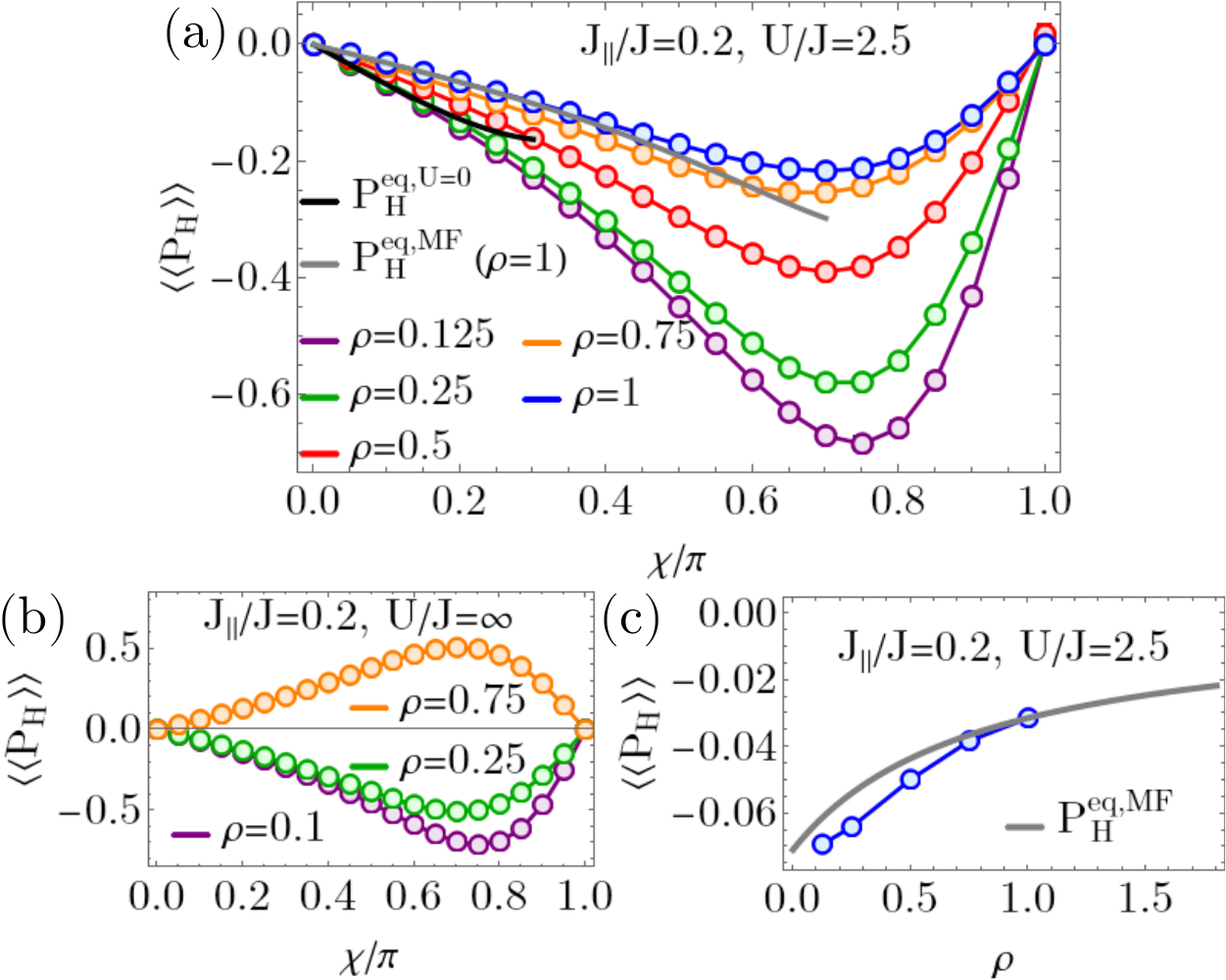}
\caption{
Time-averaged Hall polarization $\langle\langle P_H\rangle\rangle$ in the Meissner superfluid phase as a function of $\chi$ for (a) $J_\|/J=0.2$, $U/J=2.5$, (b) $J_\|/J=0.2$, $U/J=\infty$, for different values of the filling $\rho$, and (c) as a function of $\rho$ for $J_\|/J=0.2$, $U/J=2.5$. 
The system size used is (a), (c) $L=60$ and (b) $L=90$, and the strength of the linear potential (a), (c) $\mu/J=0.01$ and (b) $\mu/J=0.001$.
The black curve at small values of the flux corresponds to the analytical value $P_H^{\text{eq},U=0}$, Eq.~(\ref{eq:P_Hall_analytic_U0}), and the gray curve corresponds to the analytical value $P_H^{\text{eq,MF}}(\rho=1)$, Eq.~(\ref{eq:P_Hall_analytic_mf}).
We note that in panel (b) the values shown for $\rho=0.25$, $U/J=\infty$ are taken from Ref.~\cite{HalatiGiamarchi2024}.
 }
\label{fig:PH_vs_density}
\end{figure}

The good agreement between the analytical expression of $P_H^{\text{eq},U=0}$, Eq.~(\ref{eq:P_Hall_analytic_U0}), and the numerically determined $\langle\langle P_H\rangle\rangle$ for small values of the interaction, $U/J=0.1$ and $U/J=0.2$, can be very well seen in Fig.~\ref{fig:PH_vs_Jp}, where we depict $\langle\langle P_H\rangle\rangle$ as a function of $J_\|/J$ for small values of $\chi$.
The weak dependence of the interaction is confined only to small $J_\|/J$, while for larger values we have a much stronger influence of the interaction, for example see also Fig.~\ref{fig:PH_vs_U}(c) for $J_\|/J=0.5$ in contrast to Fig.~\ref{fig:PH_vs_U}(d) for $J_\|/J=2.5$.
Furthermore, we observe in Fig.~\ref{fig:PH_vs_Jp} that the value of $U/J$ is crucial for the dependence of $\langle\langle P_H\rangle\rangle$ as a function on $J_\|/J$ at larger values of $J_\|/J$. While for hardcore bosons the Hall polarization is decreasing with $J_\|/J$, for 
$U/J\lesssim10$ $\langle\langle P_H\rangle\rangle$ is increasing with $J_\|/J$ for the interval shown.
We associate this with the fact that a phase transition to the vortex superfluid phase might occur for larger values of $J_\|/J$ even for the values of $\chi$ shown in Fig.~\ref{fig:PH_vs_Jp}, e.g.~for the non-interacting case the phase transition occurs for $J_\|/J\approx10$ for $\chi/\pi=0.1$, see Sec.~\ref{sec:response_PT} for more details about the behavior approaching the phase transition.

We depict the dynamics of $P_H(t)$ in the Meissner superfluid phase in Fig.~\ref{fig:PH_dynamics_U} for different values of the interaction strength and in Fig.~\ref{fig:PH_dynamics_Jp}(a) varying the strength of $J_\|/J$.
Throughout the Meissner phase we observe a similar dynamical behavior, a fast increase of the magnitude of $P_H$ followed by damped oscillations towards the transient steady value. This steady behavior at long times justifies the study of the time averaged Hall polarization $\langle\langle P_H\rangle\rangle$.
In Fig.~\ref{fig:PH_dynamics_U} we observe, for all parameters depicted, that the oscillations are more prominent for small value of the interaction strength, their damping increasing with the value of $U/J$, which also has a slight impact on the frequency of the oscillations.

In Fig.~\ref{fig:PH_dynamics_Jp} we monitor the state of the system following the quench of the linear potential for $\chi=0.1\pi$, $U/J=2.5$ and a wide range of the tunneling $0.1\leq J_\|/J\leq 5$, for times up to $tJ=25$.
We observe a well defined plateau in the Hall polarization [Fig.~\ref{fig:PH_dynamics_Jp}(a)] for times considered, with the value of $\langle\langle P_H\rangle\rangle$ increasing with $J_\|/J$ for these parameters.
Following the quench of the linear potential the magnitude of both the density imbalance $\langle P_y \rangle$ and the current $\langle\boldsymbol{J}_x\rangle$ exhibits a mostly linear increase. For larger values of $J_\|/J$, we can see a deviation from the linear trend for times $tJ\gtrsim 20$, stemming from the finite size of the ladder considered here. For larger system size, or smaller values of the linear potential $\mu/J$, the deviation from the linear evolution would occur at later times. Interestingly, for the parameters and times considered we see that the plateau value of $P_H$ is not affected.

\begin{figure}[!hbtp]
\centering
\includegraphics[width=.43\textwidth]{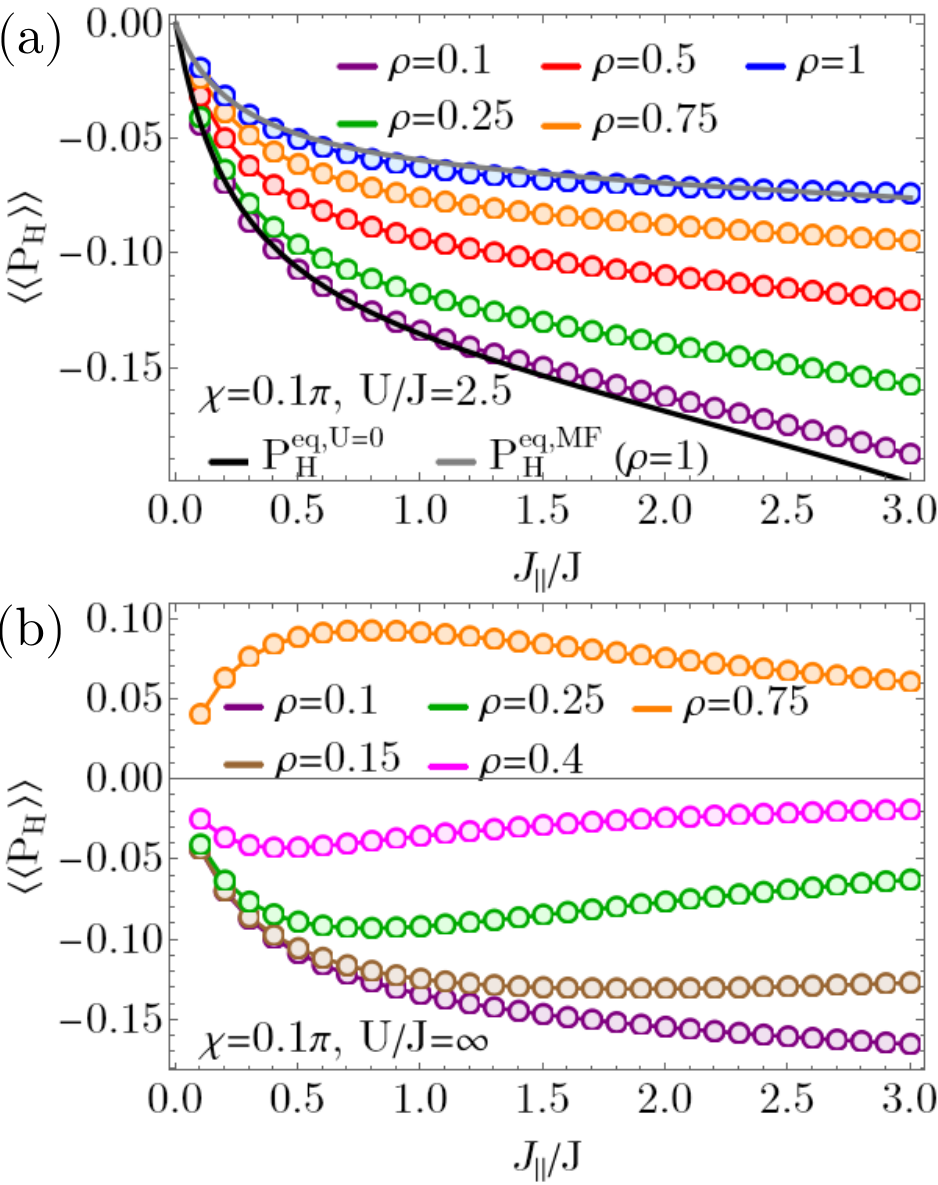}
\caption{
Time-averaged Hall polarization $\langle\langle P_H\rangle\rangle$ in the Meissner superfluid phase as a function of $J_\|/J$ for $\chi=0.1\pi$ and (a) $U/J=2.5$, (b) $U/J=\infty$, for different values of the filling $\rho$. 
The system size used is $L=90$, and the strength of the linear potential $\mu/J=0.001$.
In (a), the black curve at small values of the flux corresponds to the analytical value $P_H^{\text{eq},U=0}$, Eq.~(\ref{eq:P_Hall_analytic_U0}), and the gray curve corresponds to the analytical value $P_H^{\text{eq,MF}}(\rho=1)$, Eq.~(\ref{eq:P_Hall_analytic_mf}).
 }
\label{fig:PH_vs_density_Jp}
\end{figure}

\begin{figure}[!hbtp]
\centering
\includegraphics[width=.48\textwidth]{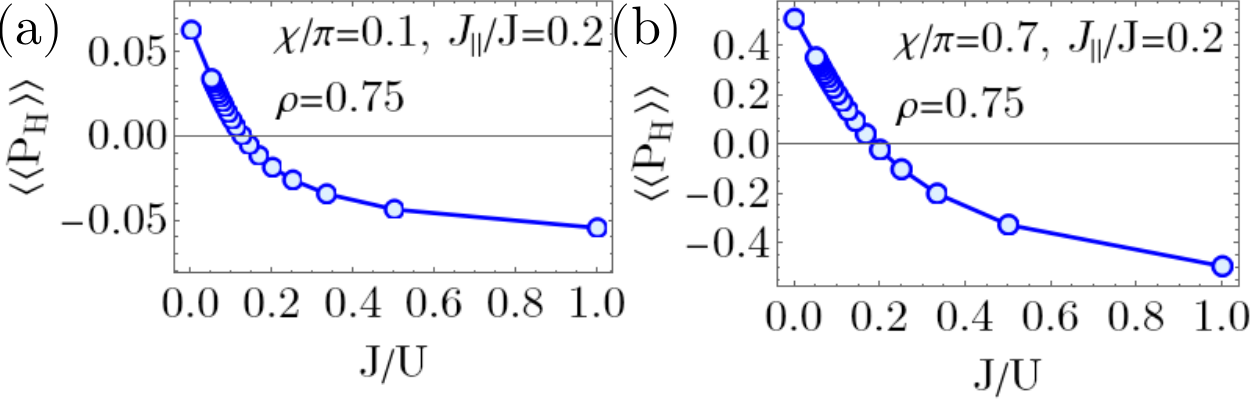}
\caption{
Time-averaged Hall polarization $\langle\langle P_H\rangle\rangle$ in the Meissner superfluid phase as a function of $J/U$ for (a) $\chi=0.1\pi$, $J_\|/J=0.2$, (b) $\chi=0.7\pi$, $J_\|/J=0.2$. The system size used is $L=90$, filling $\rho=0.75$ and the strength of the linear potential $\mu/J=0.001$.
 }
\label{fig:PH_vs_U_0.75}
\end{figure}

In order to analyze the nature of the state during the evolution we compute the dynamics of the chiral current, defined in Eq.~(\ref{eq:cur}).
We can see that $J_c(t)$ remains constant in time and equal to its ground state value, Fig.~\ref{fig:PH_dynamics_Jp}(d). This implies that the system maintains its Meissner superfluid character also after the quench during the time interval from which we extract its Hall response.
Furthermore, in Fig.~\ref{fig:PH_dynamics_Jp}(e) we show the evolution of the bond dimension used to represent the state of the system as a MPS, which roughly quantifies the amount of entanglement present \cite{Schollwoeck2011}.
We observe that by keeping the truncation error fixed to $10^{-12}$ the bond dimension decreases abruptly at short times and remains almost constant during the time interval which exhibits a linear evolution of $P_y$ and $\boldsymbol{J}_x$. We associate the increase at late times with the growing importance of the boundary effects, thus, offering us a further handle to the estimation of the influence of finite size.

In the final part of the section regarding the Hall response in the Meissner superfluid phase, we analyze the role of the atomic filling $\rho$ with the results presented in Fig.~\ref{fig:PH_vs_density} and Fig.~\ref{fig:PH_vs_density_Jp}.
For a finite interaction strength of $U/J=2.5$ increasing the atomic filling decreases the magnitude of the Hall polarization, as seen in Fig.~\ref{fig:PH_vs_density}(a) as a function of $\chi$ for $J_\|/J=0.2$ and in Fig.~\ref{fig:PH_vs_density_Jp}(a) as a function of $J_\|/J$ for $\chi=0.1\pi$.
In the case of small values of $\rho$ we compare our numerical results with the analytical result $P_H^{\text{eq},U=0}$, Eq.~(\ref{eq:P_Hall_analytic_U0}), [black curves in Fig.~\ref{fig:PH_vs_density}(a) and Fig.~\ref{fig:PH_vs_density_Jp}(a)], as the non-interacting Hall polarization also corresponds to the single particle limit.
We obtain a good agreement with $\langle\langle P_H\rangle\rangle$ for $\rho=0.125$ for a magnetic flux up to $\chi\lesssim0.15\pi$ [Fig.~\ref{fig:PH_vs_density}(a)] and for $\rho=0.1$ and $\chi=0.1\pi$ for the dependence on $J_\|/J$ [Fig.~\ref{fig:PH_vs_density_Jp}(a)].
The second comparison we perform is for larger values of the atomic filling, where we expect the mean-field approach presented in Sec.~\ref{sec:mf_limit} to hold.
For $\rho=1$, $P_H^{\text{eq,MF}}(\rho=1)$, Eq.~(\ref{eq:P_Hall_analytic_mf}), agrees well with the numerical results for $\chi\lesssim0.4\pi$ and the dependence on $J_\|/J$, see gray curves in Fig.~\ref{fig:PH_vs_density}(a) and Fig.~\ref{fig:PH_vs_density_Jp}(a).
Furthermore, in Fig.~\ref{fig:PH_vs_density}(c) we can see that the agreement with the mean-field result becomes better as we increase $\rho$.

\begin{figure}[!hbtp]
\centering
\includegraphics[width=.43\textwidth]{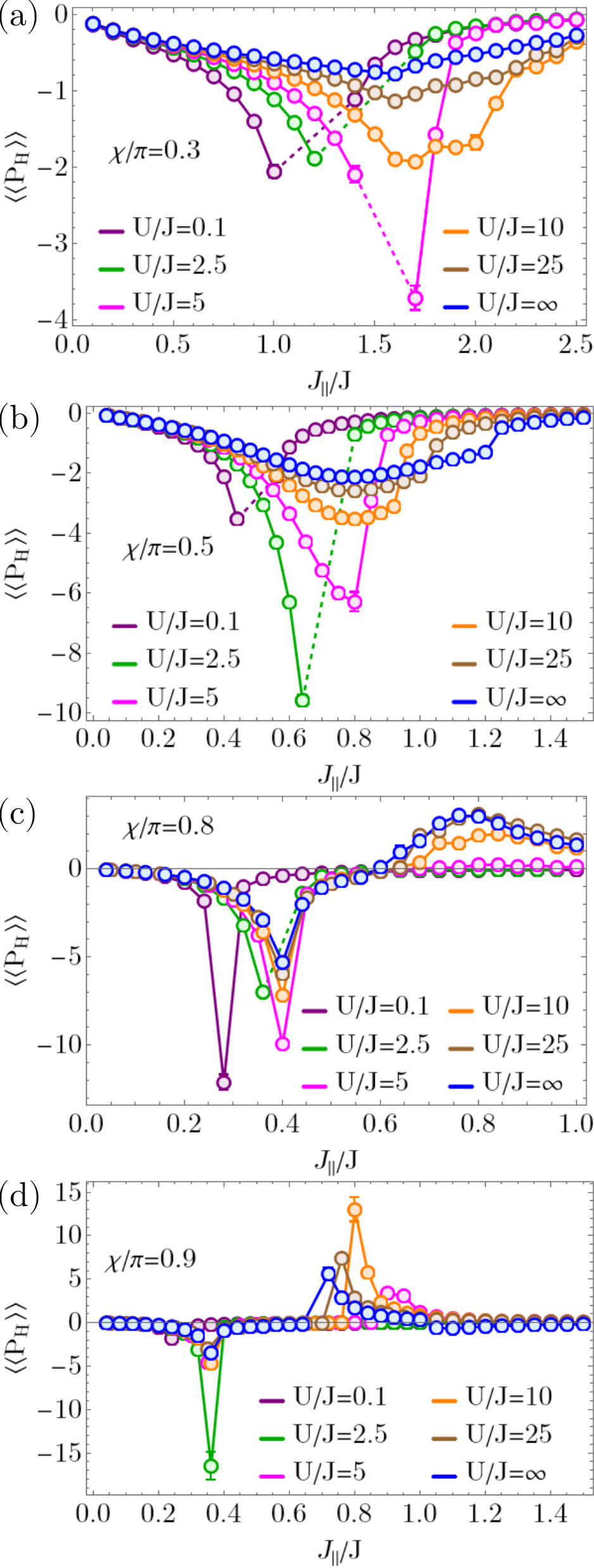}
\caption{
Time-averaged Hall polarization $\langle\langle P_H\rangle\rangle$ across phase transitions as a function of $J_\|/J$ for (a) $\chi=0.3\pi$, (b) $\chi=0.5\pi$, (c) $\chi=0.8\pi$, (d) $\chi=0.9\pi$, for different values of the interaction strength $U$. 
The system size used is $L=90$, filling $\rho=0.25$ and the strength of the linear potential $\mu/J=0.001$.
Dashed lines denote the region in which we could not define the value of $\langle\langle P_H\rangle\rangle$ as the current crosses zero during the time-evolution.
 }
\label{fig:PH_vs_Jp_transitions}
\end{figure}

\begin{figure}[!hbtp]
\centering
\includegraphics[width=.48\textwidth]{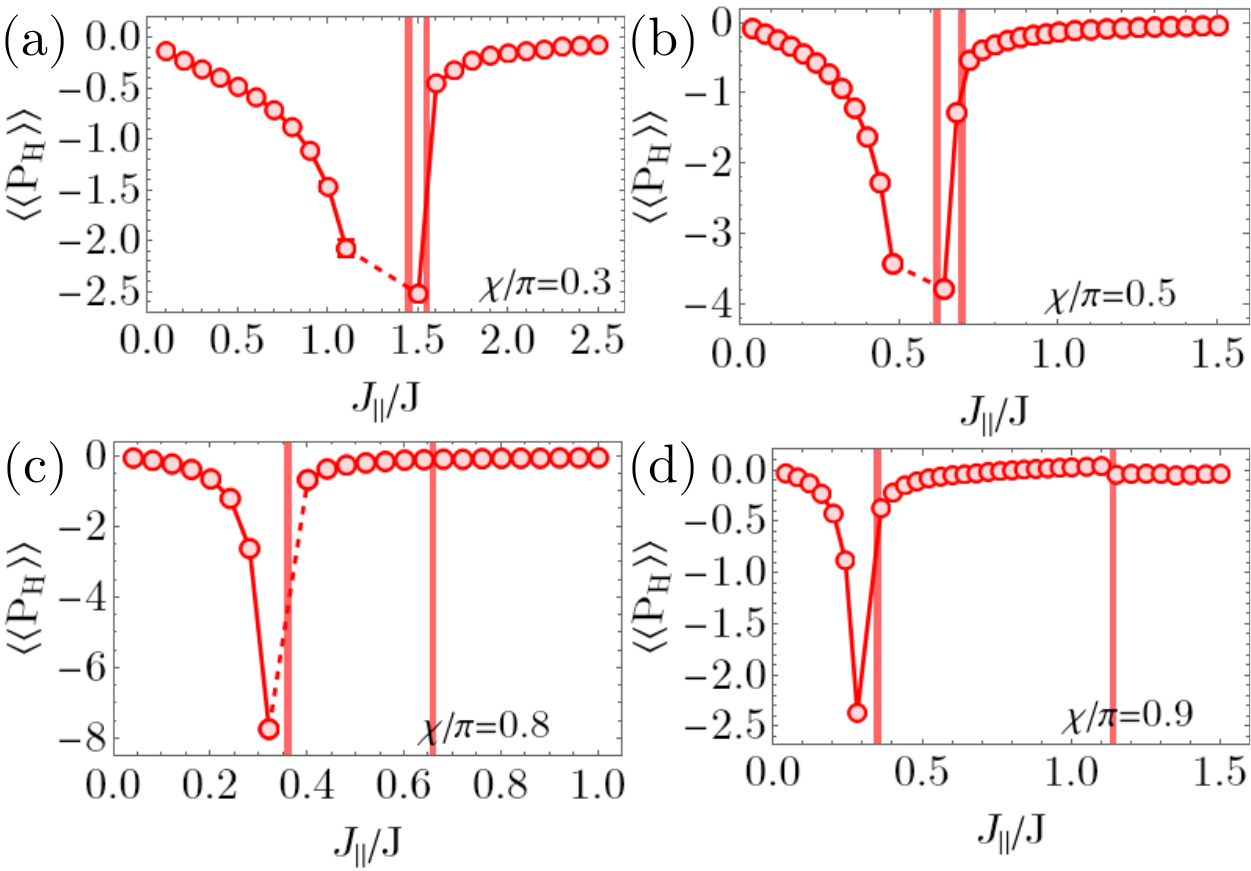}
\caption{
Time-averaged Hall polarization $\langle\langle P_H\rangle\rangle$ across phase transitions as a function of $J_\|/J$ for (a) $\chi=0.3\pi$, (b) $\chi=0.5\pi$, (c) $\chi=0.8\pi$, (d) $\chi=0.9\pi$, for $U/J=1$. 
The system size used is $L=90$, filling $\rho=0.25$ and the strength of the linear potential $\mu/J=0.001$.
The vertical lines denote the phase transition threshold values as marked in the ground state phase diagrams in Fig.~\ref{fig:phasediag}.
Dashed lines denote the region in which we could not define the value of $\langle\langle P_H\rangle\rangle$ as the current crosses zero during the time-evolution.}
\label{fig:PH_vs_Jp_transitions_U1}
\end{figure}

As for hardcore bosons a particle-hole symmetry is present in the system for $\rho=0.5$, it is interesting to investigate the dependence on the filling also in this case. 
We expect that from small fillings the magnitude of the Hall response will decrease with $\rho$ until it vanishes for $\rho=0.5$ and changes signs for larger fillings, with the same magnitude and opposite signs for $\rho$ and $1-\rho$.
We observe this behavior in Fig.~\ref{fig:PH_vs_density}(b) and Fig.~\ref{fig:PH_vs_density_Jp}(b), in particular we obtain the same value $|\langle\langle P_H\rangle\rangle|$ for $\rho=0.25$ and $\rho=0.75$, but $\langle\langle P_H\rangle\rangle$ has opposite sign for the two values of the filling.
The change in sign of the Hall polarization offers us an interesting opportunity when $\rho>0.5$, as in weak interactions $\langle\langle P_H\rangle\rangle$ is negative in the Meissner phase, while for hardcore bosons $\langle\langle P_H\rangle\rangle$ is positive.
This implies that by varying the on-site interaction strength from weakly to strongly interacting regimes we can change the sign of the Hall polarization and have a value of $U/J$ for which the Hall response vanishes.
For example, we depict this behavior in Fig.~\ref{fig:PH_vs_U_0.75} for $\rho=0.75$ for two sets of parameters in the Meissner superfluid phase.
It is an interesting open question if a symmetry emerges at the particular value of $U/J$ for which the Hall response vanishes.

\begin{figure}[!hbtp]
\centering
\includegraphics[width=.43\textwidth]{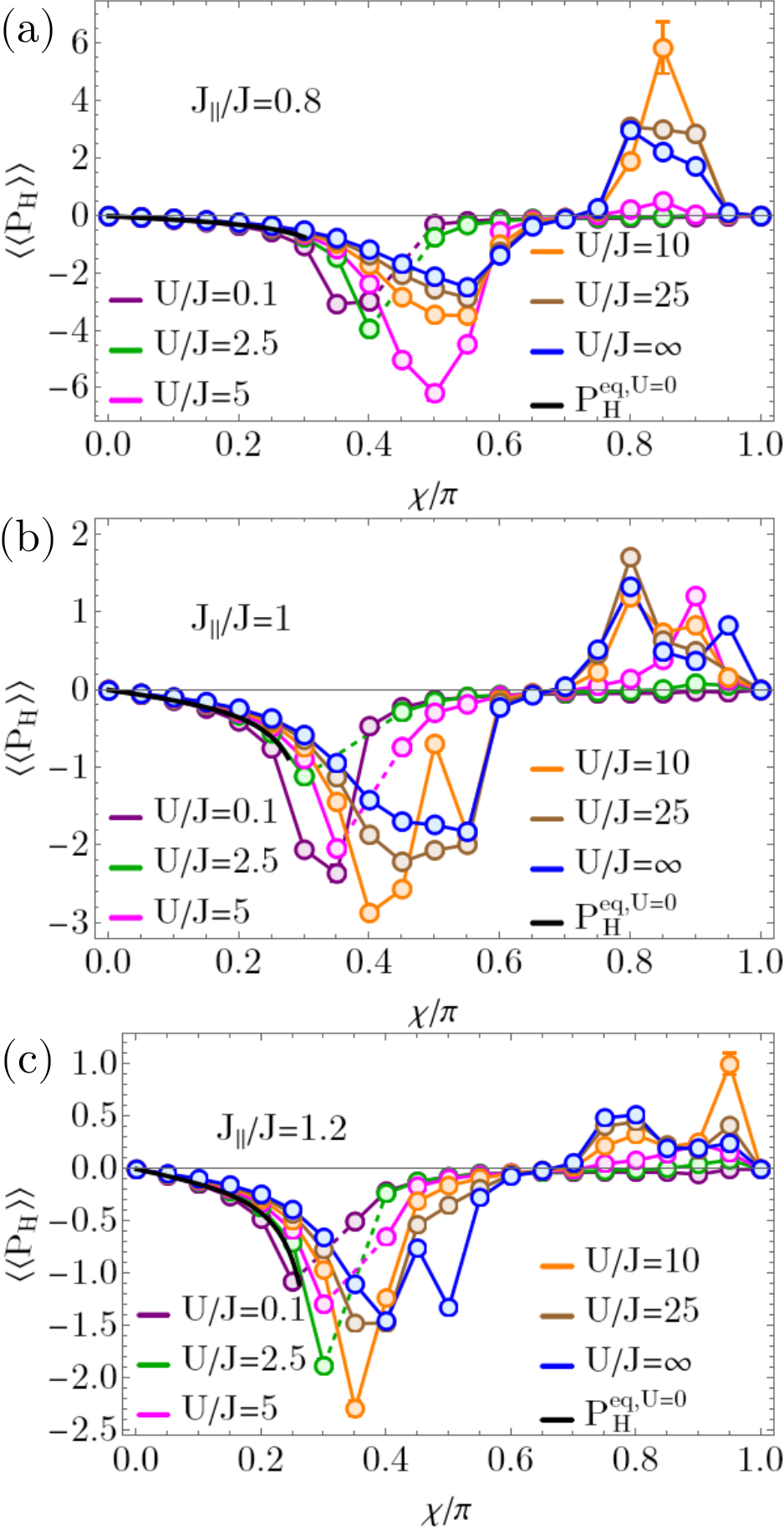}
\caption{
Time-averaged Hall polarization $\langle\langle P_H\rangle\rangle$ as a function of $\chi$ for (a) $J_\|/J=0.8$, (b) $J_\|/J=1$, (c) $J_\|/J=1.2$, for different values of the interaction strength $U$. The system size used is $L=90$, filling $\rho=0.25$ and the strength of the linear potential $\mu/J=0.001$.
The black curve at small values of the flux corresponds to the analytical value $P_H^{\text{eq},U=0}$, Eq.~(\ref{eq:P_Hall_analytic_U0}).
We note that in panel (a) the values shown for hardcore bosons, $U/J=\infty$, are taken from Ref.~\cite{HalatiGiamarchi2024}.
Dashed lines denote the region in which we could not define the value of $\langle\langle P_H\rangle\rangle$ as the current crosses zero during the time-evolution.
 }
\label{fig:PH_vs_flux_transitions}
\end{figure}

\begin{figure}[!hbtp]
\centering
\includegraphics[width=.48\textwidth]{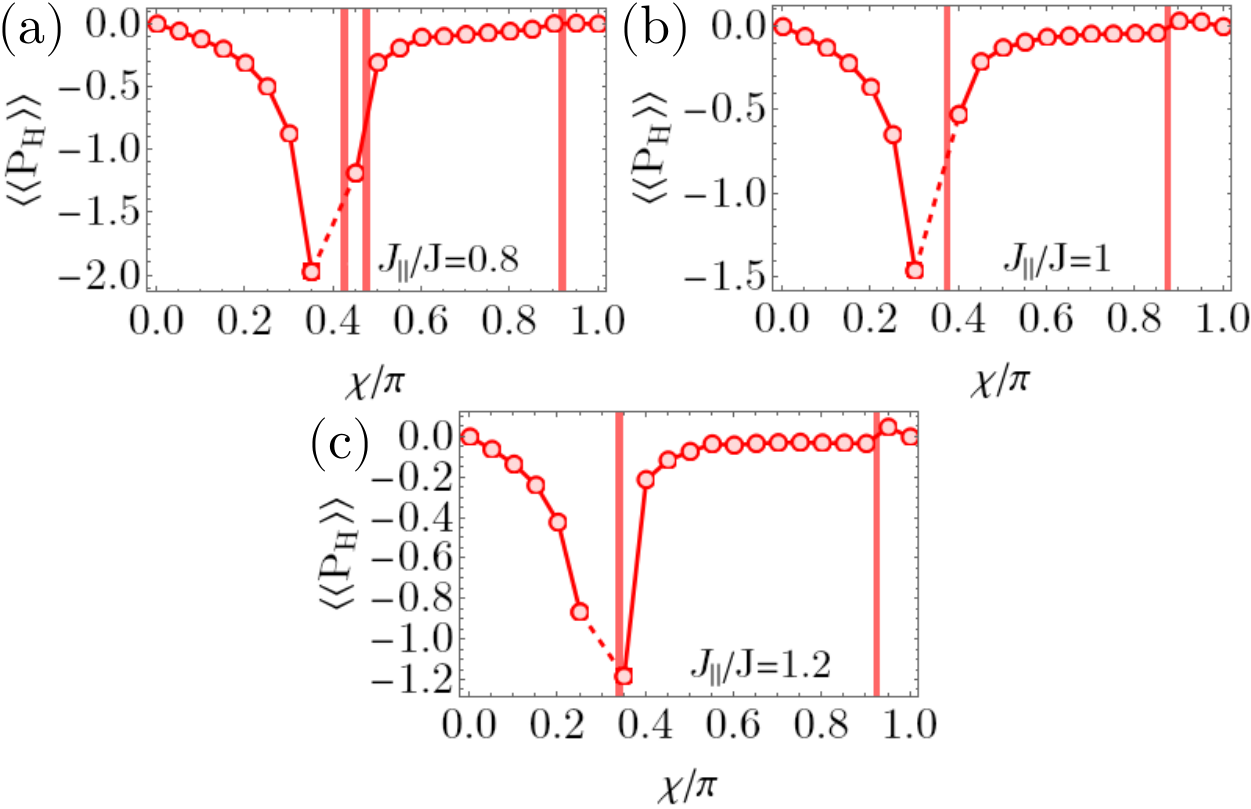}
\caption{
Time-averaged Hall polarization $\langle\langle P_H\rangle\rangle$ as a function of $\chi$ for (a) $J_\|/J=0.8$, (b) $J_\|/J=1$, (c) $J_\|/J=1.2$, for $U/J=1$. 
The system size used is $L=90$, filling $\rho=0.25$ and the strength of the linear potential $\mu/J=0.001$.
The vertical lines denote the phase transition threshold values as marked in the ground state phase diagrams in Fig.~\ref{fig:phasediag}.
Dashed lines denote the region in which we could not define the value of $\langle\langle P_H\rangle\rangle$ as the current crosses zero during the time-evolution.
 }
\label{fig:PH_vs_flux_transitions_U1}
\end{figure}

\subsection{Hall response across phase transitions \label{sec:response_PT}}

\begin{figure}[!hbtp]
\centering
\includegraphics[width=.48\textwidth]{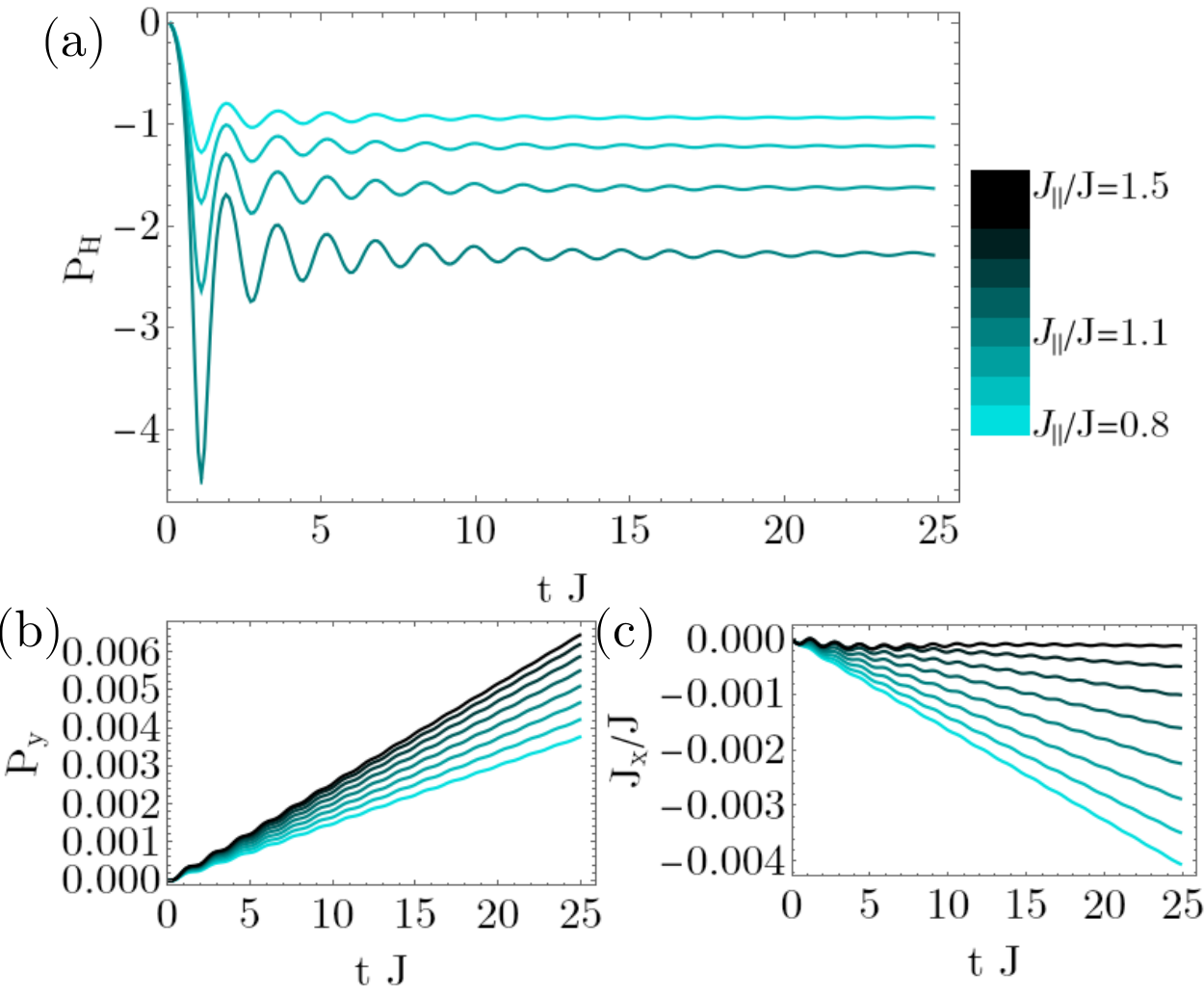}
\caption{
Time evolution in the Meissner superfluid phase towards the phase transition threshold of the (a) Hall polarization $P_H$, (b) density imbalance $P_y$, (c) current $\boldsymbol{J}_x/J$, for $\chi=0.3\pi$, $U/J=1$, for different values of the tunneling amplitude $J_\|/J$, in between $J_\|/J=0.8$ and $J_\|/J=1.1$ for $P_H$ and up to $J_\|/J=1.5$ for $P_y$ and $\boldsymbol{J}_x/J$. 
The system size used is $L=90$, and the strength of the linear potential $\mu/J=0.001$.
 }
\label{fig:PH_dynamics_Jp_transition_U1}
\end{figure}

\begin{figure}[!hbtp]
\centering
\includegraphics[width=.48\textwidth]{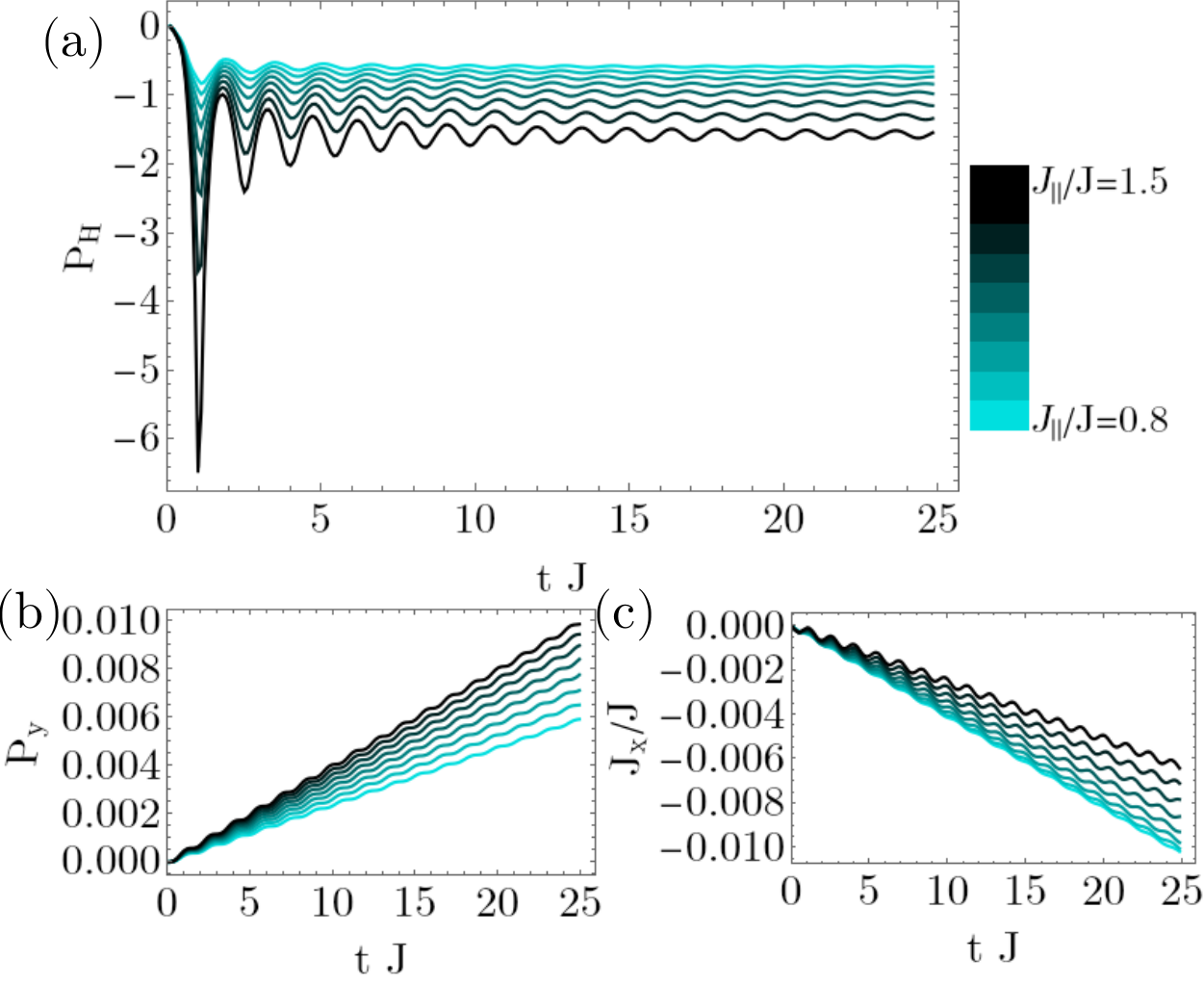}
\caption{
Time evolution in the Meissner superfluid phase towards the phase transition threshold of the (a) Hall polarization $P_H$, (b) density imbalance $P_y$, (c) current $\boldsymbol{J}_x/J$, for $\chi=0.3\pi$, $U/J=10$, for different values of the tunneling amplitude $J_\|/J$, in between $J_\|/J=0.8$ and $J_\|/J=1.5$. 
The system size used is $L=90$, and the strength of the linear potential $\mu/J=0.001$.
 }
\label{fig:PH_dynamics_Jp_transition_U10}
\end{figure}

In the following section we analyze the behavior of the Hall polarization as we vary the parameters of the model to cross from the Meissner superfluid to the biased-chiral superfluid and the vortex superfluid.
In Fig.~\ref{fig:PH_vs_Jp_transitions} we show $\langle\langle P_H\rangle\rangle$ as a function of $J_\|/J$ for $\chi\in\{0.3\pi,0.5\pi,0.8\pi,0.9\pi\}$ for several values of the interaction, for which the system crosses at least one phase boundary.
We can observe a very rich behavior, with large values of the Hall response, either negative, or positive, implying the change of sign of the Hall polarization, and a strong dependence on the value of $U/J$.

We focus first on the divergence-like feature we observe as we approach the phase transition to the vortex or biased phases from the Meissner phase, as we increase $J_\|/J$, e.g.~around $J_\|/J\approx0.7-1.5$ in Fig.~\ref{fig:PH_vs_Jp_transitions}(a), or around $J_\|/J\approx0.2-1.4$ in Fig.~\ref{fig:PH_vs_Jp_transitions}(c). In Fig.~\ref{fig:PH_vs_Jp_transitions_U1} we show $\langle\langle P_H\rangle\rangle$ as a function of $J_\|/J$ for $U/J=1$ for the same values of the flux as in Fig.~\ref{fig:PH_vs_Jp_transitions} and the phase transition thresholds are marked with vertical lines. We observe that the large increase of the Hall response in the M-SF corresponds to the presence of a phase transition.
We can understand this behavior already in the non-interacting limit, as in the following.
For $U/J=0$, the dispersion relation of the Hamiltonian is given by the expression in Eq.~(\ref{eq:dispersion}) for $\Tilde{\Phi}=0$, the lower band exhibits either a single, or a double minimum structure depending on the chosen parameters \cite{HalatiGiamarchi2023}. 
The Meissner phase is characterized by a single minimum, while the vortex and biased-chiral phase by two minima, with the transition threshold being defined by the parameters for which the lower band has a quartic minimum and satisfy the condition $1+4(J_\|/J)\cos(\chi)-4(J_\|/J)^2\sin(\chi)^2=0$.
However, for the parameters satisfying this condition also the current $\left\langle \boldsymbol{J}_x\right\rangle^\text{eq}$ vanishes, which implies that $P_H^{\text{eq},U=0}$ diverges as we approach the phase boundary from the Meissner phase, due to the change in the structure of the dispersion relation.
In Fig.~\ref{fig:PH_vs_flux_transitions} we can see the good agreement as we approach the divergence of $P_H^{\text{eq},U=0}$ (black curves) as a function of $\chi$, with the numerical result for weak interactions of $U/J=0.1$ (purple points).

\begin{figure}[!hbtp]
\centering
\includegraphics[width=.48\textwidth]{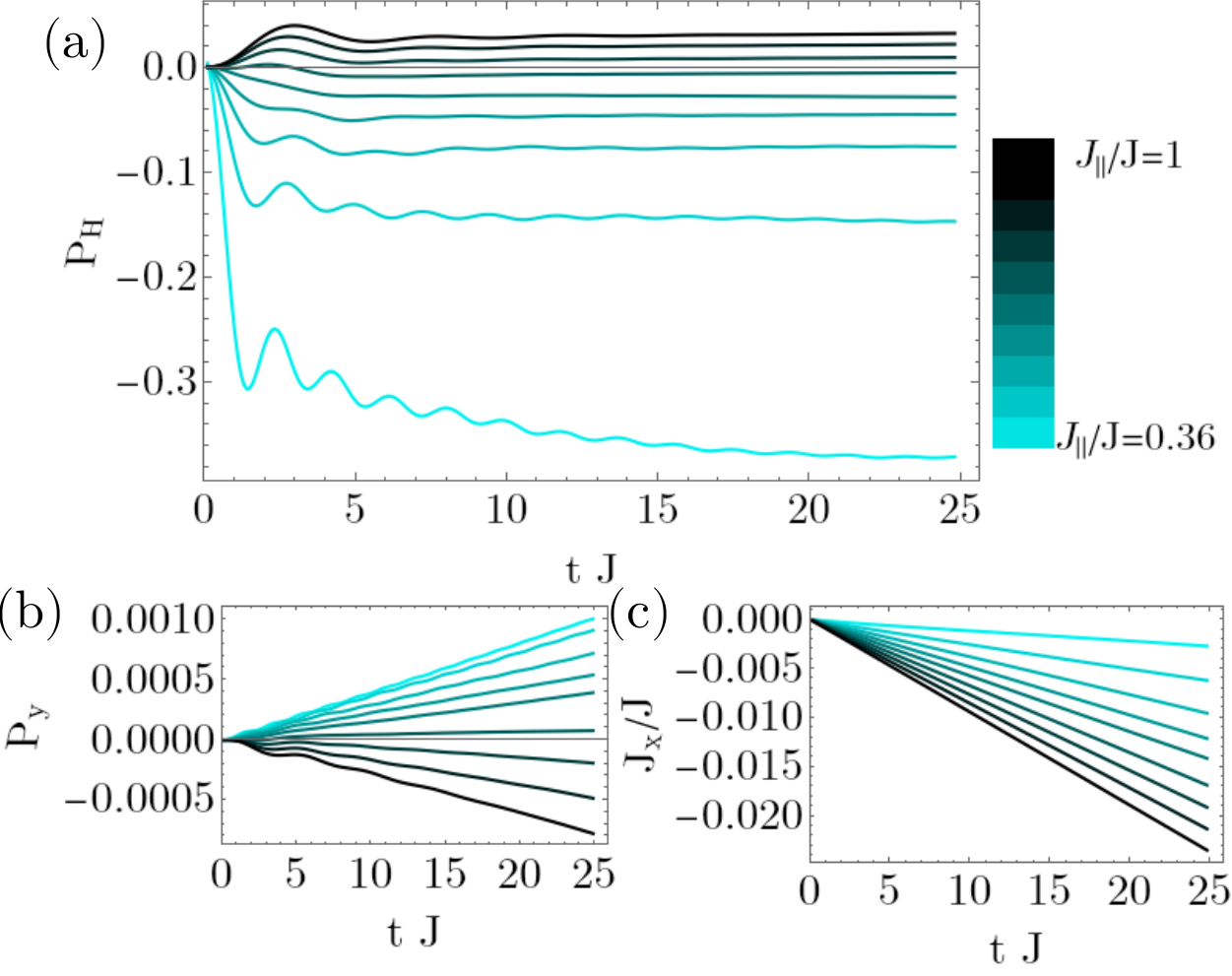}
\caption{
Time evolution in the biased-chiral superfluid phase of the (a) Hall polarization $P_H$, (b) density imbalance $P_y$, (c) current $\boldsymbol{J}_x/J$, for $\chi=0.9\pi$, $U/J=1$, for different values of the tunneling amplitude $J_\|/J$, in between $J_\|/J=0.36$ and $J_\|/J=1$. 
The system size used is $L=90$, and the strength of the linear potential $\mu/J=0.001$.
 }
\label{fig:PH_dynamics_Jp_BCSF_U1}
\end{figure}

In the numerical results at finite interaction strengths we do not expect that the current will vanish at the phase transition to obtain a divergence. However, for our protocol in which we quench a linear potential, presented in Sec.~\ref{sec:model}, even if the arising current is non-zero it can have values comparable with the amplitude of the oscillations present a short times. 
Thus, for weak interactions close to the phase boundary between M-SF and V-SF, or BC-SF, the short-time oscillations of current can determine a zero value of the current at certain points in time, which can prevent a well defined Hall polarization, Eq.~(\ref{eq:PH}).
For example, in Fig.~\ref{fig:PH_dynamics_Jp_transition_U1}(a) for $U/J=1$, we have a nicely behaved $P_H(t)$ for $0.8\leq J_\|/J\leq 1.1$, but for larger values of the tunneling $1.2\leq J_\|/J\leq 1.5$, the current crosses zero at several points in time, as seen in Fig.~\ref{fig:PH_dynamics_Jp_transition_U1}(c).
This is the reason why in Fig.~\ref{fig:PH_vs_Jp_transitions} and Fig.~\ref{fig:PH_vs_Jp_transitions_U1} for small values of the interactions, $U/J\lesssim 2$, we have points missing for certain values of the tunneling amplitude $J_\|/J$ for which we could not properly define a time-averaged $\langle\langle P_H\rangle\rangle$ (regions marked by dashed lines).
Roughly for values larger than $U/J=2.5$ we do not see the current crossing zero in its time-evolution for the parameters consider.
For $U/J=10$, results shown in Fig.~\ref{fig:PH_dynamics_Jp_transition_U10}, we observe oscillations in both the current and the density imbalance [Fig.~\ref{fig:PH_dynamics_Jp_transition_U10}(b),(c)], but as the magnitude of the current is large enough, such that we can extract a meaningful Hall polarization up to the phase boundary.
Even if for strong on-site interactions we do not have a divergence of the Hall polarization, we still see an influence of this single particle effect. 

Thus, we observe in Fig.~\ref{fig:PH_vs_Jp_transitions_U1} and Fig.~\ref{fig:PH_vs_flux_transitions_U1} for $U/J=1$, that by approaching the phase boundary marking the end of the Meissner phase $\langle\langle P_H\rangle\rangle$ increases rapidly with its maximum close to the critical point and, in most cases, followed by an abrupt decrease in its magnitude in the subsequent phase. A strong Hall response is seen also for larger values of the interactions (Fig.~\ref{fig:PH_vs_Jp_transitions} and Fig.~\ref{fig:PH_vs_flux_transitions}) close to the phase transition threshold, even in the absence of a divergence.
Interestingly, in Fig.~\ref{fig:PH_vs_Jp_transitions}(b), where the flux is $\chi=0.5\pi$, for $U/J\geq10$ we observe that the maximum of $\langle\langle P_H\rangle\rangle$ as a function of $J_\|/J$ does not correspond to the phase transition threshold, but rather is in the Meissner phase. However, also for these parameters after the phase transition we see a discontinuous jump in the value of $\langle\langle P_H\rangle\rangle$.

\begin{figure}[!hbtp]
\centering
\includegraphics[width=.48\textwidth]{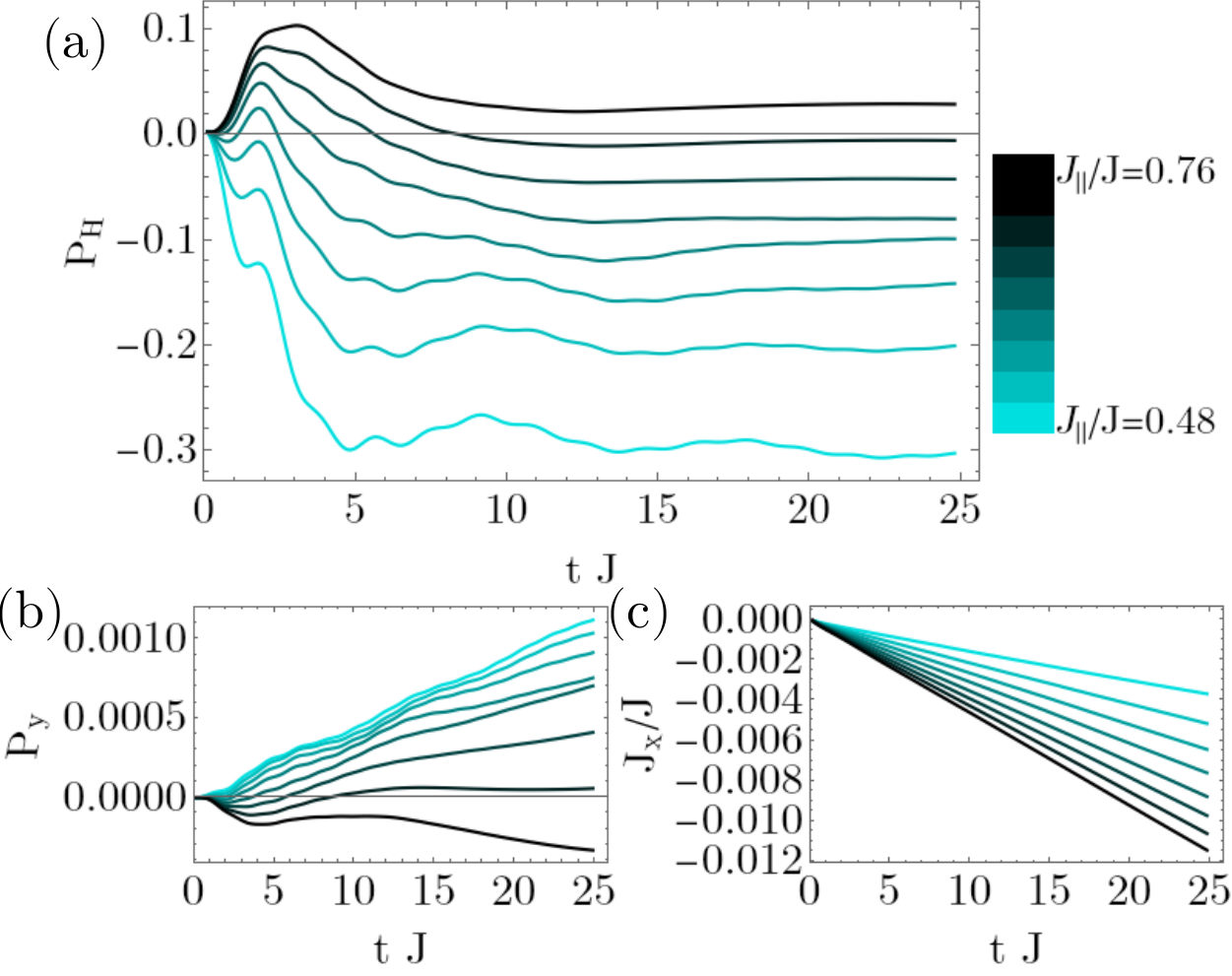}
\caption{
Time evolution in the biased-chiral superfluid phase of the (a) Hall polarization $P_H$, (b) density imbalance $P_y$, (c) current $\boldsymbol{J}_x/J$, for $\chi=0.9\pi$, $U/J=10$, for different values of the tunneling amplitude $J_\|/J$, in between $J_\|/J=0.48$ and $J_\|/J=0.76$. 
The system size used is $L=90$, and the strength of the linear potential $\mu/J=0.001$.}
\label{fig:PH_dynamics_Jp_BCSF_U10}
\end{figure}

\subsection{The Hall response of the biased-chiral superfluid phase \label{sec:response_BCSF}}

In this section, we investigate the behavior of the Hall polarization in the biased-chiral superfluid, phase characteristic to the triangular ladder having its origin in the frustrated nature of the system at larger values of the flux \cite{HalatiGiamarchi2023}.
We focus on the parameter regime for which the BC-SF phase has a larger extent, e.g.~for $\chi=0.8\pi$ in Fig.~\ref{fig:PH_vs_Jp_transitions_U1}(c) for $U/J=1$ we are in the BC-SF for $0.36\lesssim J_\|/J\lesssim0.66$, or for $\chi=0.9\pi$ in Fig.~\ref{fig:PH_vs_Jp_transitions_U1}(d) for $U/J=1$ we are in the BC-SF for $0.35\lesssim J_\|/J\lesssim1.14$ and for $U/J=10$ we are in the BC-SF for $0.46\lesssim J_\|/J\lesssim0.78$
(see also the phase diagrams in Fig.~\ref{fig:phasediag}).
$\langle\langle P_H\rangle\rangle$ in these regimes has a smooth behavior and relatively small values, in contrast to the behavior close to the transition thresholds, or in the vortex phase (see Sec.~\ref{sec:commensurability}).
Interestingly, we observe a change of sign of the Hall polarization as we increase the tunneling amplitude $J_\|/J$.
For this change of sign we lack an explanation similar to the one in Sec.~\ref{sec:Meissner} based on the change of the nature of carriers in the hardcore regime.
At the phase transition between V-SF and BC-SF (marked by the vertical line in Fig.~\ref{fig:PH_vs_flux_transitions_U1} present for $\chi>0.8\pi$) we observe a small discontinuity in $\langle\langle P_H\rangle\rangle$ as a function of the flux.
Furthermore, we obtain the same value of the Hall polarization for both symmetry broken states that span the ground state manifold of the BC-SF, regardless of their sign of the ground state density imbalance.

\begin{figure}[!hbtp]
\centering
\includegraphics[width=.48\textwidth]{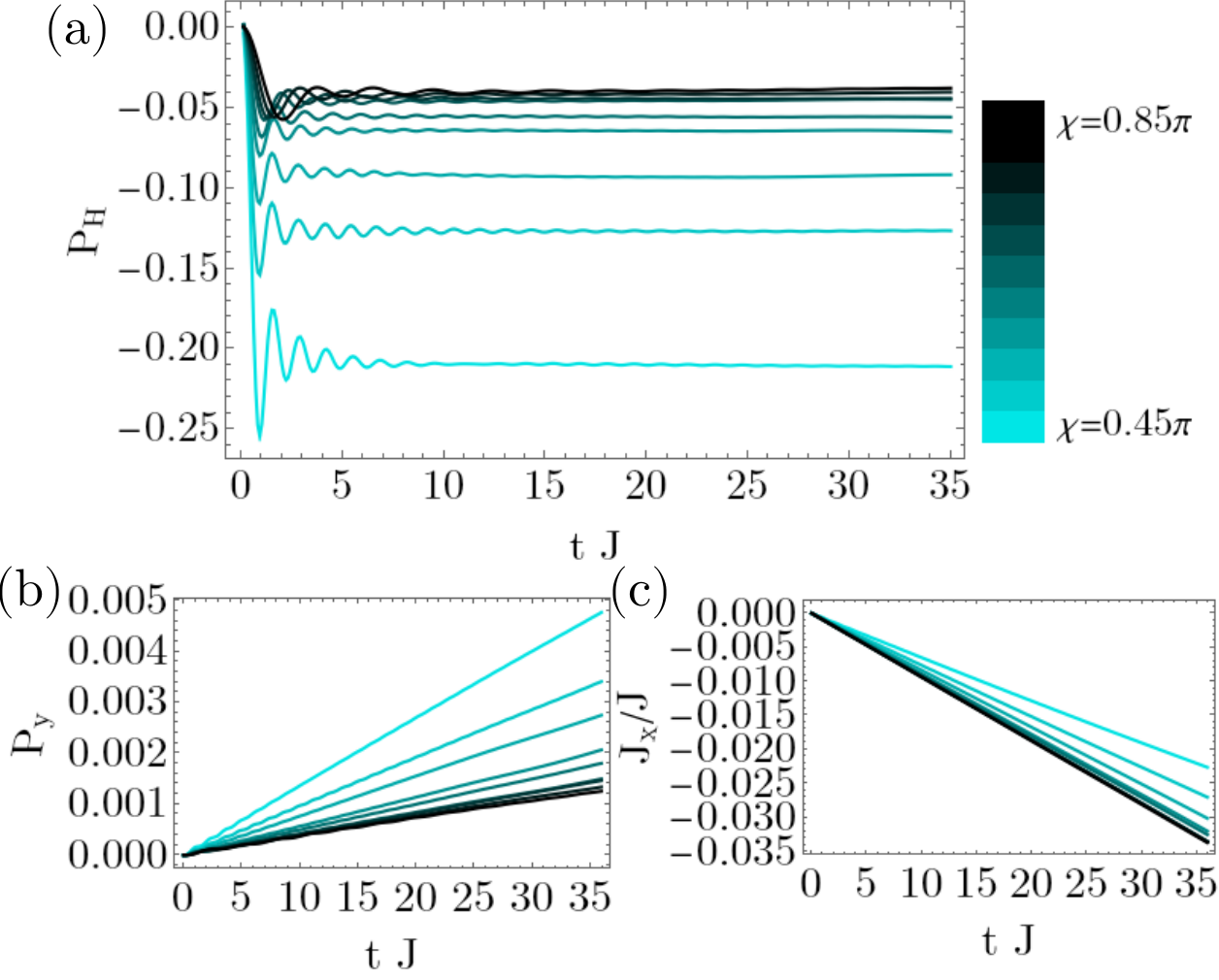}
\caption{
Time evolution in the vortex superfluid phase of the (a) Hall polarization $P_H$, (b) density imbalance $P_y$, (c) current $\boldsymbol{J}_x/J$, for $J_\|/J=1$ , $U/J=1$, for different values of the flux $\chi$, in between $\chi=0.45\pi$ and $\chi=0.85\pi$. 
The system size used is $L=90$, and the strength of the linear potential $\mu/J=0.001$.
 }
\label{fig:PH_dynamics_ch_VSF_U1}
\end{figure}

The dynamics of $P_H$ in the BC-SF is shown in Fig.~\ref{fig:PH_dynamics_Jp_BCSF_U1} for $U/J=1$ and in Fig.~\ref{fig:PH_dynamics_Jp_BCSF_U10} for $U/J=10$, together with the time-dependence of the density imbalance $P_y$ and current $\boldsymbol{J}_x/J$.
We observe that for both weak and strong interactions the dynamical behavior in the biased phase is different compared to what we saw in the previous section for the Meissner phase, giving a further motivation to investigate also the dynamics of the Hall polarization and not only its steady value.
In the case of $U/J=1$, Fig.~\ref{fig:PH_dynamics_Jp_BCSF_U1}(a), close to the phase boundary for $J_\|/J=0.36$ the Hall polarization reaches a steady value only for times $tJ\gtrsim 20$, while for larger values $J_\|/J$, deeper in the BC-SF, we reach faster the transient steady value, close to the parameters for which $\langle\langle P_H\rangle\rangle$ changes sign the fast initial oscillations are replaced with a slower oscillatory behavior.
The main features of the dynamics, and the change of sign, are due to the time-dependence of the density imbalance $P_y$, as seen in Fig.~\ref{fig:PH_dynamics_Jp_BCSF_U1}(b), while the current has a mostly linear time evolution, Fig.~\ref{fig:PH_dynamics_Jp_BCSF_U1}(c).
Similarly, also for stronger interactions, $U/J=10$ shown in Fig.~\ref{fig:PH_dynamics_Jp_BCSF_U10}, it seems that the oscillatory features of $P_H(t)$ stem from the dynamics of $P_y$. In this regime, we observe that the sign of $P_H$ is not necessarily determined at short times, as we can have an initial increase to positive values after which a relaxation to either positive, $J_\|/J=0.76$, or negative, $J_\|/J\lesssim0.72$, values occurs, see Fig.~\ref{fig:PH_dynamics_Jp_BCSF_U10}(a).

\subsection{Emergence of commensurability effects at large interaction \label{sec:commensurability}}

\begin{figure}[!hbtp]
\centering
\includegraphics[width=.48\textwidth]{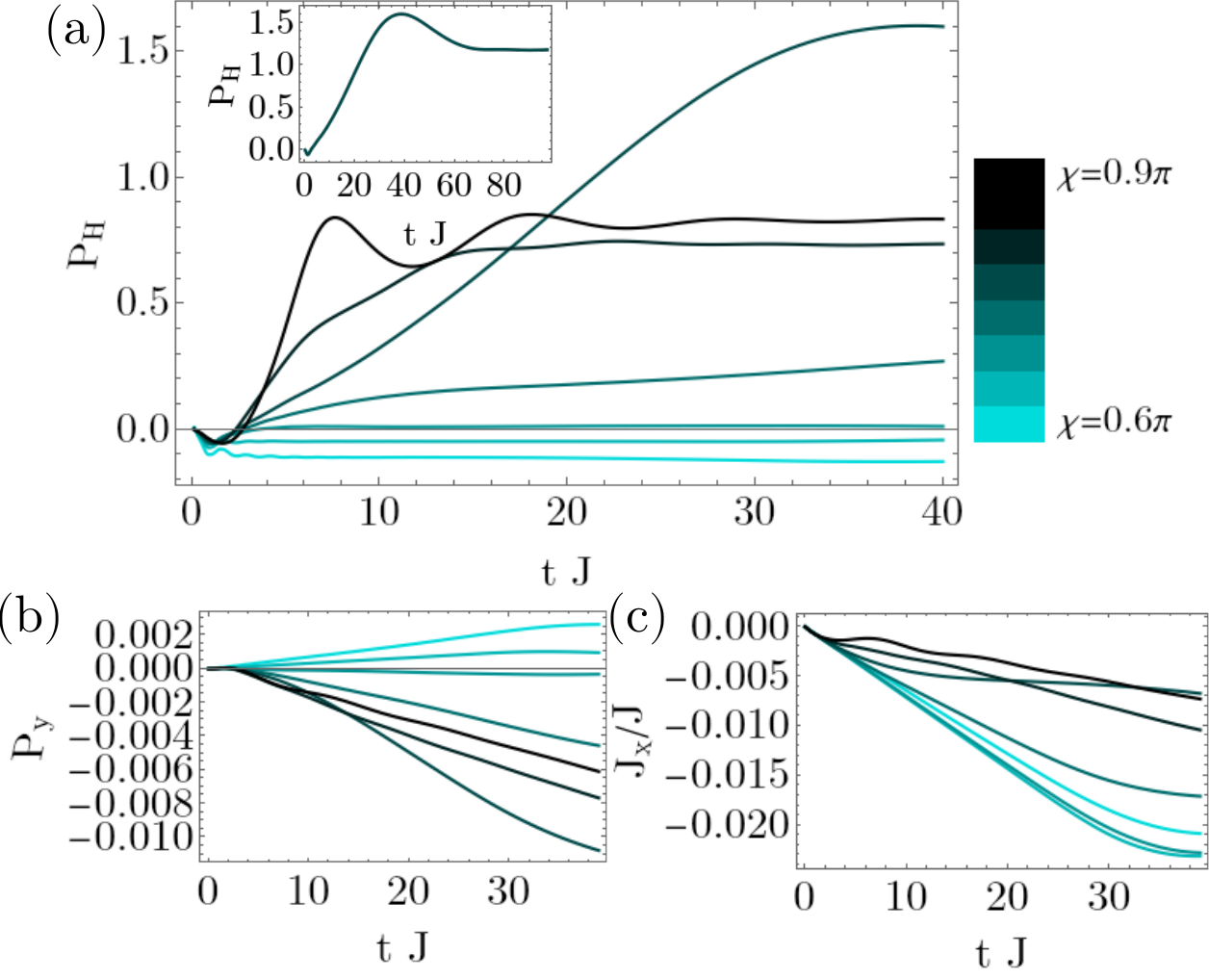}
\caption{
Time evolution in the vortex superfluid phase of the (a) Hall polarization $P_H$, (b) density imbalance $P_y$, (c) current $\boldsymbol{J}_x/J$, for $J_\|/J=1$ , $U/J=10$, for different values of the flux $\chi$, in between $\chi=0.6\pi$ and $\chi=0.9\pi$. 
The inset in panel (a) contains the dynamics of $P_H$ for $\chi=0.8\pi$ up to longer times.
The system size used is $L=90$, and the strength of the linear potential $\mu/J=0.001$.
}
\label{fig:PH_dynamics_ch_VSF_U10}
\end{figure}

\begin{figure}[!hbtp]
\centering
\includegraphics[width=.48\textwidth]{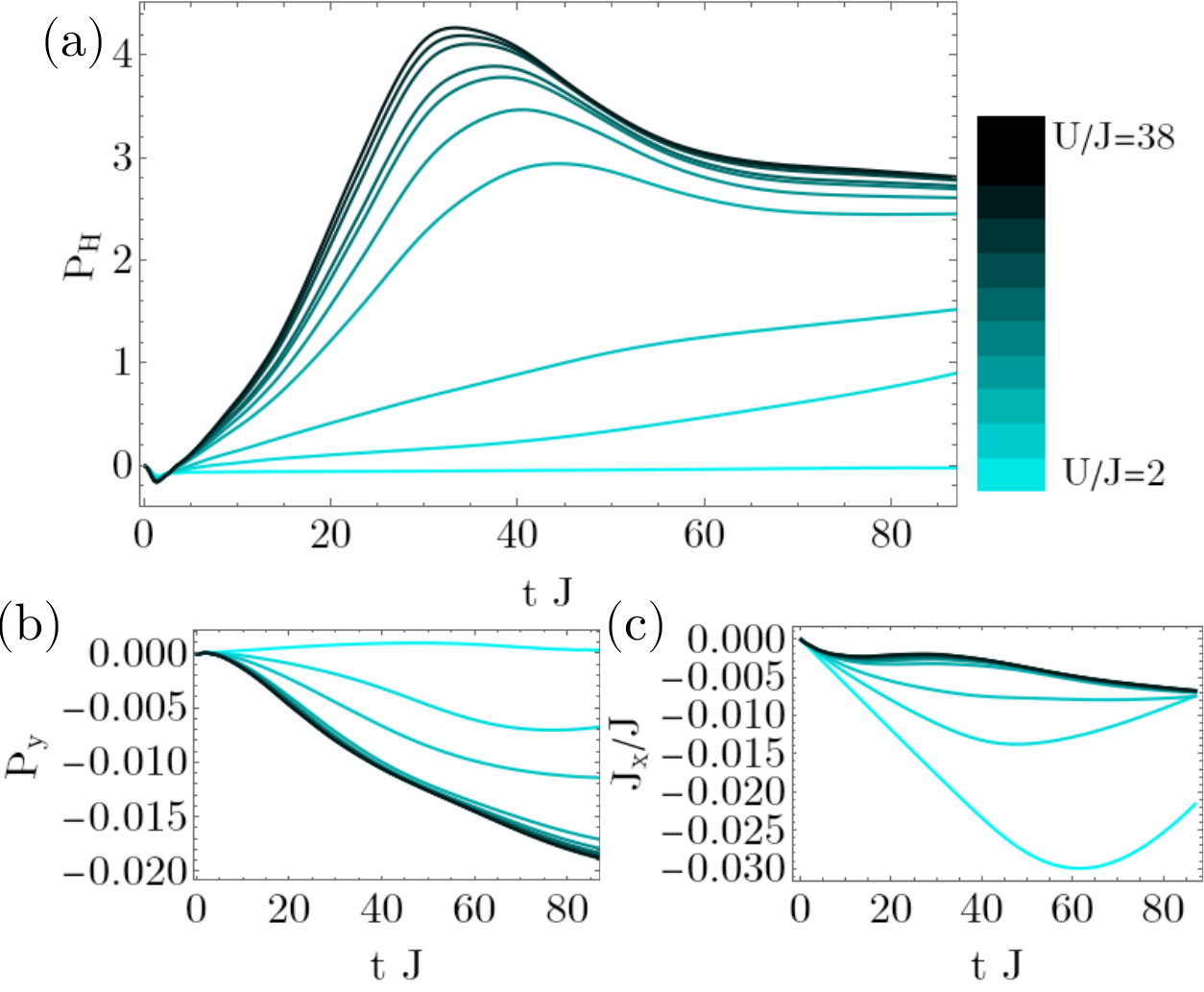}
\caption{
Time evolution in the vortex superfluid phase of the (a) Hall polarization $P_H$, (b) density imbalance $P_y$, (c) current $\boldsymbol{J}_x/J$, for $J_\|/J=0.76$, $\chi=0.8\pi$, for different values of the interaction strength $U$, in between $U/J=2$ and $U/J=38$. 
The system size used is $L=90$, and the strength of the linear potential $\mu/J=0.001$.
 }
\label{fig:PH_dynamics_U_vortex_ch08}
\end{figure}

In this section, we analyze the Hall polarization in the vortex superfluid phase, with a particular focus on the commensurability effects which are the cause of the strong positive response observed in Fig.~\ref{fig:PH_vs_Jp_transitions} and Fig.~\ref{fig:PH_vs_flux_transitions}.
We focus on understanding on how the strong positive Hall response emerges as we increase the strength of the interactions.
For example, in Fig.~\ref{fig:PH_vs_flux_transitions}(a) and Fig.~\ref{fig:PH_vs_flux_transitions_U1}(a) for $J_\|/J=1$, we observe that for weak interactions in the vortex superfluid phase ($0.37\pi\lesssim\chi\lesssim0.87\pi$ for $U/J=1$) the stationary value of the Hall polarization has a smooth behavior with a rather small magnitudes as a function of the flux.
In the case of $U/J=1$ the values remain of $\langle\langle P_H\rangle\rangle$ remain negative, by increasing the interactions to $U/J=2.5$ the magnitude of $\langle\langle P_H\rangle\rangle$ is still relatively small, but it exhibits a change of sign. Increasing the interactions to even larger values we observe a large positive response developing, with a peak around $\chi\approx0.8\pi$ for $U/J\gtrsim10$.
We can also see this behavior in the time-dependence of the Hall polarization, by contrasting the results for $U/J=1$ shown in Fig.~\ref{fig:PH_dynamics_ch_VSF_U1}, with the ones for $U/J=10$ shown in Fig.~\ref{fig:PH_dynamics_ch_VSF_U10}.
In Fig.~\ref{fig:PH_dynamics_ch_VSF_U1}(a) the dynamics of $P_H$ the dynamics resembles the one in the Meissner phase, e.g.~Fig.\ref{fig:PH_dynamics_Jp}(a), with oscillations at short times that are damped to a steady values, even though for $\chi\gtrsim0.7\pi$, as we approach the transition to the BC-SF, the period of the oscillations is larger and they persist to longer times.
This dynamical behavior is in contrast to what we observe in Fig.~\ref{fig:PH_dynamics_ch_VSF_U10}(a) for large values of the on-site interactions.
Here after the sign change of the late time value around $\chi\approx0.7\pi$, the dynamics is quite different, at very short times $P_H$ is negative, but afterwards $P_H$ starts increasing and becomes positive, due to a sign change of the polarization $P_y$ for these values of the flux, Fig.~\ref{fig:PH_dynamics_ch_VSF_U10}(b).
At even longer times we see a much slower dynamics towards a steady value than compared to the other phases investigated so far, in particular, for the largest value of $\langle\langle P_H\rangle\rangle$ occurring at $\chi=0.8\pi$ a stationary behavior can be be identified only for times larger than $tJ\gtrsim60$ [see inset of Fig.~\ref{fig:PH_dynamics_ch_VSF_U10}(a)].

\begin{figure}[!hbtp]
\centering
\includegraphics[width=0.48\textwidth]{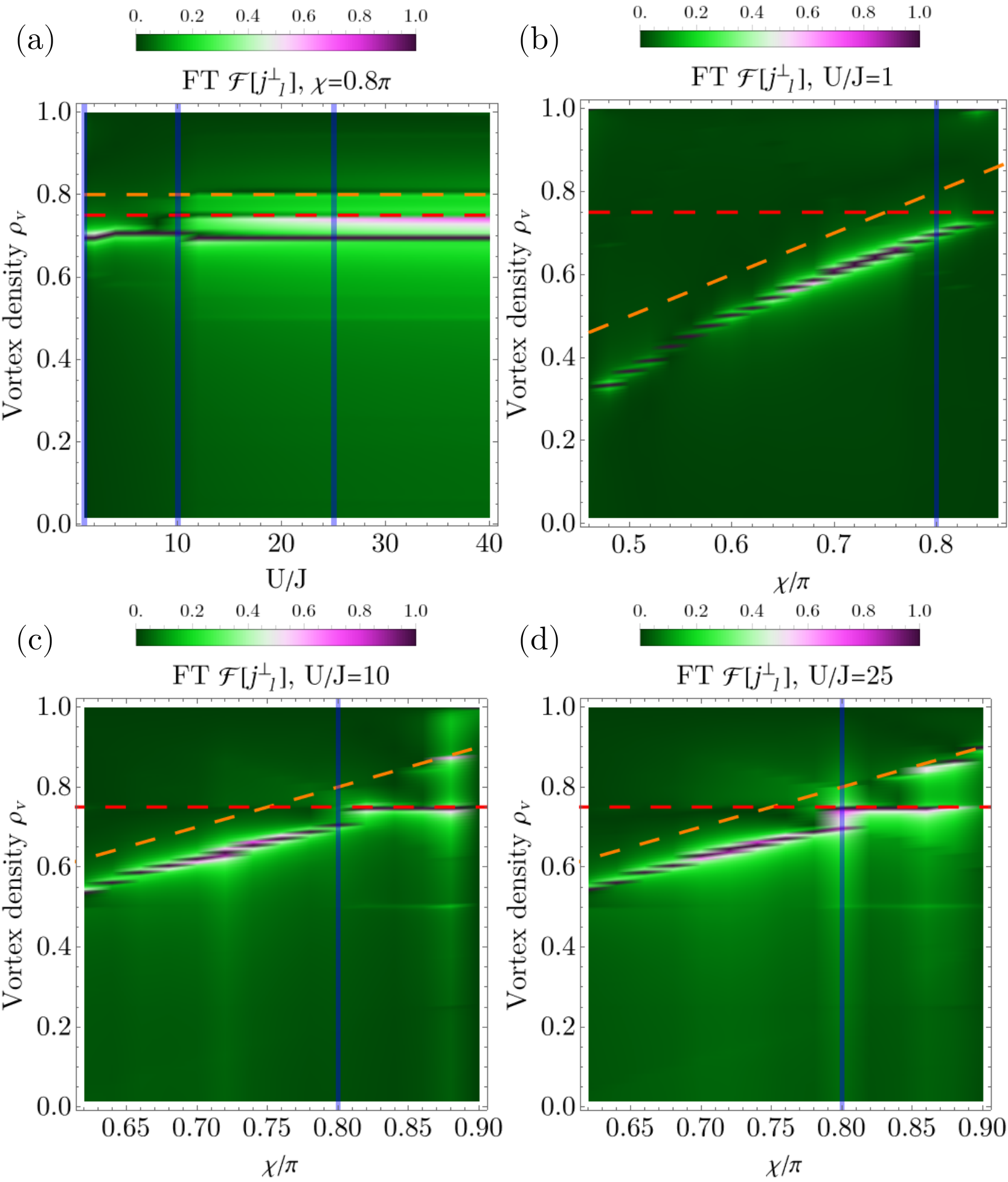}
\caption{\label{fig:rho_v_Jp076}
The Fourier transform of the ground state local rung currents, $j^\perp_j$, as a function of (a) the on-site interactions $U/J$, (b)-(d) the flux $\chi$ for $J_\|/J=0.76$, $\rho=0.25$ and (a) $\chi=0.8\pi$, (b) $U/J=1$, (c) $U/J=10$, (d) $U/J=25$. 
The vertical axis has been scaled in terms of the vortex density $\rho_v$.
The orange dashed lines corresponds to the expectation of the vortex superfluid phase of $\rho_v=\chi/\pi$, the red dashed lines corresponds to the value $\rho_v=1-\rho=0.75$. 
In the vertical blue lines marks in (a) the interactions strengths used in panels (b)-(d), and in (b)-(d) the values of the flux used in panel (a).
The system size used is (a) $L=90$, (b)-(d) $L=120$.
We normalize the Fourier transform such that its maximum is equal to one for each column.}
\end{figure}

In Ref.~\cite{HalatiGiamarchi2024} we have identified a similar behavior of saturation to large positive values after a slow dynamics in the case of hardcore bosons, which we attributed to the presence of a vortex density commensurate fixed by the value of the atomic filling. 
We found that the peak of the positive response strongly correlates with the parameters for which the commensurate vortex density dominates the expected incommensurate value of V-SF, for a wide range of parameters and atomic fillings.
As this is similar to what we observe at strong finite interaction values, i.e.~the strong positive Hall response appears as we increase the on-site interactions at the same time as the commensurate vortex density, we briefly sketch the origin of the second vortex density value \cite{HalatiGiamarchi2024}.
One approach to deal with the Hamiltonian given in Eq.~(\ref{eq:Hamiltonian}) for the case of hardcore, $U\to\infty$, interactions is to employ a Jordan-Wigner transformation to fermionic operators $c_j$, $b_j=\prod_{l=1}^{j-1}e^{i\pi c^\dagger_l c_l}c_j$. For a chain geometry, without other interactions, this transformation maps the hardcore bosons to free fermions. In contrast, for our triangular geometry we obtain an interacting fermionic model as the Jordan-Wigner string does not cancel and we have in the Hamiltonian a term with four fermionic operators.
In Ref.~\cite{HalatiGiamarchi2024} we showed that by varying this term we interpolate between a free fermions equivalent of Eq.~(\ref{eq:Hamiltonian}) and the hardcore bosons model, and for intermediate values of the fermionic interaction we obtain a vortex lattice superfluid with the vortex density $\rho_v=1-\rho$ determined by the atomic filling.
Interestingly, even if we have a phase transition to the incommensurate vortex superfluid as we approach the hardcore bosons model we can still identify a peak corresponding to $\rho_v=1-\rho$ in the Fourier transform of the rung currents, which appears to determine a large positive value of the Hall polarization.

\begin{figure}[!hbtp]
\centering
\includegraphics[width=.48\textwidth]{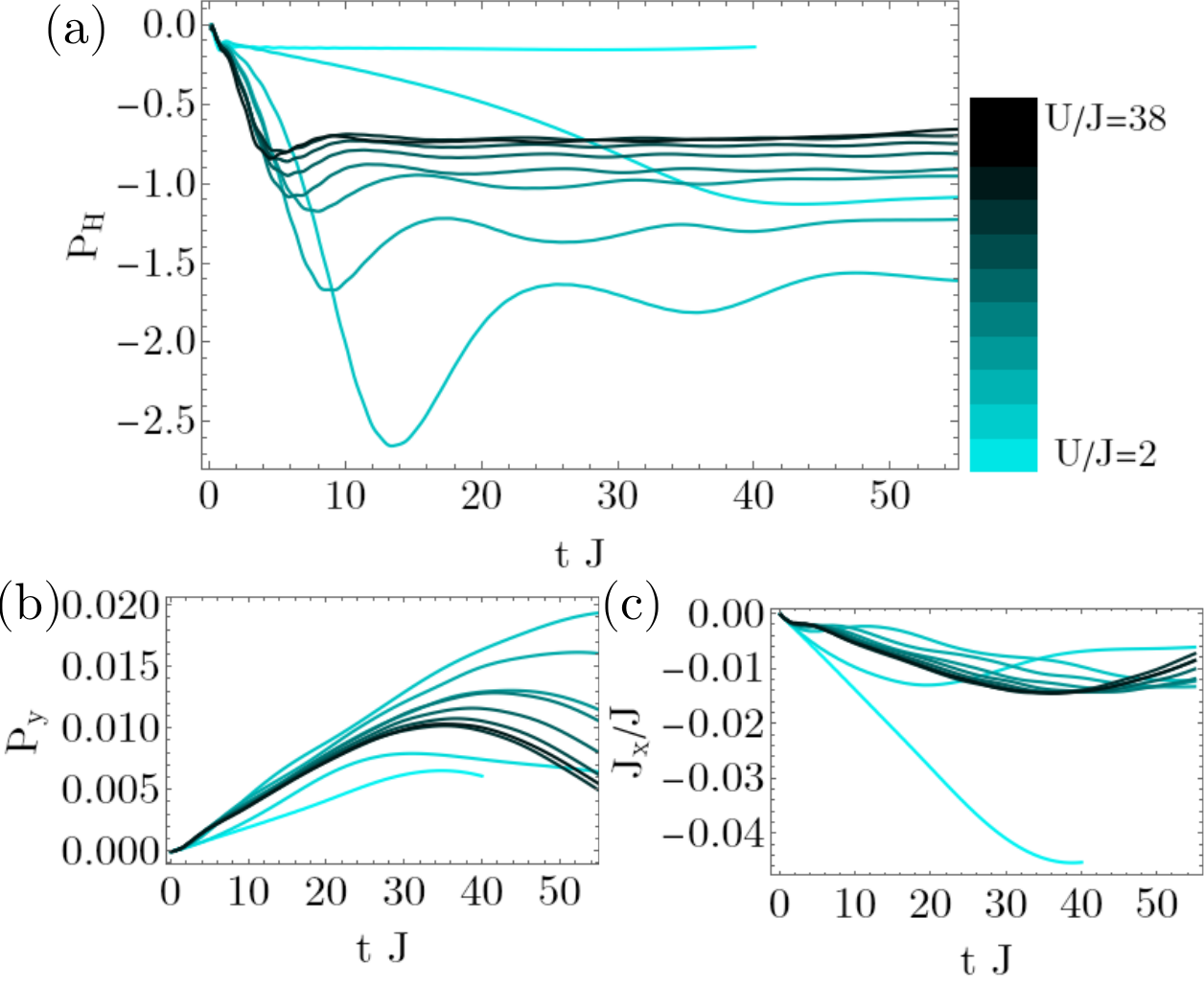}
\caption{
Time evolution in the vortex superfluid phase of the (a) Hall polarization $P_H$, (b) density imbalance $P_y$, (c) current $\boldsymbol{J}_x/J$, for $J_\|/J=2$, $\chi=0.3\pi$, for different values of the interaction strength $U$, in between $U/J=2$ and $U/J=38$. 
The system size used is $L=90$, and the strength of the linear potential $\mu/J=0.001$.
 }
\label{fig:PH_dynamics_U_vortex_ch03}
\end{figure}

\begin{figure}[!hbtp]
\centering
\includegraphics[width=0.48\textwidth]{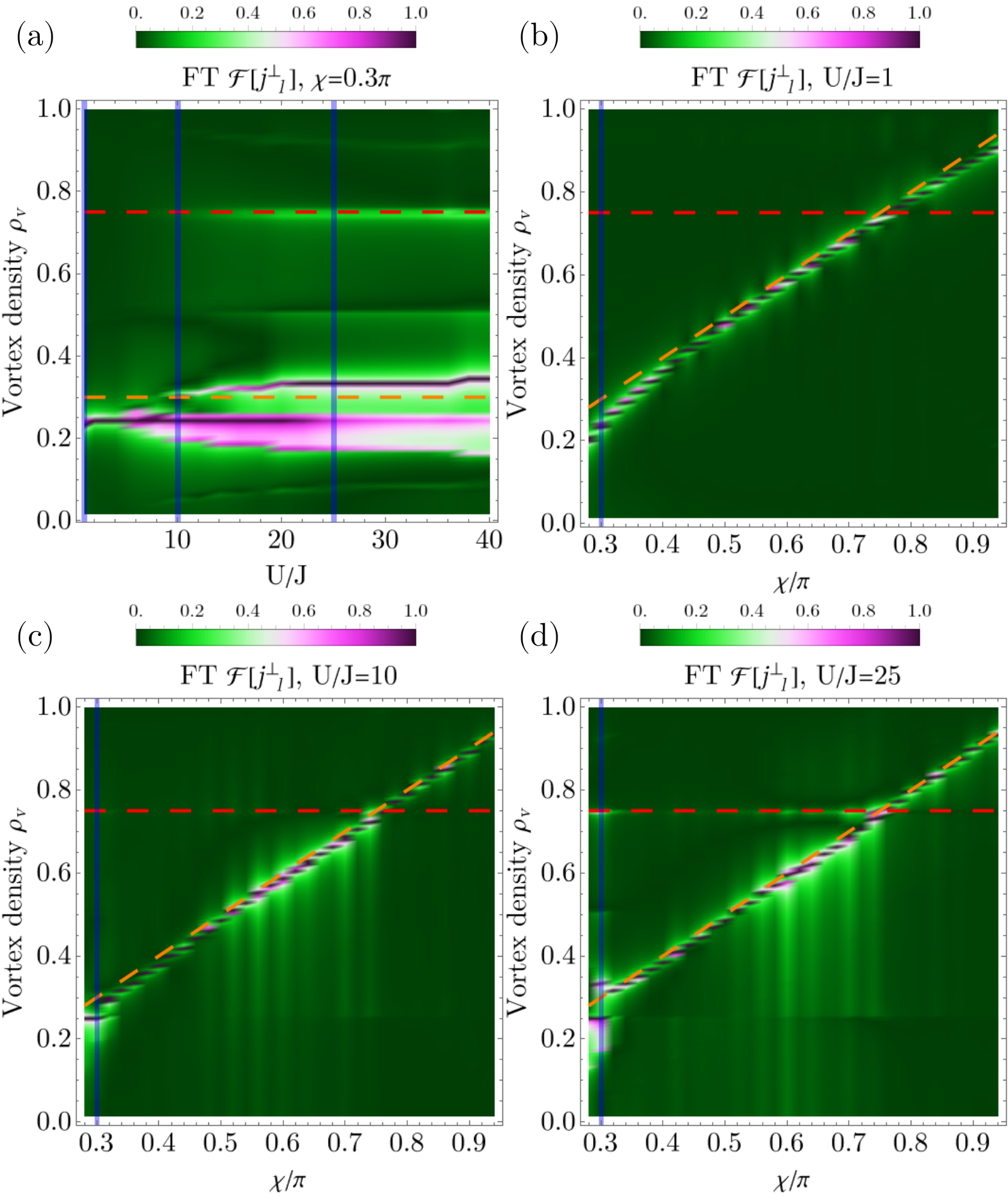}
\caption{\label{fig:rho_v_Jp2}
The Fourier transform of the ground state local rung currents, $j^\perp_j$, as a function of (a) the on-site interactions $U/J$, (b)-(d) the flux $\chi$ for $J_\|/J=2$, $\rho=0.25$ and (a) $\chi=0.3\pi$, (b) $U/J=1$, (c) $U/J=10$, (d) $U/J=25$. 
The vertical axis has been scaled in terms of the vortex density $\rho_v$.
The orange dashed lines corresponds to the expectation of the vortex superfluid phase of $\rho_v=\chi/\pi$, the red dashed lines corresponds to the value $\rho_v=1-\rho=0.75$. 
In the vertical blue lines marks in (a) the interactions strengths used in panels (b)-(d), and in (b)-(d) the values of the flux used in panel (a).
The system size used is (a) $L=90$, (b)-(d) $L=120$.
We normalize the Fourier transform such that its maximum is equal to one for each column.}
\end{figure}

In Fig.~\ref{fig:PH_dynamics_U_vortex_ch08}(a) we show the time dependence of $P_H$ for a wide range of the interactions $2\leq U/J\leq 38$ up to long times, $tJ\gtrsim80$, for parameters for which in the strongly interacting limit we observe a strong positive response, for $J_\|/J=0.76$ and $\chi=0.8\pi$.
For $U/J=2$ we obtain a small negative stationary value, however we can only follow the dynamics up to times $tJ\approx50$ before the finite size effects become important, as seen in the change of the monotony of the current in Fig.~\ref{fig:PH_dynamics_U_vortex_ch08}(c). 
Similarly, for the next two interaction strength values, $U/J=6$ and $U/J=10$, the finite size effects are relevant before we can identify a stationary Hall polarization.
However, at even larger values $U/J>10$ we can reliable compute the Hall polarization up to long times of $tJ\approx80$ and identify at plateau for $tJ\gtrsim60$ to which $P_H$ stabilizes after a slow increase and an intermediate maximum.
Next, we correlate the behavior of the Hall polarization with the ground state vortex density.
We define the vortex density of the V-SF as the values for which we have a well defined peak in the Fourier transform of the ground state local rung currents, as shown in Fig.~\ref{fig:rho_v_Jp076}.
In Fig.~\ref{fig:rho_v_Jp076}(a) we show the Fourier transform of the rung currents as a function of $U/J$, we see that for weak interaction we have a single vortex density $\rho_v\approx0.7$, while around $U/J\approx 10$ multiple peaks appear in the Fourier transform, leading to the identification of multiple vortex densities. 
This corresponds exactly the the parameter regime of the large positive values of the Hall polarization.
In order to understand better the vortex densities values present in the Fourier transform we plot the dependence on the flux for different $U/J$ in Fig.~\ref{fig:rho_v_Jp076}(b)-(d).
For $U/J=1$, Fig.~\ref{fig:rho_v_Jp076}(b), we have a single vortex density varying linearly with the flux, corresponding to the expected incommensurate value of the vortex superfluid.
We note the relation $\rho_v=\chi/\pi$ is only valid for large values of $J_\|/J$ \cite{HalatiGiamarchi2023}, but in Fig.~\ref{fig:rho_v_Jp076} we consider $J_\|/J=0.76$, explaining why even if $\rho_v$ has a linear dependence, it does not exactly agree with the value $\chi/\pi$.
If we increase the interactions to $U/J=10$ and $U/J=25$ in Fig.~\ref{fig:rho_v_Jp076}(c)-(d), we observe an additional peak in the Fourier transform from which we can identify a vortex density $\rho_v=0.75=1-\rho$, for which the value is related to the atomic filling \cite{HalatiGiamarchi2024}.
In particular, we can see that this commensurate vortex density dominates around $0.8\pi\lesssim\chi\lesssim0.85\pi$, but without changing the two-mode gapless nature of the vortex phase  \cite{HalatiGiamarchi2023}. 
For these values the incommensurate vortex density show a behavior similar to an avoided crossing, which explains why in Fig.~\ref{fig:rho_v_Jp076}(a) we see three peaks in the Fourier transform at $\chi=0.8\pi$.
We observe the same phenomenology also for the other parameters for which we see a large positive value of $\langle\langle P_H\rangle\rangle$ [see Fig.~\ref{fig:PH_vs_Jp_transitions}(c)-(d) and Fig.~\ref{fig:PH_vs_flux_transitions}(a)-(c)], with a similar correspondence to the appearance of a dominant peak in the Fourier transform of the rung currents at $\rho_v=1-\rho$.

In the last part of this section we discuss the behavior of the Hall polarization for $\chi=0.3\pi$ as we enter in the vortex superfluid phase, shown in Fig.~\ref{fig:PH_vs_Jp_transitions}(a). 
Compared to our previous discussion (see Sec.~\ref{sec:response_PT}), we see an unusual behavior for $U/J=10$ where $\langle\langle P_H\rangle\rangle$ does not exhibit a drop on its magnitude after crossing the phase boundary at $J_\|/J\approx1.65$, as observed for the other parameters.
One explanation for this could be that for $U/J=10$ and $\chi=0.3\pi$ by plotting $\langle\langle P_H\rangle\rangle$ as a function of $J_\|/J$  we are really close the phase boundary for $J_\|/J\gtrsim1.65$ [see Fig.~\ref{fig:phasediag}(b)], while for the other interaction values the transition threshold is a slightly lower values of the flux.
However, it is still interesting to analyze the time dependence of $P_H$, Fig.~\ref{fig:PH_dynamics_U_vortex_ch03}, and the behavior of the ground state vortex density, Fig.~\ref{fig:rho_v_Jp2}, for these parameters to gain additional insight.
We can observe in Fig.~\ref{fig:PH_dynamics_U_vortex_ch03}(a) that increasing the interaction strength from $U/J=2$ to $U/J=6$ results in a drastic increase in the time scale for reaching the stationary value. For even stronger interactions, $U/J\geq 10$, the steady plateau is reached earlier, with the minimal value of $\langle\langle P_H\rangle\rangle$ reached for $U/J= 10$.
The dynamics in this regime resembles the one showed in Fig.~\ref{fig:PH_dynamics_ch_VSF_U10}(a) for $\chi=0.9\pi$, but with an opposite sign.
By analyzing the behavior of the vortex density, we obtain that $U/J= 10$ is also the value for which multiple peaks in the Fourier transform of the rung currents appear, Fig.~\ref{fig:rho_v_Jp2}(a).
For weak interactions [$U/J=1$ in Fig.~\ref{fig:rho_v_Jp2}(b)], the vortex density is dominated by the incommensurate response $\rho_v=\chi/\pi$, with small deviations for smaller values of the flux close to the transition point to the Meissner phase. 
However, starting from $U/J=10$ two additional peaks are present in the Fourier transform, corresponding the the vortex densities $\rho_v=1-\rho$ and $\rho_v=\rho$, which seem to lead to a strong Hall response.
We note that we did not check for multiple parameter sets that the commensurate vortex density of $\rho_v=\rho$ within the vortex superfluid leads to a large negative value of $\langle\langle P_H\rangle\rangle$, as for the parameters considered in this work the vortex phase did not extent to low enough values of the flux.
This is in contrast with our analysis for the large positive $\langle\langle P_H\rangle\rangle$ being correlated to the presence of $\rho_v=1-\rho$ appearing at larger values of $\chi$, which we observed for many parameter sets.

\section{Discussions and Conclusions \label{sec:conclusions}}

To summarize, we have investigated the behavior of the Hall polarization for a triangular Bose-Hubbard ladder in a magnetic field, focusing on the effects of the on-site interactions from the weakly interacting regime to the hardcore limit.
We compute the time-evolution of the system following the quench of a linear potential which induces a current through the system, analyzing both the short-time non-equilibrium dynamics of the Hall polarization and its long-time saturation value.
We show that the Hall polarization can be employed to fingerprint and probe many of the features of the underlying ground state phase diagram, being particularly sensitive to the phase boundaries and the interplay of commensurate-incommensurate effects occurring at strong interactions.
In the non-interacting limit the equilibrium Hall polarization diverges as we approach the phase transition from the M-SF to the V-SF or BC-SF. Interestingly, this single particle effect can still determine a very strong negative Hall response also in the regimes of strong interactions.
Our results also show the possibility of changing the sign of the Hall polarization, for some parameter sets, e.g.~large fillings $\rho>0.5$ and strong interactions, this can be explained by the change of the character of the carries from particles to holes, while for other regimes, e.g.~$\rho=0.25$ in the V-SF and BC-SF, we do not have a similar argument.
Sign changes in the Hall response have also previously been linked to the presence of a  topological phase transition \cite{HuberLindner2011, BergLindner2015}.
We could correlate the strong positive values of the Hall polarization present for strong interactions to the presence of a commensurate vortex density in the otherwise incommensurate vortex superfluid.

We expect that our work will be experimentally relevant in the near future, as the Hall response has been measured for ultracold fermionic atoms confined to square ladders \cite{ZhouFallani2023}.
Triangular flux ladders have been realized in momentum space in Ref.~\cite{LiJia2023}, or could be achieved in real-space by employing optical lattices at the antimagic wavelength \cite{BaldelliBarbiero2024}.
Further motivation is given by the ongoing experimental interest in exploring frustration-driven quantum phenomena with ultracold atoms in triangular geometries \cite{BeckerSengstock2010,StruckSengstock2011, YangSchauss2021, MongkolkiattichaiSchauss2023, XuGreiner2023, LebratGreiner2024, PrichardBakr2024}.

\section*{ACKNOWLEDGMENTS}
We thank J.-S.~Bernier, M.~Filippone for fruitful discussions.
This work was supported by the Swiss National Science Foundation under Division II grant 200020-219400.


\begin{thebibliography}{64}%
\makeatletter
\providecommand \@ifxundefined [1]{%
 \@ifx{#1\undefined}
}%
\providecommand \@ifnum [1]{%
 \ifnum #1\expandafter \@firstoftwo
 \else \expandafter \@secondoftwo
 \fi
}%
\providecommand \@ifx [1]{%
 \ifx #1\expandafter \@firstoftwo
 \else \expandafter \@secondoftwo
 \fi
}%
\providecommand \natexlab [1]{#1}%
\providecommand \enquote  [1]{``#1''}%
\providecommand \bibnamefont  [1]{#1}%
\providecommand \bibfnamefont [1]{#1}%
\providecommand \citenamefont [1]{#1}%
\providecommand \href@noop [0]{\@secondoftwo}%
\providecommand \href [0]{\begingroup \@sanitize@url \@href}%
\providecommand \@href[1]{\@@startlink{#1}\@@href}%
\providecommand \@@href[1]{\endgroup#1\@@endlink}%
\providecommand \@sanitize@url [0]{\catcode `\\12\catcode `\$12\catcode
  `\&12\catcode `\#12\catcode `\^12\catcode `\_12\catcode `\%12\relax}%
\providecommand \@@startlink[1]{}%
\providecommand \@@endlink[0]{}%
\providecommand \url  [0]{\begingroup\@sanitize@url \@url }%
\providecommand \@url [1]{\endgroup\@href {#1}{\urlprefix }}%
\providecommand \urlprefix  [0]{URL }%
\providecommand \Eprint [0]{\href }%
\providecommand \doibase [0]{http://dx.doi.org/}%
\providecommand \selectlanguage [0]{\@gobble}%
\providecommand \bibinfo  [0]{\@secondoftwo}%
\providecommand \bibfield  [0]{\@secondoftwo}%
\providecommand \translation [1]{[#1]}%
\providecommand \BibitemOpen [0]{}%
\providecommand \bibitemStop [0]{}%
\providecommand \bibitemNoStop [0]{.\EOS\space}%
\providecommand \EOS [0]{\spacefactor3000\relax}%
\providecommand \BibitemShut  [1]{\csname bibitem#1\endcsname}%
\let\auto@bib@innerbib\@empty
\bibitem [{\citenamefont {Kitaev}(2003)}]{Kitaev2003}%
  \BibitemOpen
  \bibfield  {author} {\bibinfo {author} {\bibfnamefont {A.}~\bibnamefont
  {Kitaev}},\ }\emph {Fault-tolerant quantum computation by anyons},\ \href
  {\doibase https://doi.org/10.1016/S0003-4916(02)00018-0} {\bibfield
  {journal} {\bibinfo  {journal} {Annals of Physics}\ }\textbf {\bibinfo
  {volume} {303}},\ \bibinfo {pages} {2} (\bibinfo {year} {2003})}\BibitemShut
  {NoStop}%
\bibitem [{\citenamefont {Nayak}\ \emph {et~al.}(2008)\citenamefont {Nayak},
  \citenamefont {Simon}, \citenamefont {Stern}, \citenamefont {Freedman},\ and\
  \citenamefont {Das~Sarma}}]{NayakDasSarma2008}%
  \BibitemOpen
  \bibfield  {author} {\bibinfo {author} {\bibfnamefont {C.}~\bibnamefont
  {Nayak}}, \bibinfo {author} {\bibfnamefont {S.~H.}\ \bibnamefont {Simon}},
  \bibinfo {author} {\bibfnamefont {A.}~\bibnamefont {Stern}}, \bibinfo
  {author} {\bibfnamefont {M.}~\bibnamefont {Freedman}}, \ and\ \bibinfo
  {author} {\bibfnamefont {S.}~\bibnamefont {Das~Sarma}},\ }\emph {Non-Abelian
  anyons and topological quantum computation},\ \href {\doibase
  10.1103/RevModPhys.80.1083} {\bibfield  {journal} {\bibinfo  {journal} {Rev.
  Mod. Phys.}\ }\textbf {\bibinfo {volume} {80}},\ \bibinfo {pages} {1083}
  (\bibinfo {year} {2008})}\BibitemShut {NoStop}%
\bibitem [{\citenamefont {Tsui}\ \emph {et~al.}(1982)\citenamefont {Tsui},
  \citenamefont {Stormer},\ and\ \citenamefont {Gossard}}]{TsuiGossard1982}%
  \BibitemOpen
  \bibfield  {author} {\bibinfo {author} {\bibfnamefont {D.~C.}\ \bibnamefont
  {Tsui}}, \bibinfo {author} {\bibfnamefont {H.~L.}\ \bibnamefont {Stormer}}, \
  and\ \bibinfo {author} {\bibfnamefont {A.~C.}\ \bibnamefont {Gossard}},\
  }\emph {Two-Dimensional Magnetotransport in the Extreme Quantum Limit},\
  \href {\doibase 10.1103/PhysRevLett.48.1559} {\bibfield  {journal} {\bibinfo
  {journal} {Phys. Rev. Lett.}\ }\textbf {\bibinfo {volume} {48}},\ \bibinfo
  {pages} {1559} (\bibinfo {year} {1982})}\BibitemShut {NoStop}%
\bibitem [{\citenamefont {Laughlin}(1983)}]{Laughlin1983}%
  \BibitemOpen
  \bibfield  {author} {\bibinfo {author} {\bibfnamefont {R.~B.}\ \bibnamefont
  {Laughlin}},\ }\emph {Anomalous Quantum Hall Effect: An Incompressible
  Quantum Fluid with Fractionally Charged Excitations},\ \href {\doibase
  10.1103/PhysRevLett.50.1395} {\bibfield  {journal} {\bibinfo  {journal}
  {Phys. Rev. Lett.}\ }\textbf {\bibinfo {volume} {50}},\ \bibinfo {pages}
  {1395} (\bibinfo {year} {1983})}\BibitemShut {NoStop}%
\bibitem [{\citenamefont {Stormer}\ \emph {et~al.}(1999)\citenamefont
  {Stormer}, \citenamefont {Tsui},\ and\ \citenamefont
  {Gossard}}]{StormerGossard1999}%
  \BibitemOpen
  \bibfield  {author} {\bibinfo {author} {\bibfnamefont {H.~L.}\ \bibnamefont
  {Stormer}}, \bibinfo {author} {\bibfnamefont {D.~C.}\ \bibnamefont {Tsui}}, \
  and\ \bibinfo {author} {\bibfnamefont {A.~C.}\ \bibnamefont {Gossard}},\
  }\emph {The fractional quantum Hall effect},\ \href {\doibase
  10.1103/RevModPhys.71.S298} {\bibfield  {journal} {\bibinfo  {journal} {Rev.
  Mod. Phys.}\ }\textbf {\bibinfo {volume} {71}},\ \bibinfo {pages} {S298}
  (\bibinfo {year} {1999})}\BibitemShut {NoStop}%
\bibitem [{\citenamefont {Dalibard}\ \emph {et~al.}(2011)\citenamefont
  {Dalibard}, \citenamefont {Gerbier}, \citenamefont
  {Juzeli\ifmmode~\bar{u}\else \={u}\fi{}nas},\ and\ \citenamefont
  {\"Ohberg}}]{DalibardOehberg2011}%
  \BibitemOpen
  \bibfield  {author} {\bibinfo {author} {\bibfnamefont {J.}~\bibnamefont
  {Dalibard}}, \bibinfo {author} {\bibfnamefont {F.}~\bibnamefont {Gerbier}},
  \bibinfo {author} {\bibfnamefont {G.}~\bibnamefont
  {Juzeli\ifmmode~\bar{u}\else \={u}\fi{}nas}}, \ and\ \bibinfo {author}
  {\bibfnamefont {P.}~\bibnamefont {\"Ohberg}},\ }\emph {Colloquium: Artificial
  gauge potentials for neutral atoms},\ \href {\doibase
  10.1103/RevModPhys.83.1523} {\bibfield  {journal} {\bibinfo  {journal} {Rev.
  Mod. Phys.}\ }\textbf {\bibinfo {volume} {83}},\ \bibinfo {pages} {1523}
  (\bibinfo {year} {2011})}\BibitemShut {NoStop}%
\bibitem [{\citenamefont {Goldman}\ \emph {et~al.}(2014)\citenamefont
  {Goldman}, \citenamefont {Juzeliūnas}, \citenamefont {Öhberg},\ and\
  \citenamefont {Spielman}}]{GoldmanSpielman2014}%
  \BibitemOpen
  \bibfield  {author} {\bibinfo {author} {\bibfnamefont {N.}~\bibnamefont
  {Goldman}}, \bibinfo {author} {\bibfnamefont {G.}~\bibnamefont
  {Juzeliūnas}}, \bibinfo {author} {\bibfnamefont {P.}~\bibnamefont
  {Öhberg}}, \ and\ \bibinfo {author} {\bibfnamefont {I.~B.}\ \bibnamefont
  {Spielman}},\ }\emph {Light-induced gauge fields for ultracold atoms},\ \href
  {\doibase 10.1088/0034-4885/77/12/126401} {\bibfield  {journal} {\bibinfo
  {journal} {Reports on Progress in Physics}\ }\textbf {\bibinfo {volume}
  {77}},\ \bibinfo {pages} {126401} (\bibinfo {year} {2014})}\BibitemShut
  {NoStop}%
\bibitem [{\citenamefont {Hauke}\ and\ \citenamefont
  {Carusotto}(2022)}]{HaukeCarusotto2022}%
  \BibitemOpen
  \bibfield  {author} {\bibinfo {author} {\bibfnamefont {P.}~\bibnamefont
  {Hauke}}\ and\ \bibinfo {author} {\bibfnamefont {I.}~\bibnamefont
  {Carusotto}},\ }\href@noop {} {\emph {Quantum Hall and Synthetic
  Magnetic-Field Effects in Ultra-Cold Atomic Systems}} (\bibinfo {year}
  {2022}),\ \Eprint {http://arxiv.org/abs/2206.07727} {arXiv:2206.07727}
  \BibitemShut {NoStop}%
\bibitem [{\citenamefont {Aidelsburger}\ \emph {et~al.}(2011)\citenamefont
  {Aidelsburger}, \citenamefont {Atala}, \citenamefont {Nascimb\`ene},
  \citenamefont {Trotzky}, \citenamefont {Chen},\ and\ \citenamefont
  {Bloch}}]{AidelsburgerBloch2011}%
  \BibitemOpen
  \bibfield  {author} {\bibinfo {author} {\bibfnamefont {M.}~\bibnamefont
  {Aidelsburger}}, \bibinfo {author} {\bibfnamefont {M.}~\bibnamefont {Atala}},
  \bibinfo {author} {\bibfnamefont {S.}~\bibnamefont {Nascimb\`ene}}, \bibinfo
  {author} {\bibfnamefont {S.}~\bibnamefont {Trotzky}}, \bibinfo {author}
  {\bibfnamefont {Y.-A.}\ \bibnamefont {Chen}}, \ and\ \bibinfo {author}
  {\bibfnamefont {I.}~\bibnamefont {Bloch}},\ }\emph {Experimental Realization
  of Strong Effective Magnetic Fields in an Optical Lattice},\ \href {\doibase
  10.1103/PhysRevLett.107.255301} {\bibfield  {journal} {\bibinfo  {journal}
  {Phys. Rev. Lett.}\ }\textbf {\bibinfo {volume} {107}},\ \bibinfo {pages}
  {255301} (\bibinfo {year} {2011})}\BibitemShut {NoStop}%
\bibitem [{\citenamefont {Struck}\ \emph {et~al.}(2012)\citenamefont {Struck},
  \citenamefont {\"Olschl\"ager}, \citenamefont {Weinberg}, \citenamefont
  {Hauke}, \citenamefont {Simonet}, \citenamefont {Eckardt}, \citenamefont
  {Lewenstein}, \citenamefont {Sengstock},\ and\ \citenamefont
  {Windpassinger}}]{StruckWindpassinger2012}%
  \BibitemOpen
  \bibfield  {author} {\bibinfo {author} {\bibfnamefont {J.}~\bibnamefont
  {Struck}}, \bibinfo {author} {\bibfnamefont {C.}~\bibnamefont
  {\"Olschl\"ager}}, \bibinfo {author} {\bibfnamefont {M.}~\bibnamefont
  {Weinberg}}, \bibinfo {author} {\bibfnamefont {P.}~\bibnamefont {Hauke}},
  \bibinfo {author} {\bibfnamefont {J.}~\bibnamefont {Simonet}}, \bibinfo
  {author} {\bibfnamefont {A.}~\bibnamefont {Eckardt}}, \bibinfo {author}
  {\bibfnamefont {M.}~\bibnamefont {Lewenstein}}, \bibinfo {author}
  {\bibfnamefont {K.}~\bibnamefont {Sengstock}}, \ and\ \bibinfo {author}
  {\bibfnamefont {P.}~\bibnamefont {Windpassinger}},\ }\emph {Tunable Gauge
  Potential for Neutral and Spinless Particles in Driven Optical Lattices},\
  \href {\doibase 10.1103/PhysRevLett.108.225304} {\bibfield  {journal}
  {\bibinfo  {journal} {Phys. Rev. Lett.}\ }\textbf {\bibinfo {volume} {108}},\
  \bibinfo {pages} {225304} (\bibinfo {year} {2012})}\BibitemShut {NoStop}%
\bibitem [{\citenamefont {Aidelsburger}\ \emph {et~al.}(2013)\citenamefont
  {Aidelsburger}, \citenamefont {Atala}, \citenamefont {Lohse}, \citenamefont
  {Barreiro}, \citenamefont {Paredes},\ and\ \citenamefont
  {Bloch}}]{AidelsburgerBloch2013}%
  \BibitemOpen
  \bibfield  {author} {\bibinfo {author} {\bibfnamefont {M.}~\bibnamefont
  {Aidelsburger}}, \bibinfo {author} {\bibfnamefont {M.}~\bibnamefont {Atala}},
  \bibinfo {author} {\bibfnamefont {M.}~\bibnamefont {Lohse}}, \bibinfo
  {author} {\bibfnamefont {J.~T.}\ \bibnamefont {Barreiro}}, \bibinfo {author}
  {\bibfnamefont {B.}~\bibnamefont {Paredes}}, \ and\ \bibinfo {author}
  {\bibfnamefont {I.}~\bibnamefont {Bloch}},\ }\emph {Realization of the
  Hofstadter Hamiltonian with Ultracold Atoms in Optical Lattices},\ \href
  {\doibase 10.1103/PhysRevLett.111.185301} {\bibfield  {journal} {\bibinfo
  {journal} {Phys. Rev. Lett.}\ }\textbf {\bibinfo {volume} {111}},\ \bibinfo
  {pages} {185301} (\bibinfo {year} {2013})}\BibitemShut {NoStop}%
\bibitem [{\citenamefont {Miyake}\ \emph {et~al.}(2013)\citenamefont {Miyake},
  \citenamefont {Siviloglou}, \citenamefont {Kennedy}, \citenamefont {Burton},\
  and\ \citenamefont {Ketterle}}]{MiyakeKetterle2013}%
  \BibitemOpen
  \bibfield  {author} {\bibinfo {author} {\bibfnamefont {H.}~\bibnamefont
  {Miyake}}, \bibinfo {author} {\bibfnamefont {G.~A.}\ \bibnamefont
  {Siviloglou}}, \bibinfo {author} {\bibfnamefont {C.~J.}\ \bibnamefont
  {Kennedy}}, \bibinfo {author} {\bibfnamefont {W.~C.}\ \bibnamefont {Burton}},
  \ and\ \bibinfo {author} {\bibfnamefont {W.}~\bibnamefont {Ketterle}},\
  }\emph {Realizing the Harper Hamiltonian with Laser-Assisted Tunneling in
  Optical Lattices},\ \href {\doibase 10.1103/PhysRevLett.111.185302}
  {\bibfield  {journal} {\bibinfo  {journal} {Phys. Rev. Lett.}\ }\textbf
  {\bibinfo {volume} {111}},\ \bibinfo {pages} {185302} (\bibinfo {year}
  {2013})}\BibitemShut {NoStop}%
\bibitem [{\citenamefont {Atala}\ \emph {et~al.}(2014)\citenamefont {Atala},
  \citenamefont {Aidelsburger}, \citenamefont {Lohse}, \citenamefont
  {Barreiro}, \citenamefont {Paredes},\ and\ \citenamefont
  {Bloch}}]{AtalaBloch2014}%
  \BibitemOpen
  \bibfield  {author} {\bibinfo {author} {\bibfnamefont {M.}~\bibnamefont
  {Atala}}, \bibinfo {author} {\bibfnamefont {M.}~\bibnamefont {Aidelsburger}},
  \bibinfo {author} {\bibfnamefont {M.}~\bibnamefont {Lohse}}, \bibinfo
  {author} {\bibfnamefont {J.~T.}\ \bibnamefont {Barreiro}}, \bibinfo {author}
  {\bibfnamefont {B.}~\bibnamefont {Paredes}}, \ and\ \bibinfo {author}
  {\bibfnamefont {I.}~\bibnamefont {Bloch}},\ }\emph {Observation of chiral
  currents with ultracold atoms in bosonic ladders},\ \href {\doibase
  10.1038/nphys2998} {\bibfield  {journal} {\bibinfo  {journal} {Nature
  Physics}\ }\textbf {\bibinfo {volume} {10}},\ \bibinfo {pages} {588}
  (\bibinfo {year} {2014})}\BibitemShut {NoStop}%
\bibitem [{\citenamefont {Aidelsburger}\ \emph {et~al.}(2015)\citenamefont
  {Aidelsburger}, \citenamefont {Lohse}, \citenamefont {Schweizer},
  \citenamefont {Atala}, \citenamefont {Barreiro}, \citenamefont
  {Nascimb{\`e}ne}, \citenamefont {Cooper}, \citenamefont {Bloch},\ and\
  \citenamefont {Goldman}}]{AidelsburgerGoldman2015}%
  \BibitemOpen
  \bibfield  {author} {\bibinfo {author} {\bibfnamefont {M.}~\bibnamefont
  {Aidelsburger}}, \bibinfo {author} {\bibfnamefont {M.}~\bibnamefont {Lohse}},
  \bibinfo {author} {\bibfnamefont {C.}~\bibnamefont {Schweizer}}, \bibinfo
  {author} {\bibfnamefont {M.}~\bibnamefont {Atala}}, \bibinfo {author}
  {\bibfnamefont {J.~T.}\ \bibnamefont {Barreiro}}, \bibinfo {author}
  {\bibfnamefont {S.}~\bibnamefont {Nascimb{\`e}ne}}, \bibinfo {author}
  {\bibfnamefont {N.~R.}\ \bibnamefont {Cooper}}, \bibinfo {author}
  {\bibfnamefont {I.}~\bibnamefont {Bloch}}, \ and\ \bibinfo {author}
  {\bibfnamefont {N.}~\bibnamefont {Goldman}},\ }\emph {Measuring the Chern
  number of Hofstadter bands with ultracold bosonic atoms},\ \href {\doibase
  10.1038/nphys3171} {\bibfield  {journal} {\bibinfo  {journal} {Nature
  Physics}\ }\textbf {\bibinfo {volume} {11}},\ \bibinfo {pages} {162}
  (\bibinfo {year} {2015})}\BibitemShut {NoStop}%
\bibitem [{\citenamefont {Mancini}\ \emph {et~al.}(2015)\citenamefont
  {Mancini}, \citenamefont {Pagano}, \citenamefont {Cappellini}, \citenamefont
  {Livi}, \citenamefont {Rider}, \citenamefont {Catani}, \citenamefont {Sias},
  \citenamefont {Zoller}, \citenamefont {Inguscio}, \citenamefont {Dalmonte},\
  and\ \citenamefont {Fallani}}]{ManciniFallani2015}%
  \BibitemOpen
  \bibfield  {author} {\bibinfo {author} {\bibfnamefont {M.}~\bibnamefont
  {Mancini}}, \bibinfo {author} {\bibfnamefont {G.}~\bibnamefont {Pagano}},
  \bibinfo {author} {\bibfnamefont {G.}~\bibnamefont {Cappellini}}, \bibinfo
  {author} {\bibfnamefont {L.}~\bibnamefont {Livi}}, \bibinfo {author}
  {\bibfnamefont {M.}~\bibnamefont {Rider}}, \bibinfo {author} {\bibfnamefont
  {J.}~\bibnamefont {Catani}}, \bibinfo {author} {\bibfnamefont
  {C.}~\bibnamefont {Sias}}, \bibinfo {author} {\bibfnamefont {P.}~\bibnamefont
  {Zoller}}, \bibinfo {author} {\bibfnamefont {M.}~\bibnamefont {Inguscio}},
  \bibinfo {author} {\bibfnamefont {M.}~\bibnamefont {Dalmonte}}, \ and\
  \bibinfo {author} {\bibfnamefont {L.}~\bibnamefont {Fallani}},\ }\emph
  {Observation of chiral edge states with neutral fermions in synthetic Hall
  ribbons},\ \href {\doibase 10.1126/science.aaa8736} {\bibfield  {journal}
  {\bibinfo  {journal} {Science}\ }\textbf {\bibinfo {volume} {349}},\ \bibinfo
  {pages} {1510} (\bibinfo {year} {2015})}\BibitemShut {NoStop}%
\bibitem [{\citenamefont {Tai}\ \emph {et~al.}(2017)\citenamefont {Tai},
  \citenamefont {Lukin}, \citenamefont {Rispoli}, \citenamefont {Schittko},
  \citenamefont {Menke}, \citenamefont {Borgnia}, \citenamefont {Preiss},
  \citenamefont {Grusdt}, \citenamefont {Kaufman},\ and\ \citenamefont
  {Greiner}}]{TaiGreiner2017}%
  \BibitemOpen
  \bibfield  {author} {\bibinfo {author} {\bibfnamefont {M.~E.}\ \bibnamefont
  {Tai}}, \bibinfo {author} {\bibfnamefont {A.}~\bibnamefont {Lukin}}, \bibinfo
  {author} {\bibfnamefont {M.}~\bibnamefont {Rispoli}}, \bibinfo {author}
  {\bibfnamefont {R.}~\bibnamefont {Schittko}}, \bibinfo {author}
  {\bibfnamefont {T.}~\bibnamefont {Menke}}, \bibinfo {author} {\bibfnamefont
  {D.}~\bibnamefont {Borgnia}}, \bibinfo {author} {\bibfnamefont {P.~M.}\
  \bibnamefont {Preiss}}, \bibinfo {author} {\bibfnamefont {F.}~\bibnamefont
  {Grusdt}}, \bibinfo {author} {\bibfnamefont {A.~M.}\ \bibnamefont {Kaufman}},
  \ and\ \bibinfo {author} {\bibfnamefont {M.}~\bibnamefont {Greiner}},\ }\emph
  {Microscopy of the interacting Harper--Hofstadter model in the two-body
  limit},\ \href {\doibase 10.1038/nature22811} {\bibfield  {journal} {\bibinfo
   {journal} {Nature}\ }\textbf {\bibinfo {volume} {546}},\ \bibinfo {pages}
  {519} (\bibinfo {year} {2017})}\BibitemShut {NoStop}%
\bibitem [{\citenamefont {Genkina}\ \emph {et~al.}(2019)\citenamefont
  {Genkina}, \citenamefont {Aycock}, \citenamefont {Lu}, \citenamefont {Lu},
  \citenamefont {Pineiro},\ and\ \citenamefont
  {Spielman}}]{GenkinaSpielman2019}%
  \BibitemOpen
  \bibfield  {author} {\bibinfo {author} {\bibfnamefont {D.}~\bibnamefont
  {Genkina}}, \bibinfo {author} {\bibfnamefont {L.~M.}\ \bibnamefont {Aycock}},
  \bibinfo {author} {\bibfnamefont {H.-I.}\ \bibnamefont {Lu}}, \bibinfo
  {author} {\bibfnamefont {M.}~\bibnamefont {Lu}}, \bibinfo {author}
  {\bibfnamefont {A.~M.}\ \bibnamefont {Pineiro}}, \ and\ \bibinfo {author}
  {\bibfnamefont {I.~B.}\ \bibnamefont {Spielman}},\ }\emph {Imaging topology
  of Hofstadter ribbons},\ \href {\doibase 10.1088/1367-2630/ab165b} {\bibfield
   {journal} {\bibinfo  {journal} {New Journal of Physics}\ }\textbf {\bibinfo
  {volume} {21}},\ \bibinfo {pages} {053021} (\bibinfo {year}
  {2019})}\BibitemShut {NoStop}%
\bibitem [{\citenamefont {Chalopin}\ \emph {et~al.}(2020)\citenamefont
  {Chalopin}, \citenamefont {Satoor}, \citenamefont {Evrard}, \citenamefont
  {Makhalov}, \citenamefont {Dalibard}, \citenamefont {Lopes},\ and\
  \citenamefont {Nascimbene}}]{ChalopinNascimbene2020}%
  \BibitemOpen
  \bibfield  {author} {\bibinfo {author} {\bibfnamefont {T.}~\bibnamefont
  {Chalopin}}, \bibinfo {author} {\bibfnamefont {T.}~\bibnamefont {Satoor}},
  \bibinfo {author} {\bibfnamefont {A.}~\bibnamefont {Evrard}}, \bibinfo
  {author} {\bibfnamefont {V.}~\bibnamefont {Makhalov}}, \bibinfo {author}
  {\bibfnamefont {J.}~\bibnamefont {Dalibard}}, \bibinfo {author}
  {\bibfnamefont {R.}~\bibnamefont {Lopes}}, \ and\ \bibinfo {author}
  {\bibfnamefont {S.}~\bibnamefont {Nascimbene}},\ }\emph {Probing chiral edge
  dynamics and bulk topology of a synthetic Hall system},\ \href {\doibase
  10.1038/s41567-020-0942-5} {\bibfield  {journal} {\bibinfo  {journal} {Nature
  Physics}\ }\textbf {\bibinfo {volume} {16}},\ \bibinfo {pages} {1017}
  (\bibinfo {year} {2020})}\BibitemShut {NoStop}%
\bibitem [{\citenamefont {Zhou}\ \emph {et~al.}(2023)\citenamefont {Zhou},
  \citenamefont {Cappellini}, \citenamefont {Tusi}, \citenamefont {Franchi},
  \citenamefont {Parravicini}, \citenamefont {Repellin}, \citenamefont
  {Greschner}, \citenamefont {Inguscio}, \citenamefont {Giamarchi},
  \citenamefont {Filippone}, \citenamefont {Catani},\ and\ \citenamefont
  {Fallani}}]{ZhouFallani2023}%
  \BibitemOpen
  \bibfield  {author} {\bibinfo {author} {\bibfnamefont {T.-W.}\ \bibnamefont
  {Zhou}}, \bibinfo {author} {\bibfnamefont {G.}~\bibnamefont {Cappellini}},
  \bibinfo {author} {\bibfnamefont {D.}~\bibnamefont {Tusi}}, \bibinfo {author}
  {\bibfnamefont {L.}~\bibnamefont {Franchi}}, \bibinfo {author} {\bibfnamefont
  {J.}~\bibnamefont {Parravicini}}, \bibinfo {author} {\bibfnamefont
  {C.}~\bibnamefont {Repellin}}, \bibinfo {author} {\bibfnamefont
  {S.}~\bibnamefont {Greschner}}, \bibinfo {author} {\bibfnamefont
  {M.}~\bibnamefont {Inguscio}}, \bibinfo {author} {\bibfnamefont
  {T.}~\bibnamefont {Giamarchi}}, \bibinfo {author} {\bibfnamefont
  {M.}~\bibnamefont {Filippone}}, \bibinfo {author} {\bibfnamefont
  {J.}~\bibnamefont {Catani}}, \ and\ \bibinfo {author} {\bibfnamefont
  {L.}~\bibnamefont {Fallani}},\ }\emph {Observation of universal Hall response
  in strongly interacting Fermions},\ \href {\doibase 10.1126/science.add1969}
  {\bibfield  {journal} {\bibinfo  {journal} {Science}\ }\textbf {\bibinfo
  {volume} {381}},\ \bibinfo {pages} {427} (\bibinfo {year}
  {2023})}\BibitemShut {NoStop}%
\bibitem [{\citenamefont {L{\'e}onard}\ \emph {et~al.}(2023)\citenamefont
  {L{\'e}onard}, \citenamefont {Kim}, \citenamefont {Kwan}, \citenamefont
  {Segura}, \citenamefont {Grusdt}, \citenamefont {Repellin}, \citenamefont
  {Goldman},\ and\ \citenamefont {Greiner}}]{LeonardGreiner2023}%
  \BibitemOpen
  \bibfield  {author} {\bibinfo {author} {\bibfnamefont {J.}~\bibnamefont
  {L{\'e}onard}}, \bibinfo {author} {\bibfnamefont {S.}~\bibnamefont {Kim}},
  \bibinfo {author} {\bibfnamefont {J.}~\bibnamefont {Kwan}}, \bibinfo {author}
  {\bibfnamefont {P.}~\bibnamefont {Segura}}, \bibinfo {author} {\bibfnamefont
  {F.}~\bibnamefont {Grusdt}}, \bibinfo {author} {\bibfnamefont
  {C.}~\bibnamefont {Repellin}}, \bibinfo {author} {\bibfnamefont
  {N.}~\bibnamefont {Goldman}}, \ and\ \bibinfo {author} {\bibfnamefont
  {M.}~\bibnamefont {Greiner}},\ }\emph {Realization of a fractional quantum
  Hall state with ultracold atoms},\ \href {\doibase
  10.1038/s41586-023-06122-4} {\bibfield  {journal} {\bibinfo  {journal}
  {Nature}\ }\textbf {\bibinfo {volume} {619}},\ \bibinfo {pages} {495}
  (\bibinfo {year} {2023})}\BibitemShut {NoStop}%
\bibitem [{\citenamefont {Repellin}\ \emph {et~al.}(2020)\citenamefont
  {Repellin}, \citenamefont {L\'eonard},\ and\ \citenamefont
  {Goldman}}]{RepellinGoldman2020}%
  \BibitemOpen
  \bibfield  {author} {\bibinfo {author} {\bibfnamefont {C.}~\bibnamefont
  {Repellin}}, \bibinfo {author} {\bibfnamefont {J.}~\bibnamefont {L\'eonard}},
  \ and\ \bibinfo {author} {\bibfnamefont {N.}~\bibnamefont {Goldman}},\ }\emph
  {Fractional Chern insulators of few bosons in a box: Hall plateaus from
  center-of-mass drifts and density profiles},\ \href {\doibase
  10.1103/PhysRevA.102.063316} {\bibfield  {journal} {\bibinfo  {journal}
  {Phys. Rev. A}\ }\textbf {\bibinfo {volume} {102}},\ \bibinfo {pages}
  {063316} (\bibinfo {year} {2020})}\BibitemShut {NoStop}%
\bibitem [{\citenamefont {Peralta~Gavensky}\ \emph {et~al.}(2023)\citenamefont
  {Peralta~Gavensky}, \citenamefont {Sachdev},\ and\ \citenamefont
  {Goldman}}]{PeraltaGavenskyGoldman2023}%
  \BibitemOpen
  \bibfield  {author} {\bibinfo {author} {\bibfnamefont {L.}~\bibnamefont
  {Peralta~Gavensky}}, \bibinfo {author} {\bibfnamefont {S.}~\bibnamefont
  {Sachdev}}, \ and\ \bibinfo {author} {\bibfnamefont {N.}~\bibnamefont
  {Goldman}},\ }\emph {Connecting the Many-Body Chern Number to Luttinger's
  Theorem through St\ifmmode \check{r}\else \v{r}\fi{}eda's Formula},\ \href
  {\doibase 10.1103/PhysRevLett.131.236601} {\bibfield  {journal} {\bibinfo
  {journal} {Phys. Rev. Lett.}\ }\textbf {\bibinfo {volume} {131}},\ \bibinfo
  {pages} {236601} (\bibinfo {year} {2023})}\BibitemShut {NoStop}%
\bibitem [{\citenamefont {Prelov\ifmmode~\check{s}\else \v{s}\fi{}ek}\ \emph
  {et~al.}(1999)\citenamefont {Prelov\ifmmode~\check{s}\else \v{s}\fi{}ek},
  \citenamefont {Long}, \citenamefont {Marke\ifmmode~\check{z}\else
  \v{z}\fi{}},\ and\ \citenamefont {Zotos}}]{PrelovsekZotos1999}%
  \BibitemOpen
  \bibfield  {author} {\bibinfo {author} {\bibfnamefont {P.}~\bibnamefont
  {Prelov\ifmmode~\check{s}\else \v{s}\fi{}ek}}, \bibinfo {author}
  {\bibfnamefont {M.}~\bibnamefont {Long}}, \bibinfo {author} {\bibfnamefont
  {T.}~\bibnamefont {Marke\ifmmode~\check{z}\else \v{z}\fi{}}}, \ and\ \bibinfo
  {author} {\bibfnamefont {X.}~\bibnamefont {Zotos}},\ }\emph {Hall Constant of
  Strongly Correlated Electrons on a Ladder},\ \href {\doibase
  10.1103/PhysRevLett.83.2785} {\bibfield  {journal} {\bibinfo  {journal}
  {Phys. Rev. Lett.}\ }\textbf {\bibinfo {volume} {83}},\ \bibinfo {pages}
  {2785} (\bibinfo {year} {1999})}\BibitemShut {NoStop}%
\bibitem [{\citenamefont {Zotos}\ \emph {et~al.}(2000)\citenamefont {Zotos},
  \citenamefont {Naef}, \citenamefont {Long},\ and\ \citenamefont
  {Prelov\ifmmode~\check{s}\else \v{s}\fi{}ek}}]{ZotosPrelovsek2000}%
  \BibitemOpen
  \bibfield  {author} {\bibinfo {author} {\bibfnamefont {X.}~\bibnamefont
  {Zotos}}, \bibinfo {author} {\bibfnamefont {F.}~\bibnamefont {Naef}},
  \bibinfo {author} {\bibfnamefont {M.}~\bibnamefont {Long}}, \ and\ \bibinfo
  {author} {\bibfnamefont {P.}~\bibnamefont {Prelov\ifmmode~\check{s}\else
  \v{s}\fi{}ek}},\ }\emph {Reactive Hall Response},\ \href {\doibase
  10.1103/PhysRevLett.85.377} {\bibfield  {journal} {\bibinfo  {journal} {Phys.
  Rev. Lett.}\ }\textbf {\bibinfo {volume} {85}},\ \bibinfo {pages} {377}
  (\bibinfo {year} {2000})}\BibitemShut {NoStop}%
\bibitem [{\citenamefont {Greschner}\ \emph {et~al.}(2019)\citenamefont
  {Greschner}, \citenamefont {Filippone},\ and\ \citenamefont
  {Giamarchi}}]{GreschnerGiamarchi2019}%
  \BibitemOpen
  \bibfield  {author} {\bibinfo {author} {\bibfnamefont {S.}~\bibnamefont
  {Greschner}}, \bibinfo {author} {\bibfnamefont {M.}~\bibnamefont
  {Filippone}}, \ and\ \bibinfo {author} {\bibfnamefont {T.}~\bibnamefont
  {Giamarchi}},\ }\emph {Universal Hall Response in Interacting Quantum
  Systems},\ \href {\doibase 10.1103/PhysRevLett.122.083402} {\bibfield
  {journal} {\bibinfo  {journal} {Phys. Rev. Lett.}\ }\textbf {\bibinfo
  {volume} {122}},\ \bibinfo {pages} {083402} (\bibinfo {year}
  {2019})}\BibitemShut {NoStop}%
\bibitem [{\citenamefont {Buser}\ \emph {et~al.}(2021)\citenamefont {Buser},
  \citenamefont {Greschner}, \citenamefont {Schollw\"ock},\ and\ \citenamefont
  {Giamarchi}}]{BuserGiamarchi2021}%
  \BibitemOpen
  \bibfield  {author} {\bibinfo {author} {\bibfnamefont {M.}~\bibnamefont
  {Buser}}, \bibinfo {author} {\bibfnamefont {S.}~\bibnamefont {Greschner}},
  \bibinfo {author} {\bibfnamefont {U.}~\bibnamefont {Schollw\"ock}}, \ and\
  \bibinfo {author} {\bibfnamefont {T.}~\bibnamefont {Giamarchi}},\ }\emph
  {Probing the Hall Voltage in Synthetic Quantum Systems},\ \href {\doibase
  10.1103/PhysRevLett.126.030501} {\bibfield  {journal} {\bibinfo  {journal}
  {Phys. Rev. Lett.}\ }\textbf {\bibinfo {volume} {126}},\ \bibinfo {pages}
  {030501} (\bibinfo {year} {2021})}\BibitemShut {NoStop}%
\bibitem [{\citenamefont {Citro}\ \emph {et~al.}(2024)\citenamefont {Citro},
  \citenamefont {Giamarchi},\ and\ \citenamefont {Orignac}}]{CitroOrignac2024}%
  \BibitemOpen
  \bibfield  {author} {\bibinfo {author} {\bibfnamefont {R.}~\bibnamefont
  {Citro}}, \bibinfo {author} {\bibfnamefont {T.}~\bibnamefont {Giamarchi}}, \
  and\ \bibinfo {author} {\bibfnamefont {E.}~\bibnamefont {Orignac}},\ }\emph
  {Hall response in interacting bosonic and fermionic ladders},\ \href@noop {}
  {\bibfield  {journal} {\bibinfo  {journal} {arXiv:2404.16973}\ } (\bibinfo
  {year} {2024})},\ \Eprint {http://arxiv.org/abs/2404.16973}
  {arXiv:2404.16973} \BibitemShut {NoStop}%
\bibitem [{\citenamefont {Halati}\ and\ \citenamefont
  {Giamarchi}(2024)}]{HalatiGiamarchi2024}%
  \BibitemOpen
  \bibfield  {author} {\bibinfo {author} {\bibfnamefont {C.-M.}\ \bibnamefont
  {Halati}}\ and\ \bibinfo {author} {\bibfnamefont {T.}~\bibnamefont
  {Giamarchi}},\ }\href@noop {} {\emph {Exploring Frustration Effects of
  Strongly Interacting Bosons via the Hall Response}} (\bibinfo {year}
  {2024}),\ \Eprint {http://arxiv.org/abs/2405.19030} {arXiv:2405.19030}
  \BibitemShut {NoStop}%
\bibitem [{\citenamefont {Lopatin}\ \emph {et~al.}(2001)\citenamefont
  {Lopatin}, \citenamefont {Georges},\ and\ \citenamefont
  {Giamarchi}}]{LopatinGiamarchi2001}%
  \BibitemOpen
  \bibfield  {author} {\bibinfo {author} {\bibfnamefont {A.}~\bibnamefont
  {Lopatin}}, \bibinfo {author} {\bibfnamefont {A.}~\bibnamefont {Georges}}, \
  and\ \bibinfo {author} {\bibfnamefont {T.}~\bibnamefont {Giamarchi}},\ }\emph
  {Hall effect and interchain magneto-optical properties of coupled Luttinger
  liquids},\ \href {\doibase 10.1103/PhysRevB.63.075109} {\bibfield  {journal}
  {\bibinfo  {journal} {Phys. Rev. B}\ }\textbf {\bibinfo {volume} {63}},\
  \bibinfo {pages} {075109} (\bibinfo {year} {2001})}\BibitemShut {NoStop}%
\bibitem [{\citenamefont {Le\'on}\ \emph {et~al.}(2007)\citenamefont {Le\'on},
  \citenamefont {Berthod},\ and\ \citenamefont
  {Giamarchi}}]{LeonGiamarchi2007}%
  \BibitemOpen
  \bibfield  {author} {\bibinfo {author} {\bibfnamefont {G.}~\bibnamefont
  {Le\'on}}, \bibinfo {author} {\bibfnamefont {C.}~\bibnamefont {Berthod}}, \
  and\ \bibinfo {author} {\bibfnamefont {T.}~\bibnamefont {Giamarchi}},\ }\emph
  {Hall effect in strongly correlated low-dimensional systems},\ \href
  {\doibase 10.1103/PhysRevB.75.195123} {\bibfield  {journal} {\bibinfo
  {journal} {Phys. Rev. B}\ }\textbf {\bibinfo {volume} {75}},\ \bibinfo
  {pages} {195123} (\bibinfo {year} {2007})}\BibitemShut {NoStop}%
\bibitem [{\citenamefont {Auerbach}(2018)}]{Auerbach2018}%
  \BibitemOpen
  \bibfield  {author} {\bibinfo {author} {\bibfnamefont {A.}~\bibnamefont
  {Auerbach}},\ }\emph {Hall Number of Strongly Correlated Metals},\ \href
  {\doibase 10.1103/PhysRevLett.121.066601} {\bibfield  {journal} {\bibinfo
  {journal} {Phys. Rev. Lett.}\ }\textbf {\bibinfo {volume} {121}},\ \bibinfo
  {pages} {066601} (\bibinfo {year} {2018})}\BibitemShut {NoStop}%
\bibitem [{\citenamefont {Filippone}\ \emph {et~al.}(2019)\citenamefont
  {Filippone}, \citenamefont {Bardyn}, \citenamefont {Greschner},\ and\
  \citenamefont {Giamarchi}}]{FilipponeGiamarchi2019}%
  \BibitemOpen
  \bibfield  {author} {\bibinfo {author} {\bibfnamefont {M.}~\bibnamefont
  {Filippone}}, \bibinfo {author} {\bibfnamefont {C.-E.}\ \bibnamefont
  {Bardyn}}, \bibinfo {author} {\bibfnamefont {S.}~\bibnamefont {Greschner}}, \
  and\ \bibinfo {author} {\bibfnamefont {T.}~\bibnamefont {Giamarchi}},\ }\emph
  {Vanishing Hall Response of Charged Fermions in a Transverse Magnetic
  Field},\ \href {\doibase 10.1103/PhysRevLett.123.086803} {\bibfield
  {journal} {\bibinfo  {journal} {Phys. Rev. Lett.}\ }\textbf {\bibinfo
  {volume} {123}},\ \bibinfo {pages} {086803} (\bibinfo {year}
  {2019})}\BibitemShut {NoStop}%
\bibitem [{\citenamefont {Mishra}\ \emph {et~al.}(2013)\citenamefont {Mishra},
  \citenamefont {Pai}, \citenamefont {Mukerjee},\ and\ \citenamefont
  {Paramekanti}}]{MishraParamekanti2013}%
  \BibitemOpen
  \bibfield  {author} {\bibinfo {author} {\bibfnamefont {T.}~\bibnamefont
  {Mishra}}, \bibinfo {author} {\bibfnamefont {R.~V.}\ \bibnamefont {Pai}},
  \bibinfo {author} {\bibfnamefont {S.}~\bibnamefont {Mukerjee}}, \ and\
  \bibinfo {author} {\bibfnamefont {A.}~\bibnamefont {Paramekanti}},\ }\emph
  {Quantum phases and phase transitions of frustrated hard-core bosons on a
  triangular ladder},\ \href {\doibase 10.1103/PhysRevB.87.174504} {\bibfield
  {journal} {\bibinfo  {journal} {Phys. Rev. B}\ }\textbf {\bibinfo {volume}
  {87}},\ \bibinfo {pages} {174504} (\bibinfo {year} {2013})}\BibitemShut
  {NoStop}%
\bibitem [{\citenamefont {Anisimovas}\ \emph {et~al.}(2016)\citenamefont
  {Anisimovas}, \citenamefont {Ra\ifmmode \check{c}\else
  \v{c}\fi{}i\ifmmode~\bar{u}\else \={u}\fi{}nas}, \citenamefont {Str\"ater},
  \citenamefont {Eckardt}, \citenamefont {Spielman},\ and\ \citenamefont
  {Juzeli\ifmmode~\bar{u}\else \={u}\fi{}nas}}]{AnisimovasJuzeliunas2016}%
  \BibitemOpen
  \bibfield  {author} {\bibinfo {author} {\bibfnamefont {E.}~\bibnamefont
  {Anisimovas}}, \bibinfo {author} {\bibfnamefont {M.}~\bibnamefont {Ra\ifmmode
  \check{c}\else \v{c}\fi{}i\ifmmode~\bar{u}\else \={u}\fi{}nas}}, \bibinfo
  {author} {\bibfnamefont {C.}~\bibnamefont {Str\"ater}}, \bibinfo {author}
  {\bibfnamefont {A.}~\bibnamefont {Eckardt}}, \bibinfo {author} {\bibfnamefont
  {I.~B.}\ \bibnamefont {Spielman}}, \ and\ \bibinfo {author} {\bibfnamefont
  {G.}~\bibnamefont {Juzeli\ifmmode~\bar{u}\else \={u}\fi{}nas}},\ }\emph
  {Semisynthetic zigzag optical lattice for ultracold bosons},\ \href {\doibase
  10.1103/PhysRevA.94.063632} {\bibfield  {journal} {\bibinfo  {journal} {Phys.
  Rev. A}\ }\textbf {\bibinfo {volume} {94}},\ \bibinfo {pages} {063632}
  (\bibinfo {year} {2016})}\BibitemShut {NoStop}%
\bibitem [{\citenamefont {An}\ \emph {et~al.}(2018)\citenamefont {An},
  \citenamefont {Meier},\ and\ \citenamefont {Gadway}}]{AnGadway2018}%
  \BibitemOpen
  \bibfield  {author} {\bibinfo {author} {\bibfnamefont {F.~A.}\ \bibnamefont
  {An}}, \bibinfo {author} {\bibfnamefont {E.~J.}\ \bibnamefont {Meier}}, \
  and\ \bibinfo {author} {\bibfnamefont {B.}~\bibnamefont {Gadway}},\ }\emph
  {Engineering a Flux-Dependent Mobility Edge in Disordered Zigzag Chains},\
  \href {\doibase 10.1103/PhysRevX.8.031045} {\bibfield  {journal} {\bibinfo
  {journal} {Phys. Rev. X}\ }\textbf {\bibinfo {volume} {8}},\ \bibinfo {pages}
  {031045} (\bibinfo {year} {2018})}\BibitemShut {NoStop}%
\bibitem [{\citenamefont {Romen}\ and\ \citenamefont
  {L\"auchli}(2018)}]{RomenLaeuchli2018}%
  \BibitemOpen
  \bibfield  {author} {\bibinfo {author} {\bibfnamefont {C.}~\bibnamefont
  {Romen}}\ and\ \bibinfo {author} {\bibfnamefont {A.~M.}\ \bibnamefont
  {L\"auchli}},\ }\emph {Chiral Mott insulators in frustrated Bose-Hubbard
  models on ladders and two-dimensional lattices: A combined perturbative and
  density matrix renormalization group study},\ \href {\doibase
  10.1103/PhysRevB.98.054519} {\bibfield  {journal} {\bibinfo  {journal} {Phys.
  Rev. B}\ }\textbf {\bibinfo {volume} {98}},\ \bibinfo {pages} {054519}
  (\bibinfo {year} {2018})}\BibitemShut {NoStop}%
\bibitem [{\citenamefont {Greschner}\ and\ \citenamefont
  {Mishra}(2019)}]{GreschnerMishra2019}%
  \BibitemOpen
  \bibfield  {author} {\bibinfo {author} {\bibfnamefont {S.}~\bibnamefont
  {Greschner}}\ and\ \bibinfo {author} {\bibfnamefont {T.}~\bibnamefont
  {Mishra}},\ }\emph {Interacting bosons in generalized zigzag and
  railroad-trestle models},\ \href {\doibase 10.1103/PhysRevB.100.144405}
  {\bibfield  {journal} {\bibinfo  {journal} {Phys. Rev. B}\ }\textbf {\bibinfo
  {volume} {100}},\ \bibinfo {pages} {144405} (\bibinfo {year}
  {2019})}\BibitemShut {NoStop}%
\bibitem [{\citenamefont {Cabedo}\ \emph {et~al.}(2020)\citenamefont {Cabedo},
  \citenamefont {Claramunt}, \citenamefont {Mompart}, \citenamefont
  {Ahufinger},\ and\ \citenamefont {Celi}}]{CabedoCeli2020}%
  \BibitemOpen
  \bibfield  {author} {\bibinfo {author} {\bibfnamefont {J.}~\bibnamefont
  {Cabedo}}, \bibinfo {author} {\bibfnamefont {J.}~\bibnamefont {Claramunt}},
  \bibinfo {author} {\bibfnamefont {J.}~\bibnamefont {Mompart}}, \bibinfo
  {author} {\bibfnamefont {V.}~\bibnamefont {Ahufinger}}, \ and\ \bibinfo
  {author} {\bibfnamefont {A.}~\bibnamefont {Celi}},\ }\emph {Effective
  triangular ladders with staggered flux from spin-orbit coupling in 1D optical
  lattices},\ \href {\doibase 10.1140/epjd/e2020-10129-1} {\bibfield  {journal}
  {\bibinfo  {journal} {The European Physical Journal D}\ }\textbf {\bibinfo
  {volume} {74}},\ \bibinfo {pages} {123} (\bibinfo {year} {2020})}\BibitemShut
  {NoStop}%
\bibitem [{\citenamefont {Li}\ \emph {et~al.}(2020)\citenamefont {Li},
  \citenamefont {Cai}, \citenamefont {Wang}, \citenamefont {Li}, \citenamefont
  {Yuan},\ and\ \citenamefont {Li}}]{LiLi2020}%
  \BibitemOpen
  \bibfield  {author} {\bibinfo {author} {\bibfnamefont {Y.}~\bibnamefont
  {Li}}, \bibinfo {author} {\bibfnamefont {H.}~\bibnamefont {Cai}}, \bibinfo
  {author} {\bibfnamefont {D.-w.}\ \bibnamefont {Wang}}, \bibinfo {author}
  {\bibfnamefont {L.}~\bibnamefont {Li}}, \bibinfo {author} {\bibfnamefont
  {J.}~\bibnamefont {Yuan}}, \ and\ \bibinfo {author} {\bibfnamefont
  {W.}~\bibnamefont {Li}},\ }\emph {Many-Body Chiral Edge Currents and Sliding
  Phases of Atomic Spin Waves in Momentum-Space Lattice},\ \href {\doibase
  10.1103/PhysRevLett.124.140401} {\bibfield  {journal} {\bibinfo  {journal}
  {Phys. Rev. Lett.}\ }\textbf {\bibinfo {volume} {124}},\ \bibinfo {pages}
  {140401} (\bibinfo {year} {2020})}\BibitemShut {NoStop}%
\bibitem [{\citenamefont {Singha~Roy}\ \emph {et~al.}(2022)\citenamefont
  {Singha~Roy}, \citenamefont {Carl},\ and\ \citenamefont
  {Hauke}}]{RoyHauke2022}%
  \BibitemOpen
  \bibfield  {author} {\bibinfo {author} {\bibfnamefont {S.}~\bibnamefont
  {Singha~Roy}}, \bibinfo {author} {\bibfnamefont {L.}~\bibnamefont {Carl}}, \
  and\ \bibinfo {author} {\bibfnamefont {P.}~\bibnamefont {Hauke}},\ }\emph
  {Genuine multipartite entanglement in a one-dimensional Bose-Hubbard model
  with frustrated hopping},\ \href {\doibase 10.1103/PhysRevB.106.195158}
  {\bibfield  {journal} {\bibinfo  {journal} {Phys. Rev. B}\ }\textbf {\bibinfo
  {volume} {106}},\ \bibinfo {pages} {195158} (\bibinfo {year}
  {2022})}\BibitemShut {NoStop}%
\bibitem [{\citenamefont {Halati}\ and\ \citenamefont
  {Giamarchi}(2023)}]{HalatiGiamarchi2023}%
  \BibitemOpen
  \bibfield  {author} {\bibinfo {author} {\bibfnamefont {C.-M.}\ \bibnamefont
  {Halati}}\ and\ \bibinfo {author} {\bibfnamefont {T.}~\bibnamefont
  {Giamarchi}},\ }\emph {Bose-Hubbard triangular ladder in an artificial gauge
  field},\ \href {\doibase 10.1103/PhysRevResearch.5.013126} {\bibfield
  {journal} {\bibinfo  {journal} {Phys. Rev. Res.}\ }\textbf {\bibinfo {volume}
  {5}},\ \bibinfo {pages} {013126} (\bibinfo {year} {2023})}\BibitemShut
  {NoStop}%
\bibitem [{\citenamefont {Barbiero}\ \emph {et~al.}(2023)\citenamefont
  {Barbiero}, \citenamefont {Cabedo}, \citenamefont {Lewenstein}, \citenamefont
  {Tarruell},\ and\ \citenamefont {Celi}}]{BarbieroCeli2023}%
  \BibitemOpen
  \bibfield  {author} {\bibinfo {author} {\bibfnamefont {L.}~\bibnamefont
  {Barbiero}}, \bibinfo {author} {\bibfnamefont {J.}~\bibnamefont {Cabedo}},
  \bibinfo {author} {\bibfnamefont {M.}~\bibnamefont {Lewenstein}}, \bibinfo
  {author} {\bibfnamefont {L.}~\bibnamefont {Tarruell}}, \ and\ \bibinfo
  {author} {\bibfnamefont {A.}~\bibnamefont {Celi}},\ }\emph {Frustrated
  magnets without geometrical frustration in bosonic flux ladders},\ \href
  {\doibase 10.1103/PhysRevResearch.5.L042008} {\bibfield  {journal} {\bibinfo
  {journal} {Phys. Rev. Res.}\ }\textbf {\bibinfo {volume} {5}},\ \bibinfo
  {pages} {L042008} (\bibinfo {year} {2023})}\BibitemShut {NoStop}%
\bibitem [{\citenamefont {Beradze}\ and\ \citenamefont
  {Nersesyan}(2023)}]{BeradzeNersesyan2023}%
  \BibitemOpen
  \bibfield  {author} {\bibinfo {author} {\bibfnamefont {B.}~\bibnamefont
  {Beradze}}\ and\ \bibinfo {author} {\bibfnamefont {A.}~\bibnamefont
  {Nersesyan}},\ }\emph {Spectrum, Lifshitz transitions and orbital current in
  frustrated fermionic ladders with a uniform flux},\ \href {\doibase
  10.1140/epjb/s10051-022-00472-0} {\bibfield  {journal} {\bibinfo  {journal}
  {The European Physical Journal B}\ }\textbf {\bibinfo {volume} {96}},\
  \bibinfo {pages} {2} (\bibinfo {year} {2023})}\BibitemShut {NoStop}%
\bibitem [{\citenamefont {Beradze}\ \emph {et~al.}(2023)\citenamefont
  {Beradze}, \citenamefont {Tsitsishvili}, \citenamefont {Tirrito},
  \citenamefont {Dalmonte}, \citenamefont {Chanda},\ and\ \citenamefont
  {Nersesyan}}]{BeradzeNersesyan2023b}%
  \BibitemOpen
  \bibfield  {author} {\bibinfo {author} {\bibfnamefont {B.}~\bibnamefont
  {Beradze}}, \bibinfo {author} {\bibfnamefont {M.}~\bibnamefont
  {Tsitsishvili}}, \bibinfo {author} {\bibfnamefont {E.}~\bibnamefont
  {Tirrito}}, \bibinfo {author} {\bibfnamefont {M.}~\bibnamefont {Dalmonte}},
  \bibinfo {author} {\bibfnamefont {T.}~\bibnamefont {Chanda}}, \ and\ \bibinfo
  {author} {\bibfnamefont {A.}~\bibnamefont {Nersesyan}},\ }\emph {Emergence of
  non-Abelian SU(2) invariance in Abelian frustrated fermionic ladders},\ \href
  {\doibase 10.1103/PhysRevB.108.075146} {\bibfield  {journal} {\bibinfo
  {journal} {Phys. Rev. B}\ }\textbf {\bibinfo {volume} {108}},\ \bibinfo
  {pages} {075146} (\bibinfo {year} {2023})}\BibitemShut {NoStop}%
\bibitem [{\citenamefont {Baldelli}\ \emph {et~al.}(2024)\citenamefont
  {Baldelli}, \citenamefont {Cabrera}, \citenamefont {Juli\`a-Farr\'e},
  \citenamefont {Aidelsburger},\ and\ \citenamefont
  {Barbiero}}]{BaldelliBarbiero2024}%
  \BibitemOpen
  \bibfield  {author} {\bibinfo {author} {\bibfnamefont {N.}~\bibnamefont
  {Baldelli}}, \bibinfo {author} {\bibfnamefont {C.~R.}\ \bibnamefont
  {Cabrera}}, \bibinfo {author} {\bibfnamefont {S.}~\bibnamefont
  {Juli\`a-Farr\'e}}, \bibinfo {author} {\bibfnamefont {M.}~\bibnamefont
  {Aidelsburger}}, \ and\ \bibinfo {author} {\bibfnamefont {L.}~\bibnamefont
  {Barbiero}},\ }\emph {Frustrated Extended Bose-Hubbard Model and Deconfined
  Quantum Critical Points with Optical Lattices at the Antimagic Wavelength},\
  \href {\doibase 10.1103/PhysRevLett.132.153401} {\bibfield  {journal}
  {\bibinfo  {journal} {Phys. Rev. Lett.}\ }\textbf {\bibinfo {volume} {132}},\
  \bibinfo {pages} {153401} (\bibinfo {year} {2024})}\BibitemShut {NoStop}%
\bibitem [{\citenamefont {Le\'on}\ \emph {et~al.}(2008)\citenamefont {Le\'on},
  \citenamefont {Berthod}, \citenamefont {Giamarchi},\ and\ \citenamefont
  {Millis}}]{LeonMillis2008}%
  \BibitemOpen
  \bibfield  {author} {\bibinfo {author} {\bibfnamefont {G.}~\bibnamefont
  {Le\'on}}, \bibinfo {author} {\bibfnamefont {C.}~\bibnamefont {Berthod}},
  \bibinfo {author} {\bibfnamefont {T.}~\bibnamefont {Giamarchi}}, \ and\
  \bibinfo {author} {\bibfnamefont {A.~J.}\ \bibnamefont {Millis}},\ }\emph
  {Hall effect on the triangular lattice},\ \href {\doibase
  10.1103/PhysRevB.78.085105} {\bibfield  {journal} {\bibinfo  {journal} {Phys.
  Rev. B}\ }\textbf {\bibinfo {volume} {78}},\ \bibinfo {pages} {085105}
  (\bibinfo {year} {2008})}\BibitemShut {NoStop}%
\bibitem [{\citenamefont {White}(1992)}]{White1992}%
  \BibitemOpen
  \bibfield  {author} {\bibinfo {author} {\bibfnamefont {S.~R.}\ \bibnamefont
  {White}},\ }\emph {Density matrix formulation for quantum renormalization
  groups},\ \href {\doibase 10.1103/PhysRevLett.69.2863} {\bibfield  {journal}
  {\bibinfo  {journal} {Phys. Rev. Lett.}\ }\textbf {\bibinfo {volume} {69}},\
  \bibinfo {pages} {2863} (\bibinfo {year} {1992})}\BibitemShut {NoStop}%
\bibitem [{\citenamefont {Schollw\"ock}(2005)}]{Schollwoeck2005}%
  \BibitemOpen
  \bibfield  {author} {\bibinfo {author} {\bibfnamefont {U.}~\bibnamefont
  {Schollw\"ock}},\ }\emph {The density-matrix renormalization group},\ \href
  {\doibase 10.1103/RevModPhys.77.259} {\bibfield  {journal} {\bibinfo
  {journal} {Rev. Mod. Phys.}\ }\textbf {\bibinfo {volume} {77}},\ \bibinfo
  {pages} {259} (\bibinfo {year} {2005})}\BibitemShut {NoStop}%
\bibitem [{\citenamefont {Schollw{\"o}ck}(2011)}]{Schollwoeck2011}%
  \BibitemOpen
  \bibfield  {author} {\bibinfo {author} {\bibfnamefont {U.}~\bibnamefont
  {Schollw{\"o}ck}},\ }\emph {The density-matrix renormalization group in the
  age of matrix product states},\ \href {\doibase
  http://dx.doi.org/10.1016/j.aop.2010.09.012} {\bibfield  {journal} {\bibinfo
  {journal} {Annals of Physics}\ }\textbf {\bibinfo {volume} {326}},\ \bibinfo
  {pages} {96 } (\bibinfo {year} {2011})}\BibitemShut {NoStop}%
\bibitem [{\citenamefont {Hallberg}(2006)}]{Hallberg2006}%
  \BibitemOpen
  \bibfield  {author} {\bibinfo {author} {\bibfnamefont {K.~A.}\ \bibnamefont
  {Hallberg}},\ }\emph {New trends in density matrix renormalization},\ \href
  {\doibase 10.1080/00018730600766432} {\bibfield  {journal} {\bibinfo
  {journal} {Advances in Physics}\ }\textbf {\bibinfo {volume} {55}},\ \bibinfo
  {pages} {477} (\bibinfo {year} {2006})}\BibitemShut {NoStop}%
\bibitem [{\citenamefont {Jeckelmann}(2002)}]{Jeckelmann2002}%
  \BibitemOpen
  \bibfield  {author} {\bibinfo {author} {\bibfnamefont {E.}~\bibnamefont
  {Jeckelmann}},\ }\emph {Dynamical density-matrix renormalization-group
  method},\ \href {\doibase 10.1103/PhysRevB.66.045114} {\bibfield  {journal}
  {\bibinfo  {journal} {Phys. Rev. B}\ }\textbf {\bibinfo {volume} {66}},\
  \bibinfo {pages} {045114} (\bibinfo {year} {2002})}\BibitemShut {NoStop}%
\bibitem [{\citenamefont {Fishman}\ \emph {et~al.}(2022)\citenamefont
  {Fishman}, \citenamefont {White},\ and\ \citenamefont
  {Stoudenmire}}]{FishmanStoudenmire2020}%
  \BibitemOpen
  \bibfield  {author} {\bibinfo {author} {\bibfnamefont {M.}~\bibnamefont
  {Fishman}}, \bibinfo {author} {\bibfnamefont {S.~R.}\ \bibnamefont {White}},
  \ and\ \bibinfo {author} {\bibfnamefont {E.~M.}\ \bibnamefont
  {Stoudenmire}},\ }\emph {{The ITensor Software Library for Tensor Network
  Calculations}},\ \href {\doibase 10.21468/SciPostPhysCodeb.4} {\bibfield
  {journal} {\bibinfo  {journal} {SciPost Phys. Codebases}\ ,\ \bibinfo {pages}
  {4}} (\bibinfo {year} {2022})}\BibitemShut {NoStop}%
\bibitem [{\citenamefont {Daley}\ \emph {et~al.}(2004)\citenamefont {Daley},
  \citenamefont {Kollath}, \citenamefont {Schollwöck},\ and\ \citenamefont
  {Vidal}}]{DaleyVidal2004}%
  \BibitemOpen
  \bibfield  {author} {\bibinfo {author} {\bibfnamefont {A.~J.}\ \bibnamefont
  {Daley}}, \bibinfo {author} {\bibfnamefont {C.}~\bibnamefont {Kollath}},
  \bibinfo {author} {\bibfnamefont {U.}~\bibnamefont {Schollwöck}}, \ and\
  \bibinfo {author} {\bibfnamefont {G.}~\bibnamefont {Vidal}},\ }\emph
  {Time-dependent density-matrix renormalization-group using adaptive effective
  Hilbert spaces},\ \href {\doibase 10.1088/1742-5468/2004/04/P04005}
  {\bibfield  {journal} {\bibinfo  {journal} {Journal of Statistical Mechanics:
  Theory and Experiment}\ }\textbf {\bibinfo {volume} {2004}},\ \bibinfo
  {pages} {P04005} (\bibinfo {year} {2004})}\BibitemShut {NoStop}%
\bibitem [{\citenamefont {White}\ and\ \citenamefont
  {Feiguin}(2004)}]{WhiteFeiguin2004}%
  \BibitemOpen
  \bibfield  {author} {\bibinfo {author} {\bibfnamefont {S.~R.}\ \bibnamefont
  {White}}\ and\ \bibinfo {author} {\bibfnamefont {A.~E.}\ \bibnamefont
  {Feiguin}},\ }\emph {Real-Time Evolution Using the Density Matrix
  Renormalization Group},\ \href {\doibase 10.1103/PhysRevLett.93.076401}
  {\bibfield  {journal} {\bibinfo  {journal} {Phys. Rev. Lett.}\ }\textbf
  {\bibinfo {volume} {93}},\ \bibinfo {pages} {076401} (\bibinfo {year}
  {2004})}\BibitemShut {NoStop}%
\bibitem [{\citenamefont {Vidal}\ \emph {et~al.}(2003)\citenamefont {Vidal},
  \citenamefont {Latorre}, \citenamefont {Rico},\ and\ \citenamefont
  {Kitaev}}]{VidalKitaev2003}%
  \BibitemOpen
  \bibfield  {author} {\bibinfo {author} {\bibfnamefont {G.}~\bibnamefont
  {Vidal}}, \bibinfo {author} {\bibfnamefont {J.~I.}\ \bibnamefont {Latorre}},
  \bibinfo {author} {\bibfnamefont {E.}~\bibnamefont {Rico}}, \ and\ \bibinfo
  {author} {\bibfnamefont {A.}~\bibnamefont {Kitaev}},\ }\emph {Entanglement in
  Quantum Critical Phenomena},\ \href {\doibase 10.1103/PhysRevLett.90.227902}
  {\bibfield  {journal} {\bibinfo  {journal} {Phys. Rev. Lett.}\ }\textbf
  {\bibinfo {volume} {90}},\ \bibinfo {pages} {227902} (\bibinfo {year}
  {2003})}\BibitemShut {NoStop}%
\bibitem [{\citenamefont {Calabrese}\ and\ \citenamefont
  {Cardy}(2004)}]{CalabreseCardy2004}%
  \BibitemOpen
  \bibfield  {author} {\bibinfo {author} {\bibfnamefont {P.}~\bibnamefont
  {Calabrese}}\ and\ \bibinfo {author} {\bibfnamefont {J.}~\bibnamefont
  {Cardy}},\ }\emph {Entanglement entropy and quantum field theory},\ \href
  {\doibase 10.1088/1742-5468/2004/06/P06002} {\bibfield  {journal} {\bibinfo
  {journal} {Journal of Statistical Mechanics: Theory and Experiment}\ }\textbf
  {\bibinfo {volume} {2004}},\ \bibinfo {pages} {P06002} (\bibinfo {year}
  {2004})}\BibitemShut {NoStop}%
\bibitem [{\citenamefont {Holzhey}\ \emph {et~al.}(1994)\citenamefont
  {Holzhey}, \citenamefont {Larsen},\ and\ \citenamefont
  {Wilczek}}]{HolzeyWilczek1994}%
  \BibitemOpen
  \bibfield  {author} {\bibinfo {author} {\bibfnamefont {C.}~\bibnamefont
  {Holzhey}}, \bibinfo {author} {\bibfnamefont {F.}~\bibnamefont {Larsen}}, \
  and\ \bibinfo {author} {\bibfnamefont {F.}~\bibnamefont {Wilczek}},\ }\emph
  {Geometric and renormalized entropy in conformal field theory},\ \href
  {\doibase https://doi.org/10.1016/0550-3213(94)90402-2} {\bibfield  {journal}
  {\bibinfo  {journal} {Nuclear Physics B}\ }\textbf {\bibinfo {volume}
  {424}},\ \bibinfo {pages} {443} (\bibinfo {year} {1994})}\BibitemShut
  {NoStop}%
\bibitem [{\citenamefont {Huber}\ and\ \citenamefont
  {Lindner}(2011)}]{HuberLindner2011}%
  \BibitemOpen
  \bibfield  {author} {\bibinfo {author} {\bibfnamefont {S.~D.}\ \bibnamefont
  {Huber}}\ and\ \bibinfo {author} {\bibfnamefont {N.~H.}\ \bibnamefont
  {Lindner}},\ }\emph {Topological transitions for lattice bosons in a magnetic
  field},\ \href {\doibase 10.1073/pnas.1110813108} {\bibfield  {journal}
  {\bibinfo  {journal} {Proceedings of the National Academy of Sciences}\
  }\textbf {\bibinfo {volume} {108}},\ \bibinfo {pages} {19925} (\bibinfo
  {year} {2011})}\BibitemShut {NoStop}%
\bibitem [{\citenamefont {Berg}\ \emph {et~al.}(2015)\citenamefont {Berg},
  \citenamefont {Huber},\ and\ \citenamefont {Lindner}}]{BergLindner2015}%
  \BibitemOpen
  \bibfield  {author} {\bibinfo {author} {\bibfnamefont {E.}~\bibnamefont
  {Berg}}, \bibinfo {author} {\bibfnamefont {S.~D.}\ \bibnamefont {Huber}}, \
  and\ \bibinfo {author} {\bibfnamefont {N.~H.}\ \bibnamefont {Lindner}},\
  }\emph {Sign reversal of the Hall response in a crystalline superconductor},\
  \href {\doibase 10.1103/PhysRevB.91.024507} {\bibfield  {journal} {\bibinfo
  {journal} {Phys. Rev. B}\ }\textbf {\bibinfo {volume} {91}},\ \bibinfo
  {pages} {024507} (\bibinfo {year} {2015})}\BibitemShut {NoStop}%
\bibitem [{\citenamefont {Li}\ \emph {et~al.}(2023)\citenamefont {Li},
  \citenamefont {Du}, \citenamefont {Wang}, \citenamefont {Liang},
  \citenamefont {Xiao}, \citenamefont {Yi}, \citenamefont {Ma},\ and\
  \citenamefont {Jia}}]{LiJia2023}%
  \BibitemOpen
  \bibfield  {author} {\bibinfo {author} {\bibfnamefont {Y.}~\bibnamefont
  {Li}}, \bibinfo {author} {\bibfnamefont {H.}~\bibnamefont {Du}}, \bibinfo
  {author} {\bibfnamefont {Y.}~\bibnamefont {Wang}}, \bibinfo {author}
  {\bibfnamefont {J.}~\bibnamefont {Liang}}, \bibinfo {author} {\bibfnamefont
  {L.}~\bibnamefont {Xiao}}, \bibinfo {author} {\bibfnamefont {W.}~\bibnamefont
  {Yi}}, \bibinfo {author} {\bibfnamefont {J.}~\bibnamefont {Ma}}, \ and\
  \bibinfo {author} {\bibfnamefont {S.}~\bibnamefont {Jia}},\ }\emph
  {Observation of frustrated chiral dynamics in an interacting triangular flux
  ladder},\ \href {\doibase 10.1038/s41467-023-43204-3} {\bibfield  {journal}
  {\bibinfo  {journal} {Nature Communications}\ }\textbf {\bibinfo {volume}
  {14}},\ \bibinfo {pages} {7560} (\bibinfo {year} {2023})}\BibitemShut
  {NoStop}%
\bibitem [{\citenamefont {Becker}\ \emph {et~al.}(2010)\citenamefont {Becker},
  \citenamefont {Soltan-Panahi}, \citenamefont {Kronjäger}, \citenamefont
  {Dörscher}, \citenamefont {Bongs},\ and\ \citenamefont
  {Sengstock}}]{BeckerSengstock2010}%
  \BibitemOpen
  \bibfield  {author} {\bibinfo {author} {\bibfnamefont {C.}~\bibnamefont
  {Becker}}, \bibinfo {author} {\bibfnamefont {P.}~\bibnamefont
  {Soltan-Panahi}}, \bibinfo {author} {\bibfnamefont {J.}~\bibnamefont
  {Kronjäger}}, \bibinfo {author} {\bibfnamefont {S.}~\bibnamefont
  {Dörscher}}, \bibinfo {author} {\bibfnamefont {K.}~\bibnamefont {Bongs}}, \
  and\ \bibinfo {author} {\bibfnamefont {K.}~\bibnamefont {Sengstock}},\ }\emph
  {Ultracold quantum gases in triangular optical lattices},\ \href {\doibase
  10.1088/1367-2630/12/6/065025} {\bibfield  {journal} {\bibinfo  {journal}
  {New Journal of Physics}\ }\textbf {\bibinfo {volume} {12}},\ \bibinfo
  {pages} {065025} (\bibinfo {year} {2010})}\BibitemShut {NoStop}%
\bibitem [{\citenamefont {Struck}\ \emph {et~al.}(2011)\citenamefont {Struck},
  \citenamefont {Ölschläger}, \citenamefont {Targat}, \citenamefont
  {Soltan-Panahi}, \citenamefont {Eckardt}, \citenamefont {Lewenstein},
  \citenamefont {Windpassinger},\ and\ \citenamefont
  {Sengstock}}]{StruckSengstock2011}%
  \BibitemOpen
  \bibfield  {author} {\bibinfo {author} {\bibfnamefont {J.}~\bibnamefont
  {Struck}}, \bibinfo {author} {\bibfnamefont {C.}~\bibnamefont
  {Ölschläger}}, \bibinfo {author} {\bibfnamefont {R.~L.}\ \bibnamefont
  {Targat}}, \bibinfo {author} {\bibfnamefont {P.}~\bibnamefont
  {Soltan-Panahi}}, \bibinfo {author} {\bibfnamefont {A.}~\bibnamefont
  {Eckardt}}, \bibinfo {author} {\bibfnamefont {M.}~\bibnamefont {Lewenstein}},
  \bibinfo {author} {\bibfnamefont {P.}~\bibnamefont {Windpassinger}}, \ and\
  \bibinfo {author} {\bibfnamefont {K.}~\bibnamefont {Sengstock}},\ }\emph
  {Quantum Simulation of Frustrated Classical Magnetism in Triangular Optical
  Lattices},\ \href {\doibase 10.1126/science.1207239} {\bibfield  {journal}
  {\bibinfo  {journal} {Science}\ }\textbf {\bibinfo {volume} {333}},\ \bibinfo
  {pages} {996} (\bibinfo {year} {2011})}\BibitemShut {NoStop}%
\bibitem [{\citenamefont {Yang}\ \emph {et~al.}(2021)\citenamefont {Yang},
  \citenamefont {Liu}, \citenamefont {Mongkolkiattichai},\ and\ \citenamefont
  {Schauss}}]{YangSchauss2021}%
  \BibitemOpen
  \bibfield  {author} {\bibinfo {author} {\bibfnamefont {J.}~\bibnamefont
  {Yang}}, \bibinfo {author} {\bibfnamefont {L.}~\bibnamefont {Liu}}, \bibinfo
  {author} {\bibfnamefont {J.}~\bibnamefont {Mongkolkiattichai}}, \ and\
  \bibinfo {author} {\bibfnamefont {P.}~\bibnamefont {Schauss}},\ }\emph
  {Site-Resolved Imaging of Ultracold Fermions in a Triangular-Lattice Quantum
  Gas Microscope},\ \href {\doibase 10.1103/PRXQuantum.2.020344} {\bibfield
  {journal} {\bibinfo  {journal} {PRX Quantum}\ }\textbf {\bibinfo {volume}
  {2}},\ \bibinfo {pages} {020344} (\bibinfo {year} {2021})}\BibitemShut
  {NoStop}%
\bibitem [{\citenamefont {Mongkolkiattichai}\ \emph {et~al.}(2023)\citenamefont
  {Mongkolkiattichai}, \citenamefont {Liu}, \citenamefont {Garwood},
  \citenamefont {Yang},\ and\ \citenamefont
  {Schauss}}]{MongkolkiattichaiSchauss2023}%
  \BibitemOpen
  \bibfield  {author} {\bibinfo {author} {\bibfnamefont {J.}~\bibnamefont
  {Mongkolkiattichai}}, \bibinfo {author} {\bibfnamefont {L.}~\bibnamefont
  {Liu}}, \bibinfo {author} {\bibfnamefont {D.}~\bibnamefont {Garwood}},
  \bibinfo {author} {\bibfnamefont {J.}~\bibnamefont {Yang}}, \ and\ \bibinfo
  {author} {\bibfnamefont {P.}~\bibnamefont {Schauss}},\ }\emph {Quantum gas
  microscopy of fermionic triangular-lattice Mott insulators},\ \href {\doibase
  10.1103/PhysRevA.108.L061301} {\bibfield  {journal} {\bibinfo  {journal}
  {Phys. Rev. A}\ }\textbf {\bibinfo {volume} {108}},\ \bibinfo {pages}
  {L061301} (\bibinfo {year} {2023})}\BibitemShut {NoStop}%
\bibitem [{\citenamefont {Xu}\ \emph {et~al.}(2023)\citenamefont {Xu},
  \citenamefont {Kendrick}, \citenamefont {Kale}, \citenamefont {Gang},
  \citenamefont {Ji}, \citenamefont {Scalettar}, \citenamefont {Lebrat},\ and\
  \citenamefont {Greiner}}]{XuGreiner2023}%
  \BibitemOpen
  \bibfield  {author} {\bibinfo {author} {\bibfnamefont {M.}~\bibnamefont
  {Xu}}, \bibinfo {author} {\bibfnamefont {L.~H.}\ \bibnamefont {Kendrick}},
  \bibinfo {author} {\bibfnamefont {A.}~\bibnamefont {Kale}}, \bibinfo {author}
  {\bibfnamefont {Y.}~\bibnamefont {Gang}}, \bibinfo {author} {\bibfnamefont
  {G.}~\bibnamefont {Ji}}, \bibinfo {author} {\bibfnamefont {R.~T.}\
  \bibnamefont {Scalettar}}, \bibinfo {author} {\bibfnamefont {M.}~\bibnamefont
  {Lebrat}}, \ and\ \bibinfo {author} {\bibfnamefont {M.}~\bibnamefont
  {Greiner}},\ }\emph {Frustration- and doping-induced magnetism in a
  Fermi--Hubbard simulator},\ \href {\doibase 10.1038/s41586-023-06280-5}
  {\bibfield  {journal} {\bibinfo  {journal} {Nature}\ }\textbf {\bibinfo
  {volume} {620}},\ \bibinfo {pages} {971} (\bibinfo {year}
  {2023})}\BibitemShut {NoStop}%
\bibitem [{\citenamefont {Lebrat}\ \emph {et~al.}(2024)\citenamefont {Lebrat},
  \citenamefont {Xu}, \citenamefont {Kendrick}, \citenamefont {Kale},
  \citenamefont {Gang}, \citenamefont {Seetharaman}, \citenamefont {Morera},
  \citenamefont {Khatami}, \citenamefont {Demler},\ and\ \citenamefont
  {Greiner}}]{LebratGreiner2024}%
  \BibitemOpen
  \bibfield  {author} {\bibinfo {author} {\bibfnamefont {M.}~\bibnamefont
  {Lebrat}}, \bibinfo {author} {\bibfnamefont {M.}~\bibnamefont {Xu}}, \bibinfo
  {author} {\bibfnamefont {L.~H.}\ \bibnamefont {Kendrick}}, \bibinfo {author}
  {\bibfnamefont {A.}~\bibnamefont {Kale}}, \bibinfo {author} {\bibfnamefont
  {Y.}~\bibnamefont {Gang}}, \bibinfo {author} {\bibfnamefont {P.}~\bibnamefont
  {Seetharaman}}, \bibinfo {author} {\bibfnamefont {I.}~\bibnamefont {Morera}},
  \bibinfo {author} {\bibfnamefont {E.}~\bibnamefont {Khatami}}, \bibinfo
  {author} {\bibfnamefont {E.}~\bibnamefont {Demler}}, \ and\ \bibinfo {author}
  {\bibfnamefont {M.}~\bibnamefont {Greiner}},\ }\emph {Observation of Nagaoka
  polarons in a Fermi--Hubbard quantum simulator},\ \href {\doibase
  10.1038/s41586-024-07272-9} {\bibfield  {journal} {\bibinfo  {journal}
  {Nature}\ }\textbf {\bibinfo {volume} {629}},\ \bibinfo {pages} {317}
  (\bibinfo {year} {2024})}\BibitemShut {NoStop}%
\bibitem [{\citenamefont {Prichard}\ \emph {et~al.}(2024)\citenamefont
  {Prichard}, \citenamefont {Spar}, \citenamefont {Morera}, \citenamefont
  {Demler}, \citenamefont {Yan},\ and\ \citenamefont
  {Bakr}}]{PrichardBakr2024}%
  \BibitemOpen
  \bibfield  {author} {\bibinfo {author} {\bibfnamefont {M.~L.}\ \bibnamefont
  {Prichard}}, \bibinfo {author} {\bibfnamefont {B.~M.}\ \bibnamefont {Spar}},
  \bibinfo {author} {\bibfnamefont {I.}~\bibnamefont {Morera}}, \bibinfo
  {author} {\bibfnamefont {E.}~\bibnamefont {Demler}}, \bibinfo {author}
  {\bibfnamefont {Z.~Z.}\ \bibnamefont {Yan}}, \ and\ \bibinfo {author}
  {\bibfnamefont {W.~S.}\ \bibnamefont {Bakr}},\ }\emph {Directly imaging spin
  polarons in a kinetically frustrated Hubbard system},\ \href {\doibase
  10.1038/s41586-024-07356-6} {\bibfield  {journal} {\bibinfo  {journal}
  {Nature}\ }\textbf {\bibinfo {volume} {629}},\ \bibinfo {pages} {323}
  (\bibinfo {year} {2024})}\BibitemShut {NoStop}%
\end{thebibliography}
\end{document}